\documentclass[sigconf]{acmart}

\usepackage{multirow}
\usepackage{url}
\usepackage{makecell}
\usepackage{subcaption}

\newcommand{\add}[1]{\textcolor{black}{#1}}
\AtBeginDocument{%
  }

\copyrightyear{2026}
\acmYear{2026}
\setcopyright{cc}
\setcctype{by}
\acmConference[CHI '26]{Proceedings of the 2026 CHI Conference on Human Factors in Computing Systems}{April 13--17, 2026}{Barcelona, Spain}
\acmBooktitle{Proceedings of the 2026 CHI Conference on Human Factors in Computing Systems (CHI '26), April 13--17, 2026, Barcelona, Spain}
\acmDOI{10.1145/3772318.3791378}
\acmISBN{979-8-4007-2278-3/2026/04}

\begin{document}

\title{AIDED: Augmenting Interior Design with Human Experience Data for Designer–AI Co-Design}

\author{Yang Chen Lin}
\authornote{Both authors contributed equally to this research.}
\orcid{0000-0002-2477-4110}
\affiliation{%
  \department{Department of Computer Science}
  \institution{National Tsing Hua University}
  \city{Hsinchu}
  \country{Taiwan}
}
\email{yangchenlin@gapp.nthu.edu.tw}

\author{Chen-Ying Chien}
\authornotemark[1]
\orcid{0009-0005-8121-3925}
\affiliation{%
  \department{Department of Computer Science}
  \institution{National Tsing Hua University}
  \city{Hsinchu}
  \country{Taiwan}
}
\email{chenyingchien1021@gmail.com}

\author{Kai-Hsin Hou}
\orcid{0009-0008-7386-2636}
\affiliation{%
  \department{Department of Computer Science}
  \institution{National Tsing Hua University}
  \city{Hsinchu}
  \country{Taiwan}
}
\email{happyho0906@gmail.com}

\author{Hung-Yu Chen}
\orcid{0009-0008-7386-2636}
\affiliation{%
    Freelancer
    \city{Taichung}
    \country{Taiwan}
}
\email{hungyuchens@gmail.com}

\author{Po-Chih Kuo}
\orcid{0000-0003-4020-3147}
\affiliation{%
  \department{Department of Computer Science}
  \institution{National Tsing Hua University}
  \city{Hsinchu}
  \country{Taiwan}
}
\email{kuopc@cs.nthu.edu.tw}

\renewcommand{\shortauthors}{Lin et al.}

\begin{teaserfigure}
  \centering
  \includegraphics[width=0.90\textwidth]{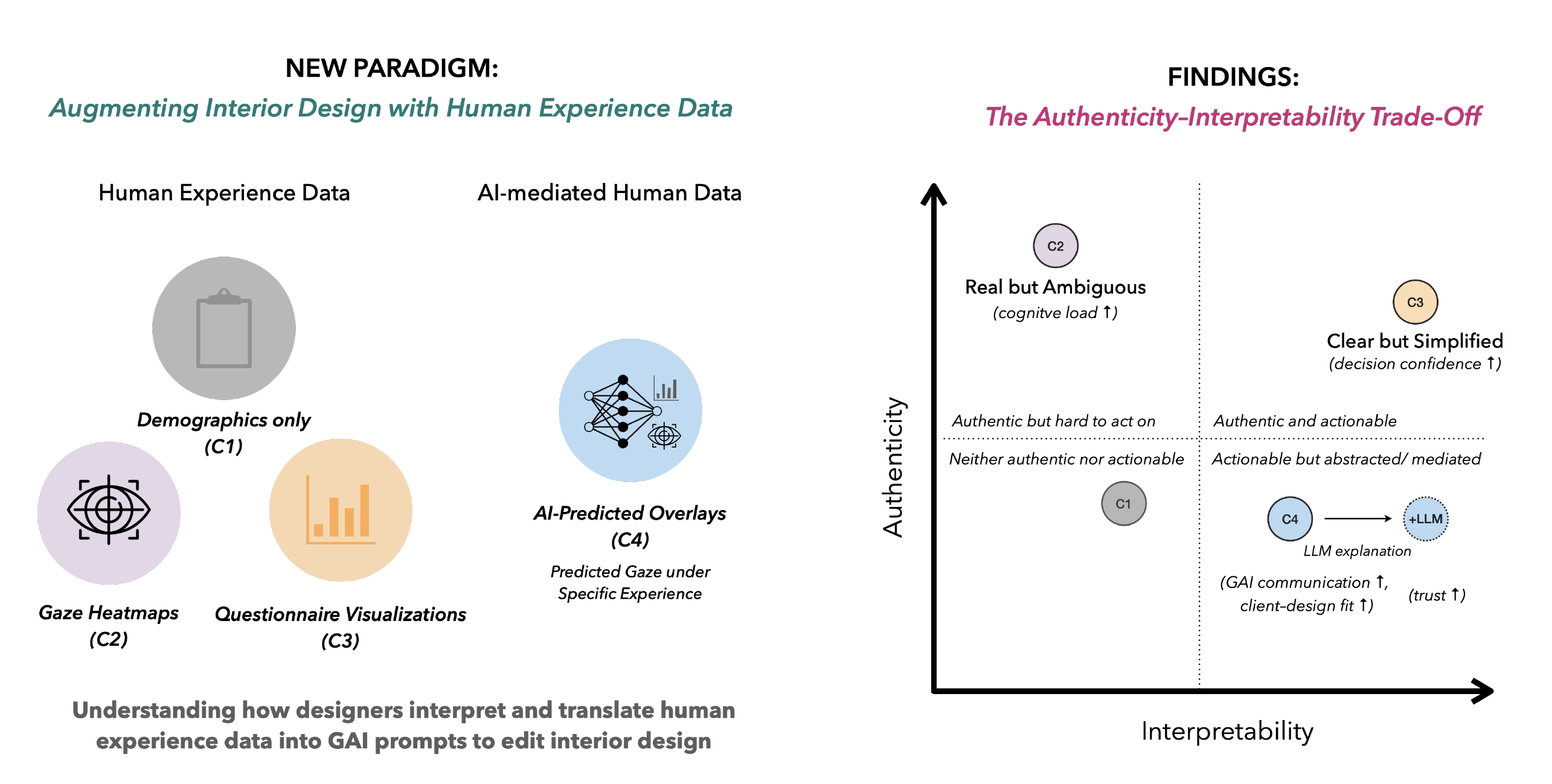}
  \caption{\add{AIDED integrates human experience and AI-mediated data into the designer–GAI co-design process. Across four modalities (C1–C4), we identify an authenticity–interpretability trade-off: authentic gaze heatmaps (C2) increase cognitive load, whereas questionnaire visualizations (C3) support design decisions, and AI-predicted overlays (C4) enhance collaboration but require LLM explanations to establish trust and actionability.}}
  \label{fig:teaser}
  \Description{The figure summarizes the AIDED concepts and its key findings. The left panel illustrates four client-information modalities: basic demographics (C1), gaze heatmaps (C2), questionnaire visualizations (C3), and AI-mediated overlays (C4), grouped into human experience data and AI-mediated data. The right panel maps these modalities onto an authenticity–interpretability space (authenticity on the vertical axis; interpretability on the horizontal axis). Gaze heatmaps are highly authentic but low in interpretability; questionnaire visualizations are more interpretable but less detailed; demographics are low on both; and AI-mediated overlays are more actionable and interpretable, particularly with LLM explanations. The diagram highlights an authenticity–interpretability trade-off.}
\end{teaserfigure}

\begin{abstract}
Interior design often struggles to capture the subtleties of client experiences, leaving gaps between what clients feel and what designers can act upon. We present AIDED, a designer–AI co-design workflow that integrates multimodal client data into generative AI (GAI) design processes. In a within-subjects study with twelve professional designers, we compared four modalities: baseline briefs, gaze heatmaps, questionnaires visualizations, and AI-predicted overlays. Results show that questionnaire data were trusted, creativity-enhancing, and satisfying; gaze heatmaps increased cognitive load; and AI-predicted overlays improved GAI communication but required natural language mediation to earn trust. Interviews confirmed that an authenticity–interpretability trade-off is central to balancing client voices with professional control. Our contributions are: (1) a system that incorporates experiential client signals into GAI design workflows, (2) empirical evidence of how different modalities affect design outcomes, and (3) implications for future AI tools that support human–data interaction in creative practice.

%last version

\end{abstract}




\maketitle

\section{Introduction}

Interior design involves a complex interaction between clients' subjective input and designers' professional expertise. While designers offer specialized knowledge in areas such as aesthetics, ergonomics, and spatial organization, clients provide essential insights derived from their lived experiences, personal preferences, and contextual limitations. However, accurately capturing the more subtle aspects of the client's experience, such as nuanced emotional responses, implicit behavioral patterns, and context-specific requirements, remains a significant challenge when using conventional methods such as mood boards or verbal accounts. Consequently, the representation of \textit{human experience data} \cite{Coburn2020, aseniero2024experiential, lin2025shaping} is often underexploited, increasing the risk that design outcomes may not fully reflect clients' authentic values or provide robust, evidence-based support \cite{lin2025shaping, DesignMindToolkit, CognitiveBIM2022, Sjovall2025NDIX}.

\add{Existing evidence-based spatial design tools have primarily focused on simulating perceptual or physical parameters. Although these systems support professional workflows effectively, they seldom incorporate direct inputs from end-user experiences. Computational tools reliably model physical factors such as ventilation, lighting, and circulation; however, aspects related to client affect, comfort, and perception are frequently omitted from the design process. This omission is significant, given that spatial environments directly influence individuals’ well-being and psychological comfort.} \add{Recent work in human–building interaction shows how everyday playful interactions with technologies and domestic spaces can be translated into spatial play potentials and design implications for architectural change \cite{CetinEr2024PlayPotentials}, further underscoring the importance of lived experience as a driver of spatial transformation beyond conventional performance metrics.}

Recent advancements in generative artificial intelligence (GAI) have introduced powerful tools that enable rapid production and stylistic experimentation. Text-to-image systems (e.g., DALL·E, Midjourney) and emerging image-editing models (e.g., Gemini Nano Banana) can generate diverse design variations within seconds, offering particular value during early ideation phases \cite{RoomDreaming2024, FashioningExpertise2024, SketchVsPrompt2024}. However, these systems typically rely on generic prompts and training data that remain disconnected from the specific human experiences that should anchor meaningful design work \cite{AIArchSurvey2023, AIDataImages2024}.

\add{This disconnect between rich generative capabilities and sparse client representations} creates a paradox: as AI-generated content becomes increasingly sophisticated, it may drift further from the experiential cues most important to clients. These gaps motivate a central practical question: \emph{How might client experience data be integrated into GAI-assisted workflows in ways that support designers’ agency, reasoning, and interpretability rather than undermine them?}

\add{To understand how professional architects and interior designers perceive the emerging role of GAI in their practice, we conducted formative interviews (section~\ref{sec:formative}) and identified three key design insights and corresponding design goals that informed the selection (section~\ref{sec:RationaleforCondtions}) and integration of different modalities of client information in our GAI workflow (section~\ref{sec:system})}.

We then pose the following research questions (RQs):

\begin{itemize}
    \item \textbf{RQ1} — \textit{How do different client-information modalities (basic demographic, gaze, questionnaire, AI-predicted overlays) influence designers’ perceived cognitive effort, trust, creativity, and overall satisfaction when working with GAI?}
    \item \textbf{RQ2} — \textit{How do designers actually engage with these modalities during the design process (e.g., identifying issues, proposing modifications, interacting with GAI), and where do they encounter friction?}
    \item \textbf{RQ3} — \textit{What is the function of AI mediation, specifically through predicted attention mechanisms and explanatory features, in facilitating the translation of client feedback into interpretable and controllable design actions, while maintaining the professional autonomy of designers?}
\end{itemize}

\add{In this study, we treat \textbf{AIDED} (Augmenting Interior Design with Human Experience Data) (Fig.~\ref{fig:system_cover}, see section~\ref{sec:system}) as an exploration of how different representations of client experience data reconfigure designer–AI collaboration in interior design (Fig.~\ref{fig:teaser}). Rather than comparing “with vs.\ without AI”, we systematically vary the \emph{type} and \emph{representation} of client information along an authenticity–interpretability spectrum, and examine how these modalities shape cognitive load, trust and usefulness, communication with GAI, and perceived design quality (see section~\ref{sec:task}).}

\add{Here, we use \emph{authenticity} to refer to how closely a representation preserves the richness and ambiguity of clients’ lived experiences (e.g., raw feedback and gaze traces), and \emph{interpretability} to capture how easily designers can read, trust, and act on that information within fast-paced GAI workflows (e.g., structured questionnaires or AI-generated overlays).}

Through a mixed-method, within-subjects study with professional interior designers, our findings show that the AIDED workflow facilitated designers in integrating client perspectives with generative AI outputs while maintaining professional authority over design decisions. By incorporating multimodal client information, participants were able to identify latent needs such as lifestyle constraints or previously unrecognized preferences that might otherwise have remained undetected, and subsequently translate these insights into tangible design modifications. Across experimental conditions, designers utilizing AIDED not only generated outcomes that more accurately reflected client profiles but also reported enhanced confidence and greater creative flexibility when iterating with generative AI support. Concurrently, the findings reveal a significant trade-off: raw gaze data increased cognitive effort and ambiguity; questionnaire visualizations were perceived as trustworthy and actionable; and AI-predicted attention overlays improved communication with the AI but necessitated accompanying natural language explanations to establish credibility. Collectively, these results underscore how experiential client data can augment co-design workflows when a careful balance is maintained between authenticity and interpretability.

This \add{exploratory} work makes three principal contributions to HCI:

\begin{enumerate}
    \item \textbf{System Contribution:} We propose AIDED, a novel workflow that broadens the range of input data utilized in interior design by incorporating multimodal experiential signals alongside traditional lifestyle and demographic information, thereby enhancing GAI-assisted design processes.
        
    \item \textbf{Theoretical Contribution:} \add{Based on empirical findings across four conditions, we articulate an authenticity–\\interpretability trade-off that characterizes how designers negotiate between preserving client perspectives and ensuring actionable interpretability during generative AI collaboration. This lens extends prior work on human–AI co-creation and interpretability by showing how such tensions emerge specifically in the context of multimodal client data.}
        
    \item \textbf{Design Implications:} By investigating how practitioners perceive, interpret, and interact with diverse types of client data within AI-augmented design workflows, this study offers practical recommendations for the development of future AI-supported design tools, emphasizing strategies to integrate client input effectively while maintaining designers’ professional autonomy in creative decision-making.
\end{enumerate}

By centering the client experience while maintaining professional control, AIDED repositions GAI from a mere style engine to a \emph{mediator} that translates subjective feedback into actionable design decisions, thereby advancing ongoing discussions in co-design, interpretability, and creative AI tools.

\section{Related Work}

\subsection{Supportive Tools for Spatial Design}

Contemporary spatial design increasingly prioritizes user-centered and evidence-based approaches. Toolkits such as DesignMind conceptualize design as cognitively informed and human-centered, while cognitive occupancy models enhance Building Information Modeling (BIM) by integrating considerations of how individuals navigate and perceive spatial environments \cite{DesignMindToolkit, CognitiveBIM2022, lee2019actoviz}. Despite these advancements, professional expertise continues to dominate the utilization of computer-aided design (CAD) tools within architectural practice \cite{lin2025shaping}. 

Recent research endeavors seek to reconcile this gap by integrating perceptual and cognitive-emotional perspectives in the study of built environments. Eye-tracking and saliency analysis techniques have been employed to identify perceptual determinants influencing spatial experience \cite{lavdas_eye-tracking_2024,karmann_saliency_2023}. Comprehensive reviews in neuroarchitecture consolidate findings on the role of cognitive-emotional responses in shaping patterns of space utilization \cite{higuera-trujillo_cognitive-emotional_2021}. Investigations into emotional attention further demonstrate that visual focus corresponds closely with affective reactions, highlighting the importance of incorporating perceptual data into design evaluation processes \cite{fan_emotional_2018}.  

Within HCI, research related to interior design extends these concepts by leveraging artificial intelligence and physiological measurements \cite{EyeTrackedColor2022, MentalGen2024}. Generative design systems have been developed to support early-stage dialogue between clients and designers  \cite{RoomDreaming2024, AIArchSurvey2023}, but lack systematic methods to integrate cognitive, emotional, and attentional signals.  

Neuroscientific studies link interior features to affective states \cite{Coburn2020}, while generative methodologies have been adapted to tailor spatial layouts according to personality profiles \cite{MBTI_AI_2024, MentalGen2024}. Although these contributions demonstrate significant potential, they leave unresolved the challenge of effectively incorporating client perception within co-creative design practices. \add{The present work builds on these efforts by treating client experience as a first-class design material rather than only as an evaluative outcome.}

\subsection{Human–Data Interaction and Data-Assisted Design}

Although designers are beginning to integrate diverse forms of personal data into their workflows, there remains a limited understanding of how to incorporate this data effectively into generative design processes. Stage-based and lived informatics models show how personal data evolve from tracking to reflection and meaning-making \cite{li_stage-based_2010,epstein_lived_2015}. Recent studies on AI-generated imagery illustrate the potential of ambiguous and alternative data representations to facilitate personal meaning-making and reflective engagement with everyday data \cite{AIDataImages2024}. Concurrently, emerging research in Cognitive Personal Informatics highlights the challenges of interpreting complex physiological and cognitive metrics, raising critical issues related to data literacy, ethical considerations, and the conceptualization of “meaningful” metrics in design contexts \cite{schneegass2023future}.

Visualization research demonstrates how perceptual cues support communication and decision-making. Strategic visibility improves wayfinding \cite{VisibilityWayfinding2024}, decision visualizations improve interpretability \cite{ArchDecisionViz2010}, behavior visualizations expose hidden patterns \cite{VideoBehaviorVis2011}, and eye-tracking informs style and color preference \cite{EyeTrackedColor2022, PrefTunedInterior2024}. 

Recent developments in gaze modeling have enabled implicit calibration and enhanced saliency-sensitive attention detection within mobile environments \cite{yang2022continuous}. However, despite these technological advancements, integration across domains remains disjointed: personal informatics primarily focuses on individual self-reflection, whereas gaze–AI systems concentrate on facilitating interactive support. Few studies have addressed the challenge of rendering client perception both interpretable and actionable to support collaborative design processes. The present work proposes a novel approach that integrates gaze analytics with saliency modeling to effectively bridge this divide.

\add{From a practice perspective, Jung et al.\ \cite{jung2022domain} describe how domain experts treat data work as a form of craft, selectively curating, transforming, and contextualizing data within their situated practices rather than simply applying generic analytics pipelines. This view aligns with our observations that interior designers likewise “craft” client information into workable design briefs, deciding which demographic details, preferences, or experiential narratives to foreground.} \add{From a representation perspective, Hilgard et al.\ argue for ``representations by humans, for humans,'' where learned encodings are explicitly optimized to be legible and helpful to people rather than only to downstream models \cite{hilgard2021learning}. Our work contributes by translating client perceptions into overlays and visualizations designed to be interpretable within designers’ existing workflows.}

Comprehensive reviews and design guidelines emphasize the importance of agency and interpretability as fundamental components for establishing trust \cite{victorelli_understanding_2020,victorelli_human-data_2020,sailaja_human-data_2021}. Our work contributes by turning client perception into actionable overlays for AI-assisted iteration.  

\add{Building on this literature, we next examine how generative AI systems have been positioned as co-creative partners, and how their interfaces structure the flow of data, control, and interpretation in design workflows.}

\subsection{Human–AI Co-Creation in Generative Design}

\add{Generative AI (GAI) has progressed beyond its initial role as a design instrument and now functions as a collaborative tool for creativity.} Text-to-image generation systems and conversational interfaces have expanded the frameworks and opportunities available for scaffolding and ideation \cite{DesignWeaver2025, RoomDreaming2024}.

\add{Research in domains such as fashion and music demonstrates that interface scaffolds are often more effective than raw prompting in supporting ideation \cite{FashioningExpertise2024, MVPrompt2024}}. Inkspire integrates GAI into early-stage product ideation, enabling designers to guide text-to-image generation through analogical sketching and a sketch–design–sketch feedback loop that reduces design fixation and enhances the comprehensibility of the AI’s evolving state. Sketching can also be used as a prompt in 3D workflows \cite{SketchVsPrompt2024}. DreamGarden treats a single designer-authored prompt as a seed and helps grow it into a playable game concept through staged AI assistance, illustrating how generative models can scaffold creative exploration over time \cite{earle2025dreamgarden}. Collaborative platforms further illustrate the impact of AI on processes such as iteration, sequencing, and the temporal dynamics of creativity \cite{selvaraju2017grad, CollabDiffusion2023}. Additional research on template-based graphic design tools explores how non-designers perceive and utilize AI-assisted layouts, revealing tensions among automation, user control, and perceived authorship \cite{nouraei2024thinking}.

Beyond generative, recent research emphasizes cognition and social experience in co-creation. Empirical studies indicate that GAI can support metacognitive reflection and suggest potential roles for support agents who help people interpret and steer AI outputs \cite{MetacognitiveAI2024, MetacogAgents2025}. Investigations into explainability demonstrate that transparency and the use of analogy can foster user trust and enable people to adjust their inputs strategically \cite{Explainability2022, AnalogyAI2024}. \add{Kim et al.\ show that explainability can help people understand how to better ``help the AI'' by exposing model reasoning and uncertainty so that users can calibrate their expectations and interaction strategies \cite{kim2023help}. Conceptual frameworks such as the COFI model delineate roles, turn-taking mechanisms, and communication patterns to structure co-creative interactions between humans and AI \cite{rezwana_designing_2023,lin_beyond_2023}. Together, this work suggests that effective human–AI co-creation depends not only on model capabilities, but also on how systems expose model reasoning, distribute agency, and scaffold iterative sensemaking.}

\subsubsection{\add{Spatial Design with GAI}}

\add{AI and generative models are increasingly utilized as supportive tools in architectural and interior design, offering evaluation, prediction, and visualization capabilities \cite{duran2025review}. InSpace introduces AI-powered workflows that enhance creativity in spatial design by enabling designers to explore alternative configurations with algorithmic assistance \cite{gunduz2024proposal}. In interior design, fine-tuning image-generation models on domain-specific datasets produces spatial visualizations that align with user-defined styles and preferences, demonstrating how style keywords and preference cues are translated into visual outputs \cite{lee2024creating}. InsideOut employs GAI models to assess the perceived quality of indoor and outdoor environments across global cities, comparing AI-generated descriptions with human reviews and revealing both partial alignment and systematic mismatches \cite{chang5357388inside}. This work suggests that AI often overemphasizes visually salient forms, whereas human reviewers prioritize interpersonal, sensory, and contextual experiences. While these systems focus on enhancing designers’ ideation, they rarely integrate concrete signals from the client experience.}

\add{Beyond performance-oriented models, work on socio-spatial comfort uses computer vision to infer how occupants feel in different indoor layouts and configurations, demonstrating how vision-based analysis can inform user-centered human–building interactions \cite{lee2021socio}. Recent work on human–agent teams similarly underscores the centrality of spatial information, showing how agents’ awareness of people’s locations and movements shapes coordination and team performance \cite{schelble2022see}. Paananen et al.\ further highlight how human-scale spatial experience emerges from subtle relationships between bodily movement, affordances, and perceived comfort in everyday environments, reinforcing the need to foreground lived experience rather than only building-wide metrics \cite{paananen2021investigating}.}

Recent technological advancements have created new opportunities, particularly through gaze-integrated systems that align generative models with user attention \cite{GazeNoter2024, GazeLog2024, rekimoto2025gazellm}. However, these systems are not yet fully integrated into design practice. The present study addresses this gap by incorporating gaze-driven client perception into generative workflows and aligning AI-derived saliency with designers' expertise. This approach frames co-creation as a tripartite collaboration among clients, designers, and GAI. \add{Recent research has also examined AI tools that prioritize human experience within spatial workflows. For example, Experiential Views employs a fine-tuned vision–language model to score architectural views along experiential dimensions, such as Social, Tranquil, and Inspirational, and visualizes these scores on floor plans and in 3D to assist architects in anticipating occupants’ emotional responses \cite{aseniero2024experiential}.}

\add{In contrast to systems that emphasize model-driven evaluation or designer-led exploration, the present study investigates how rich, multimodal \emph{client} data—including questionnaires, gaze traces, verbal feedback, and preference-informed overlays—can be represented and integrated alongside AI-edited interior images to facilitate client-specific co-design. Our approach frames these data as shared resources that designers can appropriate within their workflows, rather than as opaque inputs to a model pipeline.}

Collectively, these perspectives underscore a persistent challenge: although novel data modalities are increasingly accessible, there is a paucity of grounded knowledge about how to transform them into resources that are \textbf{\textit{both authentic and interpretable}} for professional creative practice. 

Current systems frequently navigate this tension unevenly, either maintaining fidelity to raw personal data at the expense of clarity or enhancing interpretability at the expense of authenticity. This fundamental trade-off motivates the present investigation into how varying modalities of human experience (implicitly represented in user data) influence designers’ collaboration with generative AI, thereby directly informing the research opportunity of this study.

\section{\add{Hypotheses and Study Design}}

\add{Building on the authenticity--interpretability tension highlighted in the related work and our three research questions (RQ1--RQ3), this study explores how different representations of client experience data shape designer--GAI collaboration in residential interior design. We first specify our hypotheses and then describe the study procedure.}

\subsection{\add{Hypotheses}}

\add{The following hypotheses directly correspond to our RQs and are grounded in our formative study and prior work on human--AI co-creation \cite{weisz2024design, kim2023help, naqvi2025catalyst}. They articulate expectations for how AIDED is used to examine the effects of client data on designer-GAI co-design.}

\textbf{\add{H1. Task evaluation (RQ1 \& RQ2).}} 
\add{Providing designers with richer and more structured client information will increase the perceived trustworthiness and usefulness of the data, as well as satisfaction with the resulting designs. In particular, we expect structured summaries of subjective evaluations and AI-augmented representations to receive higher ratings on task-evaluation items (e.g., trust, usefulness, design satisfaction) than those that rely solely on basic demographic profiles or raw gaze visualizations (see Insight 3 \& 4).}

\textbf{\add{H2. Communication with GAI (RQ1, RQ2 \& RQ3).}} 
\add{Access to enriched client information, especially structured or AI-augmented formats, will improve designers' confidence and fluency when generating prompts for the generative AI system, compared to working with only basic demographic information or gaze visualizations (see Insight 1).}

\textbf{\add{H3. Overall evaluation of client-information modalities (RQ1).}} 
\add{Designers will perceive structured and AI-augmented representations of client data as clearer and more actionable than gaze-only heatmaps. We further expect that all enriched modalities (e.g., gaze, questionnaires, AI-predicted attention heatmaps) will be rated as more useful and more likely to be adopted in practice than basic demographic briefs alone.}

\textbf{\add{H4. Design output quality (RQ1).}} 
\add{(a) Designers will report higher satisfaction and perceived client--design fit for their revised designs, produced with GAI assistance, compared to the original designs, with greater improvements when richer client information (e.g., structured subjective questionnaires) is available. (b) Novice evaluators will judge designs created with richer client information as more satisfactory and better aligned with client profiles than designs created with only basic demographic information or gaze-only support.}

\textbf{\add{H5. System adoption and AI mediation (RQ3).}} 
\add{Natural-language explanations and suggestions generated by the large language model will enhance the interpretability of AI-predicted attention overlays, thereby increasing participants' reported willingness to adopt the system in professional settings.}

\subsection{\add{Study Design}}

\begin{figure*}[ht]
    \centering
    \includegraphics[width=0.85\linewidth]{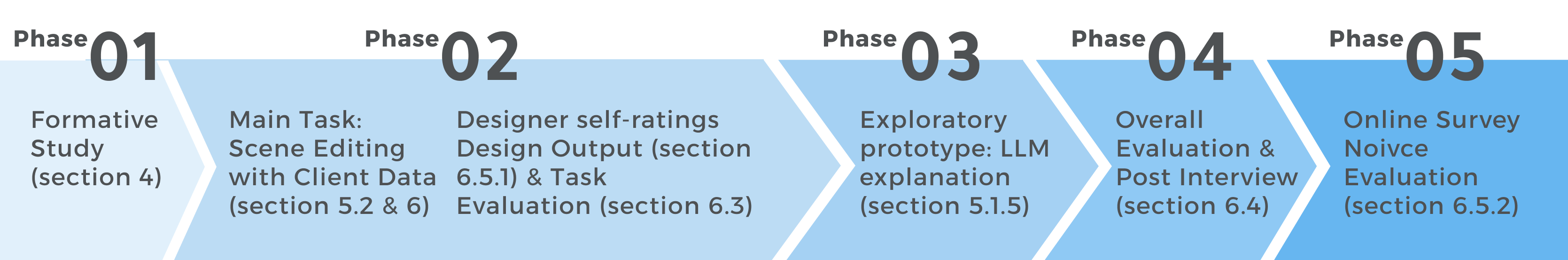}
    \caption{\add{Overview of the five-phase study procedure. Phase 1 is a formative study with professional designers (Section~\ref{sec:formative}). Phase 2 is the main scene-editing task with client data, including designer self-ratings, design outputs, and task evaluation (Sections~\ref{sec:system_workflow}, \ref{sec:task}, \ref{sec:task_evaluation}, and \ref{sec:DesignOutput_evaluation}). Phase 3 introduces an exploratory prototype that provides LLM-based explanations for the AI-predicted heatmaps (Section~\ref{sec:LLM}). Phase 4 covers the overall evaluation and post-task interviews with designers (Section~\ref{sec:overall_evaluation}). Phase 5 reports an online survey and novice evaluation of the generated designs (Section~\ref{sec:Novice_evaluation}).}}
    \label{fig:study_flow}
    \Description{A horizontal flowchart shows five connected arrow-shaped blocks labelled “Phase 01” through “Phase 05,” progressing from left to right. Phase 1 is “Formative Study (section 4).” Phase 2 is “Main Task: Scene Editing with Client Data; Designer self-ratings, Design Output, and Task Evaluation (sections 5.2, 6, 6.3, and 6.5.1).” Phase 3 is “Exploratory prototype: LLM explanation (section 5.1.5).” Phase 4 is “Overall Evaluation and Post Interview (section 6.4).” Phase 5 is “Online Survey Novice Evaluation (section 6.5.2).” The varying blue shades and arrow shapes visually emphasize the sequential order of the phases, but are redundant with the text labels.}
\end{figure*}

\add{We adopted a mixed-method, multi-phase study design (Fig.~\ref{fig:study_flow}). In Phase~1, we conducted a formative study (Section~\ref{sec:formative}) examining three key dimensions: designers’ communication practices with clients, their approaches to interpreting incomplete or ambiguous requirements, and their perspectives on incorporating physiological and psychological data into AI-assisted design tools \cite{EyeTrackedColor2022, GazeNoter2024, yan2024voila}. Insights from this phase informed the selection and design of the client-information modalities.}

\add{To examine how these information modalities influence real-world design workflows, we implemented them in \textbf{AIDED} (Augmenting Interior Design with Human Experience Data), a client-data-integrated workflow for residential interior design (Fig.~\ref{fig:system_cover}, section~\ref{sec:system}). This workflow (Fig.~\ref{fig:system_workflow}) incorporates multimodal client information (Fig.~\ref{fig:main_task}) into an iterative, GAI-assisted design process.}

\add{In Phase~2, the GAI-assisted editing task (section~\ref{sec:task}), we employed a within-subjects design with twelve professional designers across four conditions: baseline briefs (demographics and basic lifestyle information only), real gaze heatmaps, questionnaire visualizations, and AI-predicted attention heatmaps derived from questionnaire data. Each condition was paired with a unique room layout and an authentic client profile. In every condition, designers had up to 5 opportunities to modify the scene using our system, with the available client information modality tailored to that condition. After each iteration, designers evaluated their satisfaction with the current output and its fit with the client profile (section~\ref{sec:DesignOutput_evaluation}). After completing each condition, they also completed a task-level evaluation assessing cognitive effort, information load, trust, creativity, willingness to further modify the design, and overall satisfaction (section~\ref{sec:task_evaluation}).}

\add{In Phase~3, we conducted an exploratory evaluation of the Large Language Model (LLM)-based interpretation feature for AI-predicted attention overlays, which was hidden during the main editing tasks and revealed only after all four conditions were completed (section~\ref{sec:system}). Designers briefly explored this feature, which generated natural-language explanations of overlays and concrete design suggestions. Then they rated the clarity and usefulness of these LLM-assisted supports (Q11–Q12) as part of the overall evaluation (section~\ref{sec:overall_evaluation}). Their reflections on the potential role of LLM support in practice were further probed in the post-task interviews.}

\add{Upon completing all conditions and the LLM feature probe, designers provided holistic evaluations of the different modalities and the AIDED system via a broader questionnaire (Q1–Q17) and participated in semi-structured post-task interviews (Phase~4; section~\ref{sec:overall_evaluation}). These interviews captured their comprehensive experiences with the system and the perceived tradeoffs between authenticity and interpretability.}

\add{To complement expert perspectives, we further conducted an online survey with novice viewers (Phase~5; section~\ref{sec:Novice_evaluation}). Novice participants (1) ranked the satisfaction of design outputs across the four conditions and (2) evaluated whether the original or GAI-modified design better matched the client profile. This final phase links designers’ experiences and prompting strategies to external judgments of design quality and client–design fit.}

\section{Formative Study}
\label{sec:formative}

To ground the system design in authentic practices, we conducted a formative study with professional interior designers. The objective was threefold \add{(RQ2)}: (1) to examine current communication practices between designers and clients, (2) to explore how designers interpret incomplete or ambiguous client requirements, and (3) to understand their perspectives on physiological and psychological data as well as emerging AI-assisted tools.

\textbf{Participants.}
We recruited nine interior designers (five female, four male; aged 24–38) through professional networks and design studios. Participants included company founders, senior designers, assistants, and students with internship experience. All had at least two years of training or practice across residential, commercial, and public projects. This diversity ensured both seasoned expertise and early-career perspectives.

\textbf{Methods.}
Semi-structured interviews\footnote{see Appendix~A for details} (60–75 minutes each) focused on designers’ client communication, feedback interpretation, and attitudes toward new forms of data and AI tools. All interviews were audio-recorded, transcribed verbatim, and thematically coded using an inductive approach.

\subsection{Insight 1: Client Input is Fragmented, Evolves, and is Often Ambiguous}
\label{sec:Insight1}

Design professionals consistently characterized client input as fragmented and often ambiguous, ranging from precise specifications to vague stylistic preferences. Three salient patterns were identified.  

\textbf{Indirect Articulation of Requirements.} 
Clients often struggled to explicitly convey their genuine needs, with essential information emerging only indirectly. For instance, participant D008 remarked, \emph{“Some clients express desires that differ from their actual needs; they might request a more functional sliding door, yet such a solution may not be appropriate for their spatial context.”} 
Similarly, D001 recounted uncovering a client’s significant storage requirements only after learning about the client’s ownership of fifty pairs of shoes: \emph{“Some clients say one thing but actually want another… you have to keep asking until the real need emerges.”}  

\textbf{Primacy of Visual over Verbal Communication.} 
Designers underscored the predominance of visual references over verbal descriptions. As D007 articulated, \emph{“The most helpful information clients provide is when they proactively supply abundant visual materials, as visual information minimizes misunderstandings and consequently shortens communication time.”} Conversely, verbal inputs tended to emphasize stylistic considerations without adequately addressing spatial constraints. D009 noted, \emph{“Clients focus on style and ignore basic spatial dimensions… 3D is needed before they can judge.”}  

\textbf{Progressive Evolution of Client Requirements.} 
Client needs frequently evolved throughout the design process.D012 observed, \emph{“A client might initially approve retaining a particular element during one meeting, but in subsequent meetings, as surrounding design elements change and the overall composition appears unsatisfactory, that element becomes problematic.”} 
Likewise, D002 estimated that although 30--40\% of clients commence with limited clarity, many develop more specific demands that only materistages.

\subsection{Insight 2: Designers Act as Mediators and Interpreters Balancing Feasibility, Lifestyle, and Autonomy.}
\label{sec:Insight2}
Designers consistently articulated their role as intermediaries who translate clients' desires into coherent, practical outcomes while safeguarding their own creative independence. This mediating role encompassed three interconnected facets: the analysis of lifestyle, the negotiation between professional expertise and client preferences, and the translation of client needs into solutions constrained by available resources.  

\textbf{Lifestyle Analysis.} 
A comprehensive understanding of clients’ daily habits was deemed essential for effective design practice. 
For instance, participant D001 outlined a methodical approach: \emph{“Beginning from the moment clients enter their home, considering factors such as the number of shoe pairs they own, the order in which they place keys, bags, and remove footwear, as well as the quantity and types of umbrellas they keep, particularly on rainy days.”} Similarly, D008 underscored the importance of grasping the entirety of clients’ daily routines, stating, \emph{“Understanding their full lifestyle is critical, as it profoundly shapes how they engage with the space.”}  

\textbf{Professional Guidance versus Client Preference.} 
Designers emphasized the nuanced balance required between honoring client preferences and offering expert advice. D002 encapsulated this tension by stating, \emph{“It is a fifty-fifty balance: one cannot be entirely led by the client nor solely impose one’s own vision.”} As D007 articulated, \emph{“When a client suggests something I find unappealing but which, from a professional standpoint, necessitates functional adjustment, I endeavor to explain and persuade them.”} Others stressed the importance of maintaining creative autonomy: \emph{“I aim to be a designer who creates environments I personally find comfortable, leaving the purchasing decisions to the client; I no longer wish to simply accommodate others’ tastes”} (D008). 

\textbf{Translating Client Aspirations into Feasible Solutions.} 
Designers also function as facilitators who help clients reconcile their aspirations with practical limitations. D001 remarked, \emph{“Our role is to serve as a conduit when clients are uncertain about potential approaches or improvements, offering ideas and guidance.”} In a similar vein, D003 described the progression from creative freedom to pragmatic constraints throughout project phases: \emph{“Initially, I encourage clients to be imaginative, but once construction begins, I impose necessary boundaries by explaining ergonomic considerations and feasibility---sometimes even involving contractors to assist in persuasion.”}  

\subsection{Insight 3: Prioritization of Objective Data over Subjective Emotional Responses}
\label{sec:Insight3}

Designers predominantly used structured questionnaires and interviews to obtain concrete information about clients’ lifestyles.  

\textbf{Systematic Data Collection.} 
Designers formulated detailed questionnaires that addressed specific facets of daily living. 
For instance, participant D001 exemplified this method by stating: \emph{“For a three-bedroom, two-living-room layout, we inquire about the number of shoes, the sequence in which keys, bags, and shoes are handled upon entry, and details about umbrellas, including quantity and type (long-handled or folding), particularly since umbrellas cannot be stored in common stairwells.”}  In addition to questionnaires, designers employed direct observation techniques. Participant D008 remarked: \emph{“One can infer a client’s style, taste, and preferences through their existing furniture and personal belongings.”}  

\textbf{Emphasis on Functional Needs.} 
Designers placed greater emphasis on understanding practical requirements than on aesthetic preferences. Participant D006 elaborated: \emph{“We begin by identifying household members, presence of pets, daily habits such as calligraphy usage and book collections, and the types of computers used (desktop versus laptop).”} This perspective was further emphasized when the same participant noted: \emph{“We first ask about family members, pets… then habits; these matter more than style talk.”}  

\textbf{Scarcity of Emotional Data Collection.} 
Physiological or psychological data capturing clients’ emotional experiences within the designed spaces were rarely gathered in a systematic manner. Participant D009 described their focus as primarily on site-specific factors, such as geographic location, ambient sounds, and exposure considerations, alongside fundamental client needs, such as the need for a large sofa. Similarly, D006 commented: \emph{“Eye-tracking can indicate what clients pay attention to, but it remains essential to discuss the underlying reasons.”}  

\subsection{Insight 4: Cautious Optimism Toward AI Integration}
\label{sec:Insight4}

Designers acknowledged GAI's potential to expedite proposal development, synthesize information, and broaden stylistic possibilities. However, they also expressed concerns about limitations related to accuracy, controllability, and the risk of inflated expectations from perfect images. 

\textbf{Current Applications and Advantages.} 
Several designers had already incorporated AI tools into their professional practices. For example, participant D011 used ChatGPT to generate draft proposal texts, including a case study for a 30-year-old engineer client, highlighting the tool’s utility in customizing spatial recommendations. Similarly, D009 remarked on the convenience and significance of AI tools for integrating various tasks. Nonetheless, some participants noted inherent trade-offs; as D008 observed, \emph{“AI accelerates proposal development but also fosters illusory ease—projects appear simpler than they are, potentially diminishing uniqueness.”}  

\textbf{Enhancements in Visualization and Client Communication.} 
AI was regarded as particularly beneficial for improving communication with clients. 
D011 stated, \emph{“I proceed directly to 3D visualization because it is highly effective for client decision-making.”} Participant D012 emphasized AI’s exploratory capabilities, noting that while AI generates numerous options, some outputs may seem unfamiliar, possibly due to the system’s incomplete understanding of the user’s identity.  

\textbf{Concerns Regarding Quality and Control.} 
Designers consistently raised issues concerning the precision and controllability of AI-generated outputs. 
D008 highlighted discrepancies between AI renderings and real-world conditions, noting that \emph{“3D renderings inevitably differ in color and can only depict form and ambiance, whereas the real world contains imperfections such as dust and handcrafted irregularities.”} D011 underscored precision limitations, stating, \emph{“The greatest limitation is AI’s inability to achieve the precision I require, which is accurate to 0.1 centimeters.”}  

\textbf{The Necessity of Human Oversight.} 
Across cases, designers emphasized that AI should function as a supportive tool rather than a replacement for human expertise. D012 remarked, \emph{“Traditional control parameters are easier to manage, but AI control via prompts is not yet sufficiently sophisticated to direct outcomes precisely.”} Similarly, D001 characterized AI as an auxiliary resource requiring expert supervision. Designers also envisioned future developments: D012 expressed a desire for AI systems capable of automatically updating all drawings following a single modification and verifying consistency, effectively serving as an AI design assistant. Another participant, D006, suggested expanding automation capabilities, including eye-tracking to differentiate between \emph{“liked”} and \emph{“confused”} responses, as well as automating data consolidation, scheduling, and cost estimation.

\subsection{Design Objectives (DOs)}
\label{sec:do}
Building upon the aforementioned findings, we established three primary design objectives to steer the development of AIDED:

     \begin{itemize}
         \item \textbf{DO1: Elicit Authentic Client Perspectives.} The system should prioritize the elicitation and actual representation of clients’ subjective experiences in a way that designers regard as both trustworthy and practically applicable. This objective seeks to resolve the challenges posed by fragmented client feedback through the adoption of systematic methods that thoroughly capture client requirements, as well as to address the limited availability of emotional data by incorporating structured processes that more effectively reveal clients’ underlying needs.

         \item \textbf{DO2: Facilitate Interpretability.} Data representations must be designed to allow designers to efficiently identify, compare, and respond to client preferences without imposing excessive cognitive demands. This supports the designer’s intermediary role by delivering clear and actionable insights.
         \item \textbf{DO3: Enable Responsible AI Mediation.} AI-generated outputs should serve to augment, rather than replace, professional expertise by converting fragmented client signals into interpretable insights that are directly applicable to design practice. This objective responds to designers’ cautious optimism by ensuring that human oversight remains central to the process.
     \end{itemize}

\subsection{\add{Rationale for Client Information Selection}}
\label{sec:RationaleforCondtions}

\add{Building on the three design objectives (DO1–DO3) derived from the formative study, we clarify the rationale for selecting and constructing the client information modalities evaluated in our study. These goals shaped how we defined each modality and why it matters in our study.}

\add{\textbf{Demographic Briefs} represent the kind of sparse information designers often receive at the start of a project. Many participants mentioned that early briefs tend to be vague, so we treated this as our “baseline”, something that feels familiar but offers little grounding for making design decisions.}
\add{ \textbf{Raw Gaze Heatmaps} were included because designers were curious about clients’ real reactions, even though they admitted such human signals can be hard to read. They described gaze data as authentic but sometimes confusing, prompting us to test whether designers could interpret unprocessed experiential cues when no further explanation of the raw human signal was provided.}
\add{ \textbf{Questionnaire Visualizations} reflect the tools designers said they rely on in everyday work: lifestyle surveys, preference summaries, mood boards, or requirement checklists. These materials are structured and easy to interpret, which is why many designers naturally trust them more than physiological measures. Introducing this modality enabled us to examine how designers use clear, explicit preference information rather than more ambiguous signals, such as gaze.}
\add{ \textbf{AI-Predicted Attention Overlays} were added in response to mixed attitudes toward AI in the formative study. Designers were open to AI assistance but emphasized that they still want control over the interpretation. By presenting an AI-generated spatial prediction, we could assess whether such mediated information reduces the uncertainty of raw data or raises new concerns about trust and overreliance.}

\add{ Taken together, these four modalities form a spectrum from minimal and ambiguous to structured and explicit to AI-interpreted. This range reflects the realities designers described to us and provides a means to examine how different types of client information influence their reasoning and collaboration with generative tools.
}

\section{AIDED: Augmenting Interior Design with Human Experience Data}
\label{sec:system}

\subsection{System Design}

\begin{figure*}[t]
    \centering
    \includegraphics[width=0.8\linewidth]{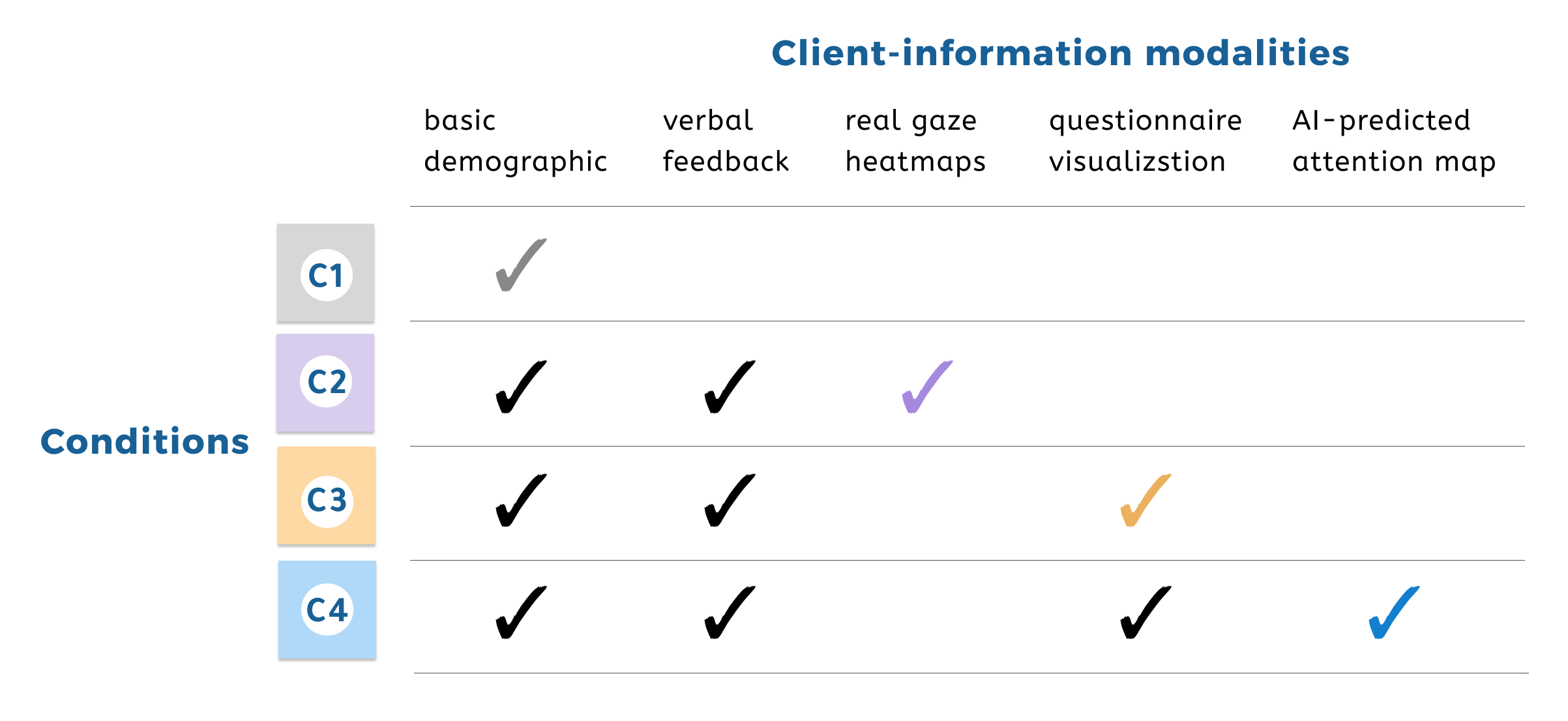}
    \caption{\add{Client-information modalities across four experimental conditions. Matrix summarizing which types of client information are available in each condition. C1 (baseline, gray) provides only basic demographic data. C2 (purple) adds verbal feedback and real gaze heatmaps. C3 (orange) includes demographics, verbal feedback, and questionnaire-based visualizations. C4 (blue) integrates all modalities, adding AI-predicted attention maps on top of demographic data, verbal feedback, and questionnaire visualizations. Checkmarks indicate which data types are available.}}
    \Description {The figure is a table titled “Client-information modalities.” Columns represent five types of client data: basic demographic information, verbal feedback, real gaze heatmaps, questionnaire visualization, and AI-predicted attention map. The rows represent four conditions, labeled C1 through C4 with colored squares: C1 gray, C2 purple, C3 orange, and C4 blue. Checkmarks indicate which data types are available. C1 has only demographics. C2 has demographics, verbal feedback, and real gaze heatmaps. C3 includes demographic data, verbal feedback, and questionnaire visualizations. C4 has demographics, verbal feedback, questionnaire visualizations, and AI-predicted attention maps. The color coding and checkmarks visually distinguish the conditions, but the labels and structure fully describe the differences.}
    \label{fig:main_task}

\end{figure*}

\begin{figure*}[t]
    \includegraphics[width=0.7\linewidth]{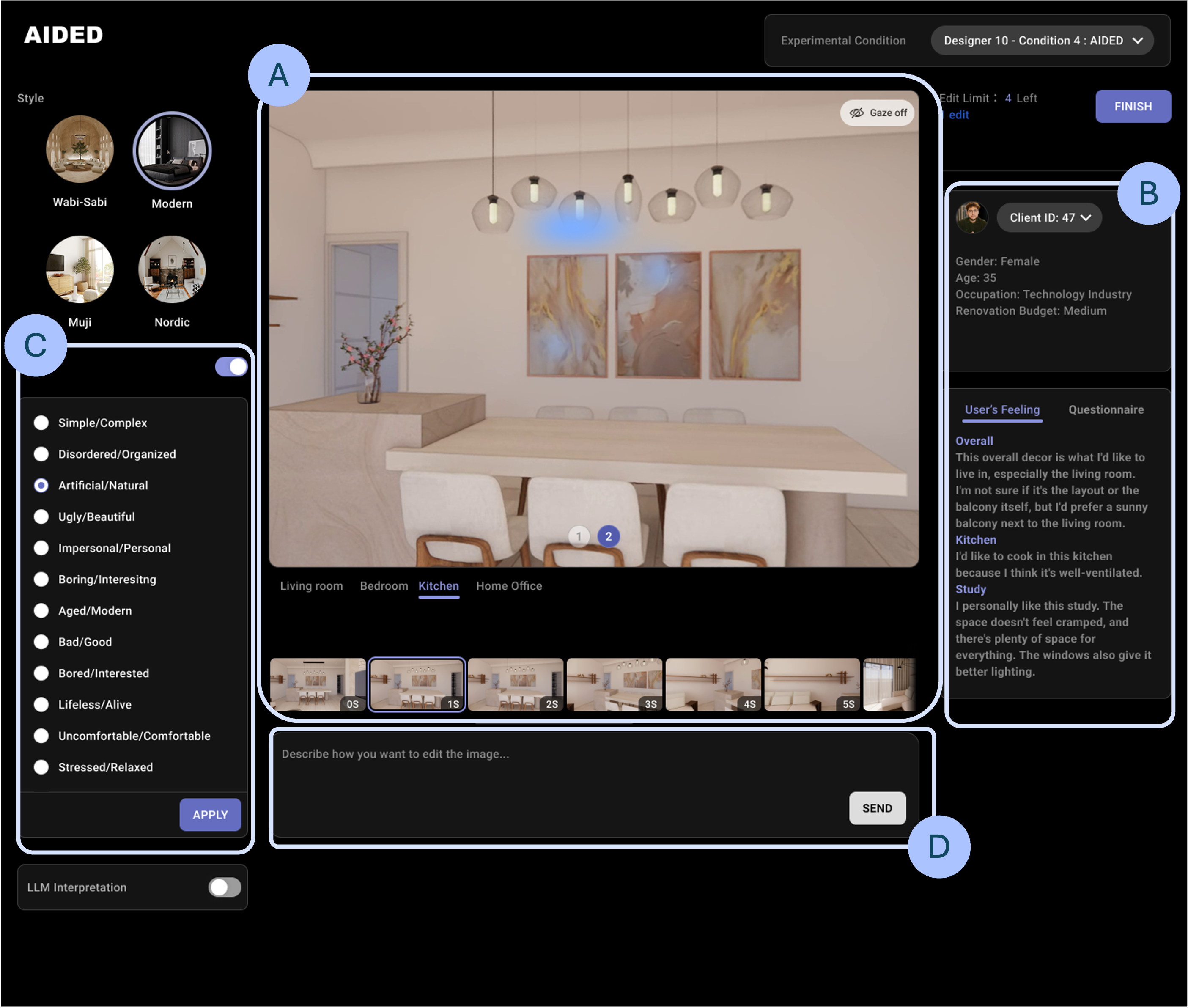}
    \caption{AIDED system interface for designer–AI co-design. (A) Main canvas showing the current interior design alongside navigation video frames spanning four scenes and a history of iterations, with heatmap overlays visible in Conditions 2 and 4. (B) Client information panel showing demographics (all conditions), verbal feedback (Conditions 2-4) and questionnaire responses (Conditions 3-4), with tabs for overall impressions and room-specific comments. (C) Control panel for AI-predicted \add{architectural experience} preference attention maps (Condition 4). (D) Prompt input box where designers describe how they want the AI to modify the scene; edited versions appear as new thumbnails for iterative refinement.}

    \Description{The figure shows the AIDED design interface on a dark background. On the left, circular style selectors (Wabi-Sabi, Modern, Muji, Nordic) and a vertical panel of questionnaire metrics are positioned beside a toggle that enables or disables client visualizations. In the center, a large image of a kitchen scene is displayed above a strip of video thumbnails and room tabs (Living room, Bedroom, Kitchen, Home Office). On the right, a tall panel displays a specific client’s ID, demographics (gender, age, occupation, renovation budget), and written comments for each room. Along the bottom, a wide text box allows the designer to type a prompt describing how to edit the image. The visual layout emphasizes how different panels support client-informed AI editing but all information is also available as on-screen text.}
    \label{fig:system_cover}
    
\end{figure*}

Our system (see Fig. \ref{fig:system_cover}) investigates how different types of client information influence interior designers when they collaborate with generative AI to create personalized residential spaces. The core question driving our study is whether richer client representations can enhance the designer-AI collaboration process. Specifically, we investigate whether combining explicit preferences with implicit visual behaviors leads to design workflows and spaces that better align with client preferences. This controlled comparison enables us to understand which information types are most valuable for personalized design and how designers adapt their prompting strategies based on available data. \add{Therefore}, we developed an experimental platform with four distinct conditions: Condition 1 (C1) baseline, Condition 2 (C2) eye-tracking (gaze) heatmaps, Condition 3 (C3)  questionnaire visualizations, and Condition 4 (C4) AI-predicted attention heatmaps derived from questionnaire data. What changes between conditions is the type and richness of available client data (see Fig. \ref{fig:main_task}). Each provides designers with varying levels of client data. At the same time, they iteratively refine designs using text-guided image modification. The system enables designers to work through the same design task across four distinct information modalities.

\subsubsection{Human Data Visualization: \add{(Demographic, Eye Gaze and Subjective Questionnaire)}}
% Questionnaire, Interview, Eye gaze
To understand how different types of client information affect design decisions, our system visualizes four categories of data from real participants (N=30)  \cite{chien2026incorporatingeyetrackingsignalsmultimodal}. viewing interior spaces (see Fig. \ref{fig:system_workflow}).

Basic demographic profiles provide essential context about clients, including age, gender, occupations, and renovation budget. This foundational information helps designers understand their intended users. Beyond demographics, verbal feedback captures clients' thoughts and reactions after watching interior design videos. These qualitative responses include impressions of the entire space, whether they would want to live there, and specific feedback on individual rooms, such as the living room, kitchen, and bedroom. These verbal descriptions provide insights into clients' reasoning and emotional responses that quantitative ratings alone cannot capture. 

Raw gaze heatmaps show where clients actually looked as they explored interior spaces. We visualize this using heatmaps, revealing unconscious attention patterns that clients may not be aware of. \add{In details, we convert gaze coordinates within each 1-second window into a 2D attention map by accumulating fixations on a 1920×1080 grid and applying a Gaussian blur kernel (approximately one degree of visual angle) to smooth individual samples. The resulting maps are normalized to the [0,1] range and rendered as a semi-transparent purple overlay on the corresponding video frame (see Fig. \ref{fig:system_workflow} Step 3 Human Experience Data), with warmer colors indicating regions that attracted more cumulative attention from the client cohort. Designers thus see, for each frame, which spatial elements consistently drew visual focus when clients explored the scene. These visualizations help designers understand which elements naturally draw attention.}

Additionally, questionnaire data present quantitative ratings across 15 \add{Architectural Experience (AE)} dimensions adapted from established perception scales \cite{Coburn2020}. These structured evaluations provide comparable metrics for how clients perceive qualities like complexity, naturalness, and comfort across different designs.

\subsubsection{AI Model \add{for generating AI-predicted Overlays}}

\add{To generate AI-predicted attention overlays, we adopt a pre-trained multimodal predictive model that estimates clients' \add{AE}  evaluations from interior walk-through videos \cite{chien2026incorporatingeyetrackingsignalsmultimodal}. The model was trained on data from 28 participants who viewed eight 80-second first-person videos of interior spaces across four design styles (Modern, Nordic, Wabi-Sabi, and MUJI). It was trained to predict binary labels for 15 \add{AE}  questions per sequence. This multimodal model achieves 0.72 accuracy on objective questions (e.g., light, organization) and 0.67 on subjective questions (e.g., comfort, hominess, uplift), outperforming strong video-only baselines on most dimensions. The underlying multimodal predictive model and training procedure are described in detail by Chien et al.~\cite{chien2026incorporatingeyetrackingsignalsmultimodal}.}

\add{In AIDED, instead of using the model's classification outputs directly, we apply Gradient-weighted Class Activation Mapping (Grad-CAM) \cite{selvaraju2017grad}. It has fused representation layers to derive spatial attention maps for each \add{AE}  dimension. For each metric dimension (e.g., comfort, hominess, naturalness), we compute a Grad-CAM map for the current frame, rescale it to the image resolution, and normalize it to [0, 1]. In the interface, designers can toggle between dimensions using a simple control panel; the selected dimension is visualized as a semi-transparent blue color overlay (see Fig. \ref{fig:system_workflow} Step 3 AI-predicted Data) with the same opacity as the raw gaze heatmaps. This consistent visual encoding allows designers to directly compare where real gaze concentrated with where the model predicts important regions for a given \add{AE} dimension (metric), using the questionnaire profile to determine which dimensions to inspect.}

\subsubsection{Stimulus} Our experimental stimuli were derived from walk-through videos of interior architectural spaces viewed from a first-person perspective \cite{lin2025shaping}. The original dataset comprised 16 videos, each showing four spatial configurations, rendered in four design styles: Modern, Nordic, Wabi-Sabi, and MUJI. These styles represent everyday residential aesthetics in contemporary design. We extracted frames from these videos at one-second intervals. Each 80-second video produced 80 static images that designers could modify using text prompts. This extraction process yielded discrete images compatible with generative AI tools while preserving the variety of viewpoints and spaces from the original videos. All extracted frames retained their original 1920 $\times$ 1080 pixel resolution and uniform lighting conditions.

We organized these frames by room type into four categories: living rooms, bedrooms, kitchens, and home offices. This organization enabled designers to focus on specific functional spaces and to understand how client preferences may vary across room types.

\subsubsection{Data Format and Integration}
Our system integrates diverse data formats to support comprehensive client representation while ensuring consistency across different visualization methods. Demographic and questionnaire responses are stored in structured tables, and questionnaire data are subsequently converted into bar chart visualizations for easier interpretation during design sessions. This approach enables designers to efficiently compare ratings across distinct \add{AE} dimensions.

Verbal feedback data initially consists of audio recordings transcribed via speech-to-text conversion, with the resulting transcripts made available directly to designers. Eye-tracking data were collected at a sampling rate of 60 Hz utilizing the Tobii Pro Spark system and subsequently processed into heatmaps aggregated over one-second intervals. These heatmaps were temporally synchronized with video timestamps to ensure precise alignment. Additionally, AI-predicted attention maps were produced as overlays at the original resolution and framed to the original videos, serving as an alternative to empirical eye-tracking data.

All data types maintain temporal and spatial alignment through a unified coordinate system, ensuring designers can relate information across different visualizations. This multimodal integration enables designers to develop a comprehensive understanding of client preferences and behaviors when crafting text prompts for design modifications.

\subsubsection{\add{Evaluation of LLM Interpretation on AI-predicted Overlays}} 
\label{sec:LLM}
\begin{figure*}[t]
    \centering
   \includegraphics[width=0.8\linewidth]{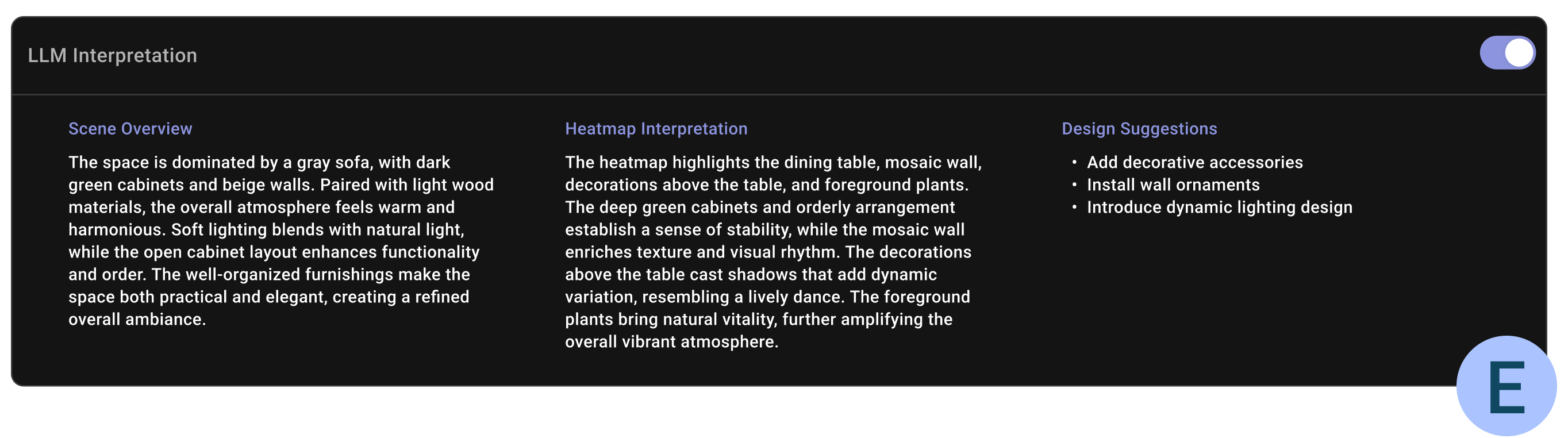}
    \caption{\add{LLM-generated interpretation of AI-predicted Overlays. When enabled, the LLM Interpretation panel (E) summarizes the current scene, explains the meaning of the AI-predicted attention overlay using the selected metric (architectural experience), and proposes concrete design suggestions. The left column provides a narrative scene overview; the middle column identifies the spatial regions highlighted in the heatmap; and the right column lists actionable design suggestions that designers can adapt for their prompts.}}
    \Description{The figure shows a wide, dark rectangular panel labeled “LLM Interpretation” with a toggle switch in the top-right corner indicating that the feature is turned on. Inside the panel, three text columns appear. The first column, titled “Scene Overview,” contains a paragraph describing the overall interior composition, furniture, materials, and atmosphere. The second column, titled “Heatmap Interpretation,” explains which parts of the scene (such as the dining table, wall art, or plants) are highlighted by the AI-predicted attention map and what they imply about the metrics (architectural experience) that were selected. The third column, titled “Design Suggestions,” provides a short bulleted list of possible edits, such as adding accessories, installing wall ornaments, or changing lighting. The panel visually conveys that the LLM translates visual data into natural-language summaries and suggestions, yet all key information is presented in text.}
    \label{fig:LM_interface}
\end{figure*}

We implemented an LLM–based interpretation feature for AI-generated attention maps as an exploratory prototype, separate from our main task. This feature integrates ChatGPT-5 to analyze AI-generated attention maps alongside the original interior image and questionnaire ratings \add{(see Appendix~D for details about how we deliver the prompt)}. The LLM generates structured design reports in three sections (see Fig.~\ref{fig:LM_interface}): a scene overview describing spatial atmosphere and functionality, a heatmap interpretation that explains why clients might focus on specific elements based on their \add{AE} evaluations, and design suggestions that provide three concrete modification ideas with explanations linking attention patterns to potential design improvements, \add{informed by principles from pattern language and living geometry \cite{postle2025llm}.}

The interpretation helps designers (at least in principle) quickly understand client preferences and translate abstract attention patterns into practical design decisions, by turning AI-generated attention overlays into a more readable narrative that connects spatial regions to client-relevant \add{AE}  qualities.

\add{Importantly, this feature was \emph{not} available during the main study tasks. Designers completed all four conditions using only the interfaces described above. Only after finishing all conditions did we reveal the LLM interface as a hidden testing feature and invite designers to explore it briefly. We did not ask them to perform any additional design or editing tasks or to make new design decisions based on this feature.}

\subsection{\add{System Work Flow}}
\label{sec:system_workflow}
\begin{figure*}[htbp]
    \centering
   \includegraphics[width=0.65\linewidth]{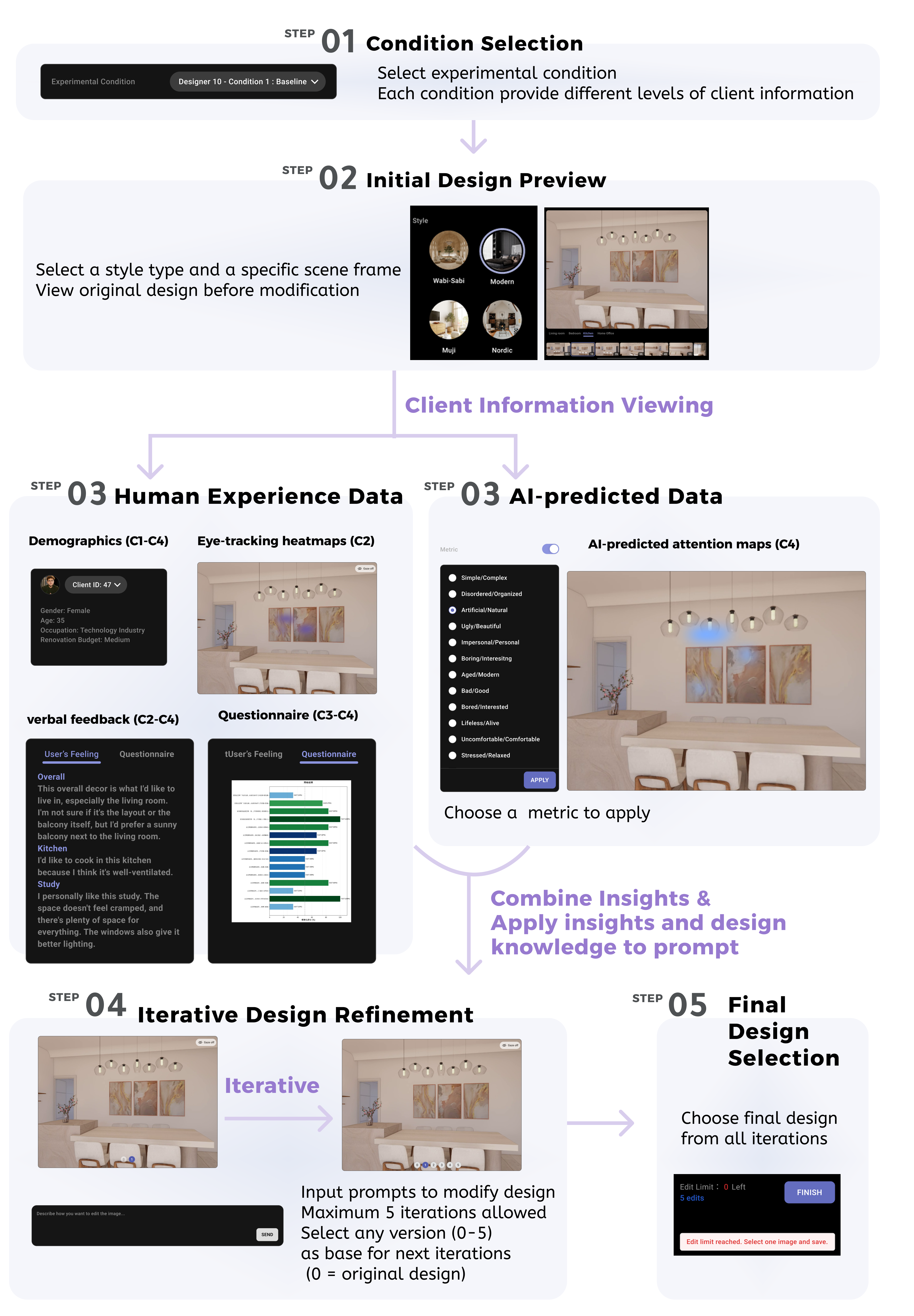}
    \caption{\add{Workflow of the AIDED generative AI design task. Step 1: Designers select one of four experimental conditions, each offering different levels of client information (C1–C4). Step 2: Designers select an interior style and a specific scene frame, and review the original design before any modification. Step 3: Designers view client information, including human experience data (demographics, gaze heatmaps, verbal feedback, and questionnaire visualizations, depending on the condition) and, in C4, AI-predicted attention maps. Step 4: Designers iteratively refine the interior by entering prompts; up to five edited versions can be generated, and any previous version can be used as the starting point for the next iteration. Step 5: Designers select a final design from all generated iterations.}}
    \Description{The figure is a vertical flowchart titled with numbered “STEP 01” to “STEP 05” boxes. Step 01 shows a dropdown labeled “Experimental Condition,” indicating that the designer can choose from conditions C1–C4. An arrow leads to Step 02, which displays the AIDED interface with style icons (Wabi-Sabi, Modern, Muji, Nordic) and an initial scene preview; this step is captioned “Initial Design Preview.” From Step 02, arrows labeled “Client Information Viewing” split into two Step-03 boxes. On the left, “Human Experience Data” displays panels for demographics, eye-tracking heatmaps, verbal feedback, and questionnaire bar charts, with notes such as “Demographics (C1–C4)” and “Eye-tracking heatmaps (C2).” On the right, “AI-predicted Data” shows the questionnaire-metric selector and an AI-predicted attention map, labeled “Choose a metric to apply” and “AI-predicted attention maps (C4).” Both branches converge to Step 04, “Iterative Design Refinement,” illustrated with a sequence of edited kitchen images and text indicating that designers can enter prompts, iterate up to 5 times, and choose any previous version as the base. The final arrow leads to Step 05, “Final Design Selection,” which displays the interface for selecting the final design. The diagram shows that designers first select a condition, then review client information, and finally use generative AI to iteratively refine and select a design.}
    \label{fig:system_workflow}
\end{figure*}

Figure~\ref{fig:system_workflow} illustrates the complete workflow from condition selection, \add{client information viewing} to the final design output. In each condition, designers view an initial interior space and review available client information. Then they iteratively modify the design by writing prompts for a generative AI model (Gemini-2.0-flash) in response to generated outputs and the client information provided.

\section{Main Task: GAI-Assisted Editing Task}  
\label{sec:task}

\subsection{Experimental Design} This task required designers to work with a text-guided image modification system (Figure \ref{fig:main_task}), using prompts to improve an existing design so that it better reflected client needs and feedback. We employed a four-condition within-subjects design with Latin-square counterbalancing. Each designer worked with authentic client data from two distinct client profiles, creating a controlled yet naturalistic interaction context. Each condition was paired with exactly one client and that client’s responses, ensuring that across the four conditions, designers encountered four unique layouts. In each condition, the designer freely selected a style preset (Modern, Nordic, Wabi-sabi, or MUJI). We did not enforce style counterbalancing across conditions or participants. This decision reflects the ecological reality of professional workflows: style selection is often intrinsic to designers’ strategies, while our primary manipulation focused on client information modalities. 

Within each condition, the designer edited two room scenes (one assigned and one self-selected) drawn from four major room categories: living/dining, bedroom, kitchen, and office. Designers were allowed to generate and iteratively refine up to five GAI-based variations per scene. At each iteration, they could either start again from the original image or continue editing from a previously generated version, allowing for non-linear exploration within the five-iteration limit.  

\add{The LLM-based interpretation interface was not part of these four experimental conditions and remained hidden during the main study. It was shown to designers only after they had completed all tasks, as an exploratory prototype for post hoc feedback.}

\subsection{Participants}
\begin{figure*}[t]
    \centering
    \includegraphics[width=0.85\linewidth]{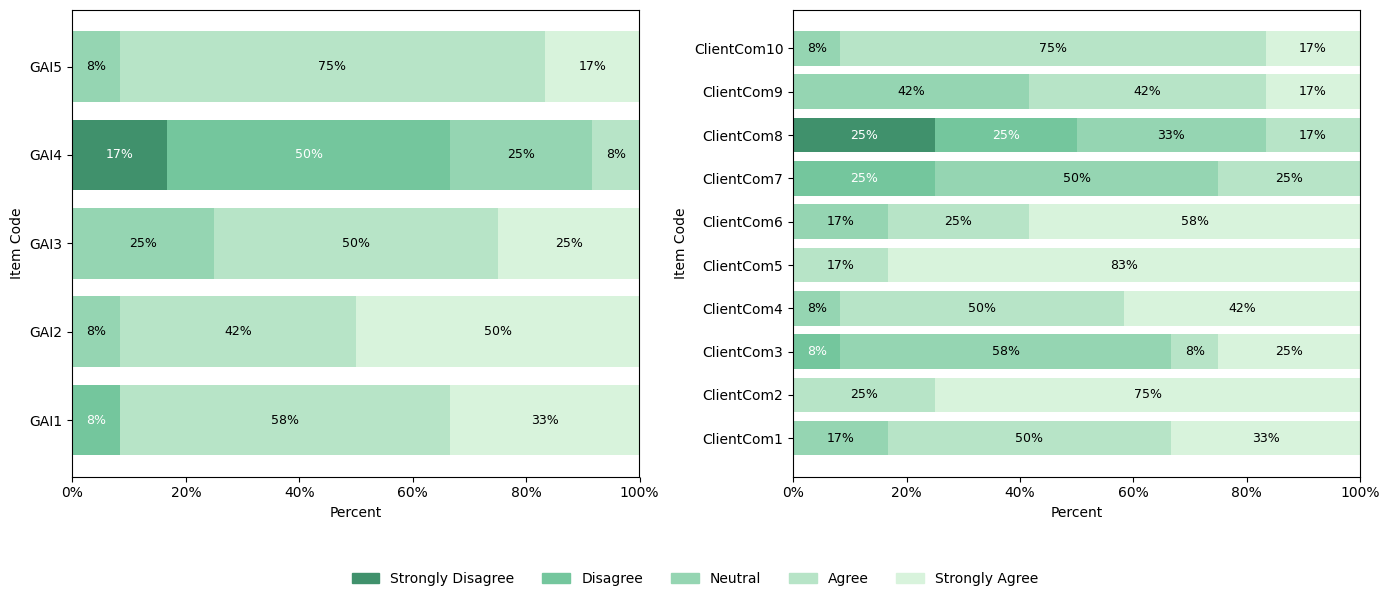}
    \caption{Pre-survey results (N=12). Items covered two domains: attitudes toward generative AI (GAI) (e.g., perceived usefulness, efficiency, and potential threats) and practices related to design–client communication (ClientCom.) (e.g., dependence on visualization, influence of budget considerations, and the balance between intuition and data-driven decision-making). Participants reported frequent interactions with clients, reliance on questionnaires and visual references, and varied prior experience with generative AI tools. These findings provide a contextual baseline of existing practices and expectations preceding the AIDED study.}
    \Description{This figure shows two horizontally stacked bar charts. The left chart (GAI1–GAI5) depicts Likert responses to questions about generative AI; most bars are dominated by the “Neutral” and “Agree” segments, with very few “Strongly Disagree” responses, indicating generally open but cautious attitudes toward GAI. The right chart (ClientCom1–ClientCom10) shows items related to client communication; responses cluster around “Agree,” suggesting that designers frequently communicate with clients, rely on visualization, and already use structured questionnaires. A legend along the bottom of the maps shades green from “Strongly Disagree” to “Strongly Agree.”}
    \label{fig:presurvey}
\end{figure*}

We recruited twelve interior and architectural designers (\(N=12\)) through professional networks, academic programs, and design firms. Most had also participated in the formative study. The sample included six interior designers, four architects, and two graduate students, with professional experience ranging from less than one year to over ten years across freelance, firm-based, and academic contexts. Recruitment emphasized diversity in age, gender, and specialization to capture varied perspectives on client communication and GAI-assisted design. All participants reported regular client interactions and completed a pre-survey\footnote{see Appendix~C} capturing baseline attitudes toward design practices, communication, and GAI (Figure~\ref{fig:presurvey}). \add{The entire task required approximately 60--80 minutes to complete. Participants received a compensation of USD\$40 upon completion. The study protocol, along with all client data used in this study, was anonymized and approved by the Research Ethics Committee of National Tsing Hua University.}

\begin{table*}[t]
\centering
\caption{Participant demographics, professional backgrounds, communication practices, and GAI tool usage (N=12).}
\label{tab:demographics}
\footnotesize
\setlength{\tabcolsep}{2.8pt}
\renewcommand{\arraystretch}{1.05}
\begin{tabular}{l l c l l l l l p{4.2cm}}
\toprule
ID & Gender & Age & Practice & Background & Client Comms. & Feedback Cycle & GAI Tools & GAI Usage Purpose \\
\midrule
D001 & Female & 36 & 10 yrs & Interior Design & $>$10 & 1--2 days & None & None \\
D002 & Male   & 25 & Student & Architecture & 5--10 & 1 week & T2I, LLMs & Ideation; Rendering; Client comms.; Narratives \\
D003 & Male   & 27 & 5 yrs & Interior Design & $>$10 & 1 week & LLMs & None \\
D004 & Female & 24 & $<$1 yr & Architecture & 5--10 & 1--2 days & LLMs, GAI ID & Ideation; Rendering; Narratives \\
D005 & Female & 28 & 5 yrs & Architecture & 5--10 & 1 week & T2I, LLMs & Quick rendering \\
D006 & Female & 38 & $>$10 yrs & Interior Design & 3--5 & 1 week & GAI ID & Quick rendering \\
D007 & Female & 32 & 5 yrs & Interior Design & 5--10 & 1--2 days & T2I, LLMs & Client comms.; Model generation \\
D008 & Female & 31 & 3 yrs & Interior Design & $>$10 & 1 week & LLMs & 2D/3D understanding; Data collection \\
D009 & Male   & 25 & 3 yrs & Interior Design & $>$10 & 1--2 days & LLMs, GAI ID & Ideation; Rendering; Client comms.; Narratives \\
D010 & Female & 25 & 3 yrs & Interior Design & $>$10 & 1--2 days & GAI ID & None \\
D011 & Male   & 31 & 3 yrs & Interior Design & $>$10 & 1 week & T2I, LLMs, GAI ID & Ideation; Client comms.; Narratives \\
D012 & Female & 26 & 5 yrs & Architecture & $>$10 & 1 week & T2I, LLMs, GAI ID & Ideation; Rendering \\
\bottomrule
\end{tabular}
\end{table*}

\subsection{Task Evaluation}  
\label{sec:task_evaluation}
To assess participants’ subjective experiences across the four information conditions, we designed a post-study questionnaire comprising 14 items (TE1–TE14). Each item was rated on a seven-point Likert scale ranging from 1 (strongly disagree) to 7 (strongly agree). The questionnaire items were adapted from existing literature on trust in client information, communication with GAI, and design satisfaction. Additionally, the instrument was pilot-tested with two professional designers to ensure both clarity and relevance.  

The items were organized into four conceptual dimensions aligned with the study’s aims (see Table~\ref{tab:task_eval_dimensions}):  
\textbf{Cognitive Load.}  
Items TE1 and TE2 measured the cognitive effort required to interpret and manage client information.  
\textbf{Trust and Usefulness of Client Information.}  
Items TE3 through TE7 assessed designers’ perceptions of the reliability, clarity, and practical utility of client data across experimental conditions. Representative items included trust in the interpretability of client information (TE4) and the degree to which the information supported improved design decisions (TE6).  
\textbf{GAI Communication.}  
Items TE8 to TE10 assessed designers’ proficiency in conveying client requirements and professional insights to the GAI system, with a focus on fluency and confidence in prompt-based interactions.  
\textbf{Design Satisfaction.}  
Items TE11 to TE14 captured designers’ evaluations of the quality and appropriateness of designs generated following GAI-assisted editing, including confidence in the results (TE11), overall satisfaction (TE12), and perceived adequacy in fulfilling client needs (TE13).

\begin{table*}[ht]
\centering
\caption{Task Evaluation Items (TE1--TE14).}
\label{tab:task_eval_dimensions}
\small
\begin{tabular}{l p{12.8cm} l}
\toprule
\textbf{Item} & \textbf{Evaluation Statement} & \textbf{Construct} \\
\midrule
TE1  & Understanding client information required significant mental effort. & Cognitive Load \\
TE2  & There was an excessive amount of client information to process at once. & Cognitive Load \\
\midrule
TE3  & Client information inspired my design creativity. & Trust/Usefulness \\
TE4  & I trust the meaning of the client information. & Trust/Usefulness \\
TE5  & I am willing to modify my design based on the client information provided. & Trust/Usefulness \\
TE6  & The information assisted me in making improved design decisions. & Trust/Usefulness \\
TE7  & The system aided me in comprehending and incorporating client needs. & Trust/Usefulness \\
\midrule
TE8  & I was able to smoothly communicate client needs to the GAI. & GAI Communication \\
TE9  & I effectively conveyed professional ideas to the GAI. & GAI Communication \\
TE10 & I felt confident in my prompts under this condition. & GAI Communication \\
\midrule
TE11 & I am confident in the design outcomes edited with GAI. & Design Satisfaction \\
TE12 & I am satisfied with the final design results. & Design Satisfaction \\
TE13 & Edited designs would meet client requirements. & Design Satisfaction \\
TE14 & While using GAI, I felt in control without losing design freedom. & Design Satisfaction \\
\bottomrule
\end{tabular}
\end{table*}

\subsection{Overall Evaluation}
\label{sec:overall_evaluation}

\textbf{Comprehensive Assessment.}  
Beyond task-specific metrics, we administered a 17-item post-study questionnaire to capture participants’ overall perceptions of the four information modalities and the system as a whole. Each item was rated on a five-point Likert scale (1 = strongly disagree, 5 = strongly agree) and organized by modality, with additional items addressing system-level evaluation (see Table~\ref{tab:overall_eval_dimensions}). Specifically:  
\begin{itemize}  
    \item \textbf{Condition 2 (Eye-tracking heatmaps):} Q1–Q3 assessed the clarity, interpretability, and practical applicability of raw heatmap visualizations.  
    \item \textbf{Condition 3 (Questionnaire visualizations):} Q4–Q6 evaluated the clarity, authenticity, and usefulness of charts derived from questionnaire data.  
    \item \textbf{Condition 4 (AI-predicted overlays):} Q7–Q10 examined the credibility and utility of AI-generated overlays for identifying potential design improvements.  
    \item \textbf{LLM explanations and suggestions:} Q11–Q12 investigated the added value of natural language support, including explanations of heatmaps and recommendations for design modifications.  
    \item \textbf{System-level impressions:} Q13–Q17 captured broader considerations such as usability and adoption potential, covering learning effort (Q13), accuracy relative to traditional methods (Q14), willingness to employ the system in real-world projects (Q15), recognition of client states (Q16), and integration of client experiences into generative AI prompts (Q17).  
\end{itemize}

\begin{table*}[ht]
\centering
\caption{Overall Evaluation Items (Q1--Q17)}
\label{tab:overall_eval_dimensions}
\begin{tabular}{p{15.5cm}c}
\toprule
\textbf{Evaluation Statement} & \textbf{Dimension} \\
\midrule
\textbf{Q1}. Heatmap visualizations were clear and easily interpretable. & C2 \\
\textbf{Q2}. The heatmap clearly showed how client attention was distributed. & C2 \\
\textbf{Q3}. I was able to identify design elements that needed adjustment based on the heatmaps shown. & C2 \\
\textbf{Q4}. Questionnaire visualizations were transparent and comprehensible. & C3 \\
\textbf{Q5}. Questionnaire authentically represented clients’ feelings. & C3 \\
\textbf{Q6}. Questionnaire aided my understanding of client preferences. & C3 \\
\textbf{Q7}. AI-predicted overlays appeared reasonable and trustworthy. & C4 \\
\textbf{Q8}. AI-predicted overlays provided plausible explanations for client gaze patterns. & C4 \\
\textbf{Q9}. Our system (AIDED) allowed for quick identification of areas and directions for design changes. & C4 \\
\textbf{Q10}. AI-predicted overlays facilitated more precise design modifications. & C4 \\
\textbf{Q11}. Explanations provided by LLM-assisted features helped me grasp the meaning behind AI-predicted overlays. & LLM \\
\textbf{Q12}. Design suggestions generated by the LLM supported me in creating prompts for GAI design modifications. & LLM \\
\textbf{Q13}. Learning curve for using this system was minimal. & SYS \\
\textbf{Q14}. Relative to traditional methodologies, our system (AIDED) more accurately reflected client needs. & SYS \\
\textbf{Q15}. I express willingness to employ this system in real-world projects to capture client experiences. & SYS \\
\textbf{Q16}. Our system (AIDED) facilitated my recognition of the importance of clients’ physiological and psychological states. & SYS \\
\textbf{Q17}. Our system (AIDED) increased my likelihood of incorporating client experiences into GAI prompts for design changes. & SYS \\
\bottomrule
\end{tabular}
\vspace{2pt}
{\footnotesize
\parbox{0.95\linewidth}{\raggedright
\textit{Note.} Dimension codes: C2 = Gaze Heatmap; C3 = Questionnaire Visualizations; C4 = AI-predicted Overlays; LLM = LLM Support; SYS = System-level Impressions.
}}
\end{table*}

\textbf{Post-Study Interview.}
To complement the quantitative survey data, semi-structured post-task interviews were conducted to capture designers’ comprehensive experiences with the AIDED system\footnote{see Appendix~B}. These interviews explored the influence of each modality on decision-making processes and identified contexts in which the system could be most effectively incorporated into professional design workflows.

\subsection{Design Output Evaluation}

\subsubsection{Designer Self-Assessment Perspective}
\label{sec:DesignOutput_evaluation}

To obtain designers’ evaluations of the quality of their work, self-assessments were collected immediately following each design task. For each scene, designers provided ratings for both the original design (prior to modification) and the revised design (subsequent to GAI-assisted editing under the specified experimental condition). \add{\footnote{All original designs, revised designs, and corresponding iteration logs are provided in the supplementary material.}} Two dimensions were measured using 5-point Likert scales:
\begin{itemize}
    \item \textbf{Design Satisfaction} — the degree to which designers felt satisfied with the visual and functional attributes of the design outcome.
    \item \textbf{Design Appropriateness (Client Fit)} — the extent to which designers perceived the design as suitable for the stated client’s needs and preferences.
\end{itemize}
By assessing both the initial and modified designs, this approach facilitated within-subject comparisons to examine how varying levels of client information (Conditions 1–4) affected designers’ professional satisfaction with their work and their perceived congruence with client requirements.

\subsubsection{General Novice Perspective}
\label{sec:Novice_evaluation}

To examine how the availability of client information influences designer-GAI co-design, we conducted an evaluation study with general novices via an online survey. We aimed to investigate whether providing designers with comprehensive client data enables them to produce more satisfactory designs when using generative AI. Thirty participants (18 female, 12 male), aged from 22 to 57, took part in the study.

\textbf{Design Satisfaction \add{(DS)} Assessment}\footnote{see Appendix~E.1} This assessment consisted of 11 questions where participants evaluated sets of four interior designs (labeled A, B, C, D) that had been generated through designer-GAI co-design in different experimental sessions. To avoid ordering bias, the presentation order was randomized, and the labels did not correspond to the experimental conditions (1-4). Participants ranked the four designs in each set from most to least satisfactory (Rank 1-4) according to their personal preferences. This assessment measured whether designers with access to comprehensive client information could produce designs that general novices found more appealing when working with generative AI.

\textbf{Design Appropriateness \add{(DA)} Assessment}\footnote{see Appendix~E.2} Twenty-six participants completed a before-and-after comparison study consisting of 7 questions to evaluate whether progressive information enrichment helped designers create more client-appropriate modifications. The scenarios were constructed using actual client data from our system implementation, specifically drawing on basic demographic profiles and interview feedback. Four participants were excluded from this analysis because their profiles (demographics, interviews, questionnaires, and eye-tracking data) had served as inputs for the design conditions. Since the scenarios in this assessment explicitly described these individuals' characteristics and preferences, we analyzed their responses separately to understand how the original clients themselves perceived the modified designs. 

In this assessment, participants read scenario descriptions derived from real client demographics and interview responses, then viewed paired comparisons showing original and GAI-modified designs for each of the four conditions. For each pair, participants selected option (a) or option (b), with the assignment of original and modified versions randomized between trials to avoid order bias. This comparison within the scenario allowed us to assess whether increasing levels of client information enabled designers to create modifications that participants perceived as more appropriate for the described client.

Both assessments were administered through an online survey platform with standardized image presentation parameters. This dual evaluation approach captures general novices' perspectives on designs produced under different information conditions, examining both overall design satisfaction and perceived alignment between modifications and client profiles. These two evaluations reveal how the general public perceives the outcomes of designer-GAI co-design when designers have access to varying levels of client information.

\subsection{Statistical Analysis}
A hierarchical analytical framework was employed in accordance with established guidelines for repeated-measures designs. All statistical tests were two-tailed, with a significance level set at \(\alpha = .05\). For the task evaluation, which comprised 12 items across four conditions, omnibus differences were examined using Friedman tests. Items demonstrating significant effects (\(p < .05\)) were subsequently analyzed through pairwise Wilcoxon signed-rank tests with adjusted \(p\)-values. Effect sizes were reported as Kendall's \(W\) for Friedman tests and as \(r = z / \sqrt{n}\) for Wilcoxon tests, with 95\% confidence intervals estimated via 5{,}000 bootstrap resamples. For overall evaluation items, one-sample Wilcoxon signed-rank tests were conducted against the neutral midpoint (\(\mu = 3\)). To address multiple comparisons, Holm-Bonferroni corrections were applied to confirmatory pairwise comparisons of task evaluation items to control the family-wise error rate. In contrast, for exploratory analyses of the overall evaluation items (Q1--Q17), the Benjamini--Hochberg false discovery rate (FDR) procedure was used, as it is better suited for correlated variables and preserves statistical power.

\add{In the novice design output evaluation, we analyzed Design Satisfaction (DS) rankings and Design Appropriateness (DA) responses using non-parametric tests. For DS, we applied a Friedman test on mean ranks across conditions, followed by Wilcoxon signed-rank tests for pairwise comparisons, reporting effect sizes as $r = z / \sqrt{n}$. For DA, we computed for each participant the proportion of scenarios (out of 7) in which they selected the modified design ($n = 26$). A Friedman test assessed overall differences between conditions, and one-sample Wilcoxon signed-rank tests against the neutral baseline of 0.5 examined whether each condition led participants to choose the modified design more often than chance. Pairwise DA comparisons also used Wilcoxon signed-rank tests, with Holm–Bonferroni corrections applied to all DS and DA pairwise tests.
}

\section{Main Findings}

\subsection{\add{Task Evaluation across conditions}: Client Data Modulate Better Design Decisions, Not Preserve Creative Autonomy}

\begin{figure*}[ht]
    \centering
    % Subfigure 1: Kendall's W Effect Size Analysis
    \includegraphics[width=0.75\textwidth]{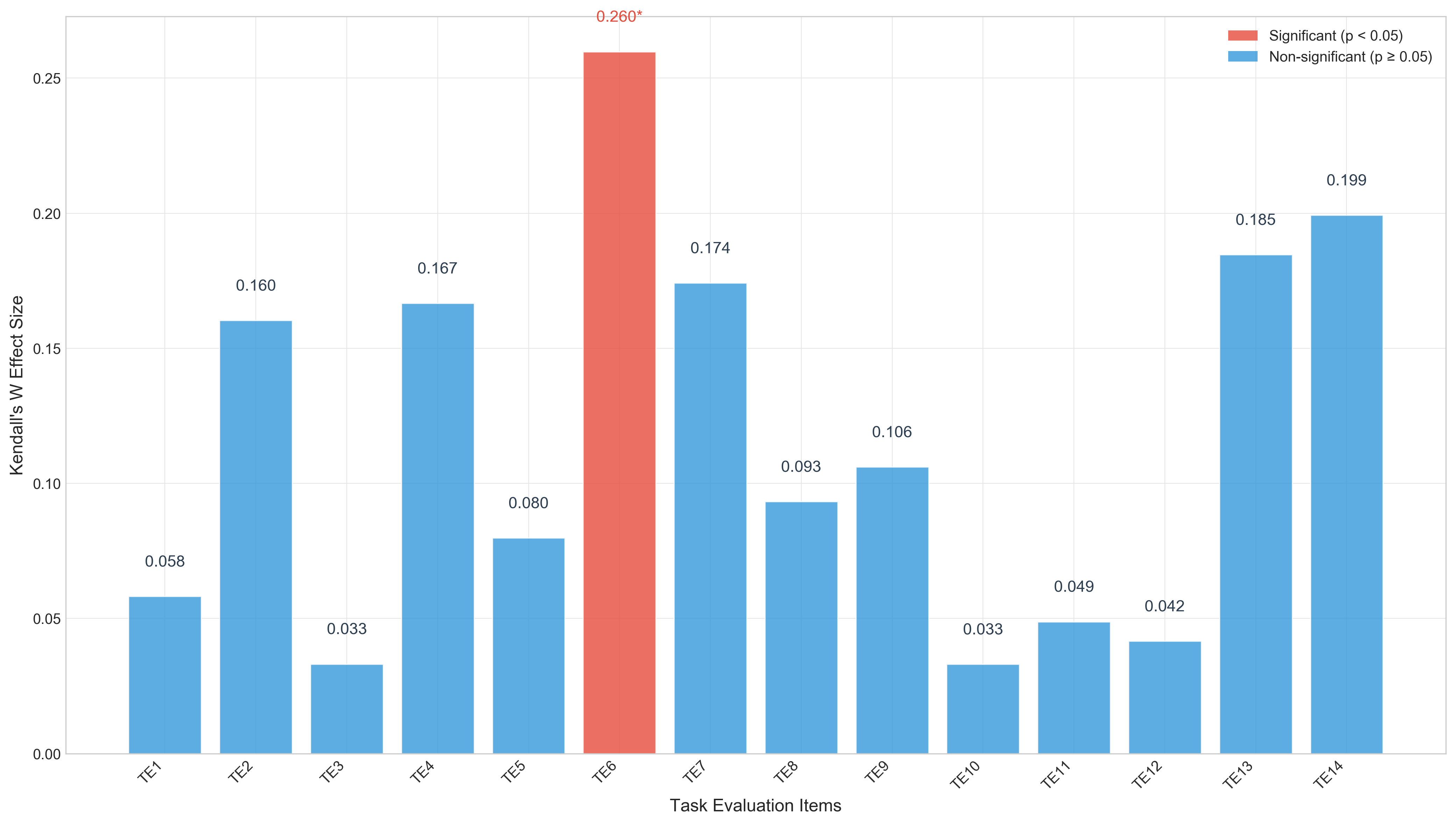}
    \caption{\add{Kendall’s $W$ effect sizes for the 14 task-evaluation items; higher indicates stronger between-condition differences). 
    The single \textcolor{red}{red} bar marks the only item reaching significance in the Friedman test: TE6 “Information assisted me in making improved design decisions.” ($\chi^{2}(3)=9.35$, $p=.025$, $W=.260$; asterisk denotes $p<.05$). 
    All remaining items (\textcolor{blue}{blue}) were non-significant; several nevertheless showed medium-sized effects (e.g., TE14 $W=.199$, TE13 $W=.185$, TE7 $W=.174$).}}

    \Description{A horizontal bar chart displays Kendall's $W$ effect sizes for task-evaluation items TE1 through TE14. The x-axis lists the items, and the y-axis ranges roughly from 0 to 0.27. Most bars are medium-height blue bars between about 0.03 and 0.20, indicating small effects. TE6 stands out as a tall red bar at approximately 0.26, labeled with an asterisk to mark statistical significance. A legend in the upper-right corner explains that the red bar indicates a significant effect (p < .05) and blue bars indicate non-significant effects (p ≥ .05).}
    \label{fig:kendalls_w_effectsize}

    \vspace{1em} % space between the two figures
    
    % Subfigure 2: Effect Size (r) Forest Plot for TE6
    \includegraphics[width=0.75\textwidth]{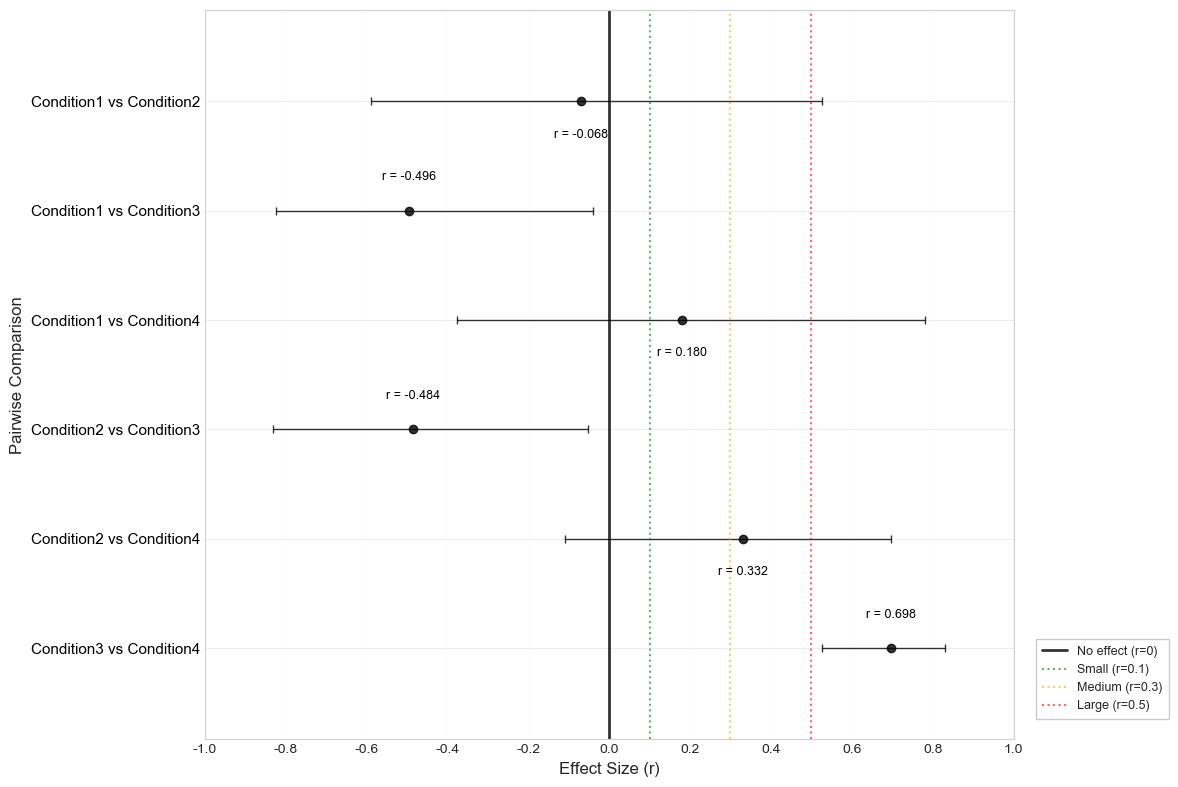}
    \caption{Pairwise Wilcoxon Effect Sizes for TE6. Forest plot showing $r$ values with 95\% confidence intervals. The strongest contrast was between C3 (questionnaire visualizations) and C4 (AI-predicted overlays), yielding a large effect  Condition~3 versus Condition~2 (gaze heatmaps) produced a moderate effect ($r \approx .48$, CI [-.84 -.05]), suggesting structured questionnaires offered clearer decision cues than gaze alone. Other comparisons yielded small, non-significant effects ($r<.25$).}

    \label{fig:te6_forest_plot}
    \Description{This forest plot has six horizontal rows, each labeled with a pair of conditions (e.g., “Condition1 vs Condition3”). The x-axis shows effect size r from -1.0 to 1.0, with a solid vertical line at 0 (no effect) and colored dashed lines at 0.1, 0.3, and 0.5 for small, medium, and large effects. Each comparison is represented by a dot (the estimated r) with a thin horizontal confidence interval. Most dots lie near zero with wide intervals crossing the no-effect line, indicating negligible differences between conditions. The C3 vs C4 comparison is centered at r = 0.7, with the confidence interval fully to the right of 0, visually standing out as the only clearly large effect. A legend in the lower right corner explains the reference lines for small, medium, and large effects.}
\end{figure*}

\add{Among all task-level measures, TE6 (\emph{“Information helped me make better design decisions”}) was the only dimension that showed statistically significant differences across the four conditions, $\chi^{2}(3) = 9.35$, $p = .025$, Kendall's $W = .26$ (see Fig.~\ref{fig:kendalls_w_effectsize}). Descriptively, designers provided the highest decision-support ratings for questionnaire visualizations (C3; $M = 6.00$, $SD = 0.95$) were rated highest, followed by gaze heatmaps (C2; $M = 5.00$, $SD = 1.71$), the baseline condition (C1; $M = 4.75$, $SD = 2.05$)}, and finally AI-predicted overlays (C4; $M = 4.50$, $SD = 1.57$). Post-hoc pairwise analyses (Fig.~\ref{fig:te6_forest_plot}) revealed the most pronounced difference between Condition~3 and Condition~4 ($W = 0.0$, $r = .69$, $p = .016$), \add{demonstrating that structured questionnaire data (C3) were perceived as substantially more supportive for decision-making than AI-predicted overlays.} However, this effect did not remain statistically significant following Holm correction for multiple comparisons ($p_{\text{adj}} = .094$). Medium-to-large effect sizes were also observed when comparing Condition~3 with Condition~2 ($W = 5.0$, $r = -.48$, $p = .094$) and Condition~3 with Condition~1 ($W = 7.5$, $r = -.50$, $p = .086$), suggesting that structured questionnaire data were generally rated more favorably than raw gaze heatmaps or the baseline condition. Despite these notable magnitudes, the differences likewise failed to reach significance after correction. No other pairwise comparisons approached statistical significance (see Table~\ref{tab:te6_wilcoxon} for full results).

\add{Post interview data help explain why TE6 uniquely among all metrics—varied significantly across conditions.} Several designers identified the questionnaire visualizations as the most reliable source of information; one participant remarked, \emph{“The questionnaire charts were the most direct, allowing me to capture the real needs”} (D001). Conversely, eye-tracking heatmaps (C2) were valued for their ability to rapidly highlight focal points, though participants often expressed uncertainty about their interpretability: \emph{“They looked at it, but I couldn’t tell whether it was good or bad—I had to guess”} (D005). The explanatory feature based on large language models (LLMs) was perceived as particularly influential. As one participant noted, \emph{“With the AI-predicted heatmaps, I had no idea how to understand it, but if the AI translated it into text, I immediately knew why”} (D002). AI-predicted overlays (C4) initially scored lower due to interpretability issues, but were perceived as far more helpful when paired with LLM-generated explanations. \add{This suggests that the effectiveness of AI-predicted overlays depends on interpretability support, revealing a gap between the model's explanation and design decisions.}

\begin{table}[ht]
\centering
\caption{Pairwise Wilcoxon signed-rank results for TE6 (effect sizes, confidence intervals, and corrected $p$-values).}
\begin{tabular}{lccccc}
\toprule
Comparison & $W$ & $p$ & $p_{\text{adj}}$ & $r$ & 95\% CI \\
\midrule
C3 vs C4 & 0.0  & .016 & .094 & .70 & [.53, .84] \\
C2 vs C3 & 5.0  & .094 & .430 & -.48 & [-.84, -.05] \\
C1 vs C3 & 7.5  & .086 & .430 & -.50 & [-.82, -.02] \\
C2 vs C4 & 2.0  & .250 & .750 & .33 & [-.11, .69] \\
C1 vs C4 & 21.0 & .533 & 1.000 & .18 & [-.37, .78] \\
C1 vs C2 & 8.5  & .813 & 1.000 & -.07 & [-.57, .54] \\
\bottomrule
\end{tabular}
\vspace{2pt}
{\footnotesize
\parbox{0.95\linewidth}{\raggedright
\textit{Note.} Pairwise comparisons were conducted using Wilcoxon signed-rank tests.
$W$ denotes the test statistic; $r = z / \sqrt{n}$.
Ninety-five percent confidence intervals were estimated via 5{,}000 bootstrap resamples.
Adjusted $p$ values ($p_{\text{adj}}$) were computed using Holm--Bonferroni corrections
to control the family-wise error rate.
}}
\label{tab:te6_wilcoxon}
\end{table}

The remaining items did not achieve statistical significance (all $p \geq .066$); nevertheless, several demonstrated small to moderate effect sizes. For TE14 (\textit{“While using GAI, I felt in control without losing design freedom.”}), the Friedman test produced $\chi^{2}(3) = 7.17$, $p = .0666$, Kendall's $W = .20$. Mean ratings increased from Condition~1 ($M = 3.00$, $SD = 1.65$) and Condition~2 ($M = 3.17$, $SD = 1.70$) to Condition~3 ($M = 3.75$, $SD = 1.76$) and Condition~4 ($M = 3.83$, $SD = 1.75$), indicating a trend toward enhanced perceived autonomy when structured or AI-augmented client data were available. Similarly, TE13 (\emph{“Edited designs would meet client requirements.”}) yielded $\chi^{2}(3) = 6.00$, $p = .0840$, Kendall's $W = .18$. Mean scores rose from Condition~1 ($M = 2.92$, $SD = 1.68$) and Condition~2 ($M = 2.92$, $SD = 1.08$) to Condition~3 ($M = 3.92$, $SD = 1.56$) and Condition~4 ($M = 3.50$, $SD = 1.51$). Although these patterns did not reach statistical significance, they suggest that richer client information modalities are associated with designers’ heightened sense of control and increased confidence that their revised designs would fulfill client requirements.

Building on the modality-specific findings, the interviews revealed overarching themes regarding the integration of client information into designers’ professional practices. Participants consistently emphasized that client data functioned not as rigid directives but as a foundation for more informed decision-making. For example, participant D002 noted, \emph{“The information helped me judge the direction more quickly, but it did not tell me exactly how to change”}, highlighting the system’s role as a supportive framework rather than a prescriptive authority. Likewise, participant D009 underscored the value of structured feedback, stating, \emph{“I would always start with basic information, but more detailed data, especially subjective preferences—helped me prioritize which modifications to make.”} These qualitative insights align with the moderate effect sizes observed in measures of control (TE14) and client satisfaction (TE13), suggesting that client information enhanced designers’ guidance and orientation without constraining their creative autonomy.

Concurrently, the lack of significant differences across most measured items highlights the resilience of designers’ professional heuristics. Even under Condition~1, in which no client feedback was provided, participants were able to generate credible design modifications by drawing on intuition and accumulated expertise. As participant D005 remarked, \emph{“Rather than looking at so many charts, I was quicker to directly adjust based on my own experience”}, indicating that professional judgment can partially mitigate the absence of structured data. This compensatory process likely explains the minimal variation observed across most task evaluation metrics.

Taken together, the results of the task evaluation indicate a dual dynamic. 
From a quantitative perspective, only TE6 (\emph{“Information assisted me in making improved design decisions.”}) reached statistical significance across modalities, highlighting the unique role of client data in enhancing decision-making confidence. From a qualitative standpoint, designers consistently described client data as supportive rather than directive, illustrating a complex interaction between professional expertise and AI-generated insights. These findings suggest that although client data (\textit{human experience data}) may not fundamentally alter established workflows, it can significantly bolster designers’ confidence, especially in translating ambiguous client requirements into concrete design actions.

\subsection{Overall Evaluation of using AIDED}

\begin{figure*}[t]
    \centering
    \includegraphics[width=0.9\linewidth]{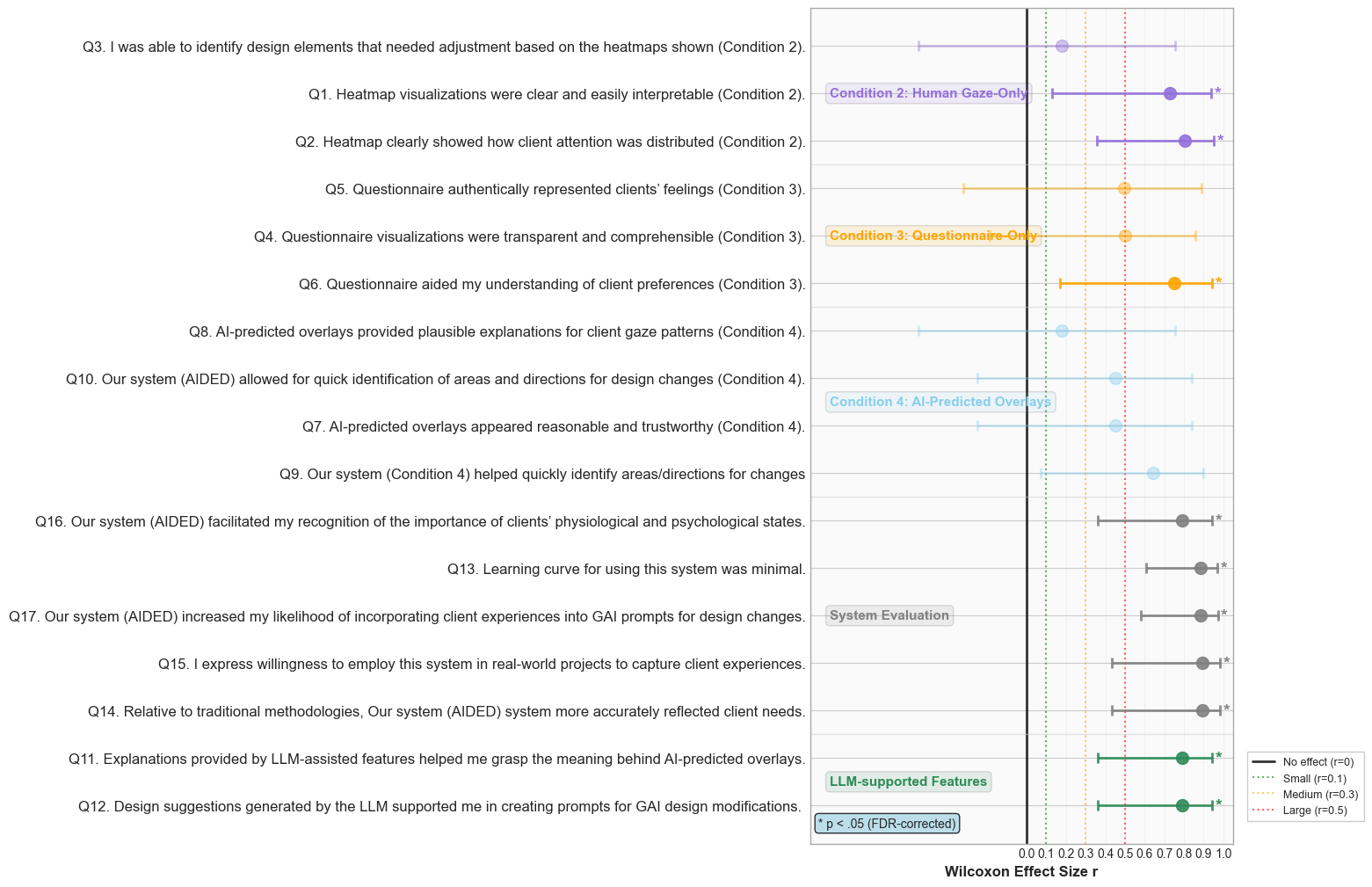}
    \caption{Wilcoxon signed-rank effect sizes ($r$) with 95\% confidence intervals for the 17 overall evaluation items (Q1–Q17), tested against the neutral midpoint. Items are grouped by information modality and system feature: gaze heatmaps (Condition~2), questionnaire visualizations (Condition~3), AI-predicted overlays (Condition~4), LLM-supported features, and system-level impressions. Gaze heatmaps (Q1–Q3) showed large effects for perceptual clarity and attention distribution (Q1–Q2), but only small effects for actionable interpretability (Q3). Questionnaire visualizations (Q4–Q6) yielded medium-to-large effects, with Q6 (understanding client preferences) reaching statistical significance. AI-predicted overlays (Q7–Q10) demonstrated small-to-moderate effects, with relatively wide confidence intervals, indicating cautious reliance on AI-mediated attention cues. In contrast, LLM-supported features (Q11–Q12) and system-level evaluations (Q13–Q17) consistently exhibited large effect sizes, indicating low learning cost, strong support for integrating client experience into GAI prompting, and high perceived adoption potential. Asterisks indicate items with FDR-corrected $p < .05$. Vertical reference lines mark benchmarks for no effect ($r=0$), small ($r=0.1$), medium ($r=0.3$), and large ($r=0.5$) effects.}
    
    %\caption{Wilcoxon effect sizes ($r$) with 95\% confidence intervals across 17 overall evaluation items. Gaze heatmaps (Q1–Q2) demonstrated large effects for clarity ($r \approx .70$ to $.80$) but limited interpretability (Q3, $r \approx .20$). Questionnaire visualizations (Q4–Q6) produced medium-to-large effects ($r \approx .50 $ to $.75$), with Q6 (preference understanding) reaching significance. AI-predicted overlays (Q7–Q10) yielded only small-to-moderate effects ($r \approx .40$ to $.65$) with wide CIs, reflecting participants’ cautious trust. \add{All system evaluation(Q13-Q17) showed large effect size.}The strongest outcomes were observed for LLM features (Q11–Q12), which showed consistently large effects ($r \approx .80$), \add{highlighting that AIDED is easy to learn, supports supports client-experience integration, and natural language mediation (LLM explanations) as a transformative mechanism for rendering client data interpretable and actionable. * indicates $p_{\text{adj}} < .05$. }}

    \Description{Forest plot summarizing Wilcoxon effect sizes ($r$) for 17 overall evaluation items. Each horizontal row corresponds to one questionnaire item (Q1–Q17), labeled with its statement and associated condition or feature. Dots indicate effect size estimates and horizontal bars denote 95\% confidence intervals. Vertical reference lines mark benchmarks for no effect ($r=0$), small ($r=0.1$), medium ($r=0.3$), and large ($r=0.5$) effects. Gaze heatmap items (Condition~2) cluster around large effects for clarity but small effects for actionable design identification. Questionnaire visualizations (Condition~3) show medium-to-large effects with narrower confidence intervals. AI-predicted overlays (Condition~4) display modest effect sizes with greater uncertainty. LLM-supported features and system-level items appear far to the right, with large effects and confidence intervals that do not cross zero, visually emphasizing strong perceived interpretability, usability, and adoption readiness.}

    \label{fig:overall_eval}
\end{figure*}

The 17-item post-study questionnaire offers an in-depth assessment of designers' evaluations of both individual modalities and the overall system. Several items achieved significance after FDR correction, with large effect sizes ($r \approx .70$–$.90$)(Table~\ref{tab:overall_eval}). Fig.~\ref{fig:overall_eval} presents a summary of Wilcoxon effect sizes accompanied by 95\% confidence intervals, illustrating a stratified hierarchy of perceived influence among the system's features.

\subsubsection{System Usability and Adoption Potential.}
Designers judged the system as easy to learn and practical for professional use. Q13 (learning curve) received the highest mean rating ($M = 4.33$, $SD = 0.65$), with a large effect size ($r = .88$, $p_{\text{adj}} = .02$). Similarly, Q15 (willingness to employ the system in real projects) was rated favorably ($M = 3.92$, $SD = 0.90$, $r = .89$, $p_{\text{adj}} = .03$), indicating strong adoption potential. As one participant explained, \emph{``This system makes me more willing to try it in real projects''} (D011). These findings underscore the perceived professional relevance and feasibility of integrating AIDED into design workflows.

\subsubsection{Authenticity and Interpretability of Client Signals.}
Structured questionnaire data were considered both clear and authentic. Q6 (understanding client preferences) scored $M = 3.83$ ($SD = 0.94$), with a large effect size ($r = .75$, $p_{\text{adj}} = .05$), while Q5 (authenticity) received $M = 3.58$ ($SD = 1.31$, $r = .50$, n.s.). These results suggest that although not all items were statistically significant, designers consistently relied on questionnaire-derived insights as concrete representations of client needs. One remarked, \emph{``The questionnaire charts were the most direct, allowing me to capture the real needs''} (D001).

\subsubsection{Heatmaps as Intuitive but Limited.}
Eye-tracking heatmaps were judged highly legible but limited in interpretability. Q1 (clarity) scored $M = 3.92$ ($SD = 1.24$, $r = .73$, $p_{\text{adj}} = .05$), and Q2 (distribution of attention) $M = 4.17$ ($SD = 0.72$, $r = .89$, $p_{\text{adj}} = .02$), both significant. However, Q3 (ability to identify design elements needing adjustment) remained weak ($M = 3.17$, $SD = 1.03$, $r = .18$, n.s.). As one participant put it, \emph{``I can see where they looked, but I don’t know if it’s good or bad—I had to guess''} (D005). These findings confirm that while gaze visualizations were easy to read, they lacked explanatory depth.

Designers consistently described client information as supportive guidance rather than prescriptive instruction. Questionnaire visualizations (C3) were trusted for their efficiency in capturing client preferences (e.g., D001, D010), whereas raw heatmaps (C2), though visually clear, required interpretation to distinguish positive from negative feedback (D005, D012).

\subsubsection{Skepticism Toward AI-predicted Overlays.}
Evaluations of AI overlays (Q7–Q10) were more ambivalent. For example, Q9 (quick identification of design areas) yielded $M = 3.83$ ($SD = 1.03$), $r = .64$, $p_{\text{adj}} = .06$ (n.s.), while Q10 (precision of modifications) was even weaker ($M = 3.50$, $SD = 1.09$, $r = .45$, $p_{\text{adj}} = .23$). Although overlays were perceived as ``reasonable,'' they were consistently rated lower than questionnaire visualizations or mediated explanations. As one participant explained, \emph{``It looks convincing at first, but I still trust the questionnaire more''} (D010).

\subsubsection{Transformative Role of LLM Explanations.}
The strongest effects emerged for items involving LLM mediation. Q11 (explanations of heatmaps) received $M = 4.17$ ($SD = 0.94$, $r = .79$, $p_{\text{adj}} = .02$), and Q17 (integration of client experiences into prompts) $M = 4.17$ ($SD = 0.72$, $r = .89$, $p_{\text{adj}} = .02$), both highly significant. Q12 (system-generated suggestions) was also rated positively ($M = 4.17$, $SD = 0.94$, $r = .79$, $p_{\text{adj}} = .02$). Designers described these features not as ancillary add-ons but as core mechanisms for translating opaque client data into actionable insights. As one participant put it, \emph{``With AI translating the data into text, I immediately knew why clients looked at certain things''} (D002). LLM mediation—through explanations and suggestions—proved particularly effective in transforming ambiguous signals into understandable rationales and actionable recommendations (D002, D011). Nevertheless, some participants emphasized the importance of retaining access to the raw underlying data to avoid over-reliance on AI-generated text (D006).

\begin{table*}[t]
\centering
\caption{Overall evaluation items (Q1--Q17): descriptive statistics, effect sizes, and $p$-values (raw and FDR-corrected).}
\label{tab:overall_eval}
\small
\begin{tabular}{llcccccc}
\toprule
Item & Evaluation & $M$ (SD) & $r$ & $p$ & $p_{\text{adj}}$ & Sig. \\
\midrule
Q13 & SYS: Learning cost & 4.33 (0.65) & .88 & .001 & .02 & ** \\
Q14 & SYS: Accuracy vs. traditional & 3.92 (0.90) & .89 & .016 & .03 & ** \\
Q15 & SYS: Willingness to adopt & 3.92 (0.90) & .89 & .016 & .03 & ** \\
Q16 & SYS: Recognize client states & 4.17 (0.94) & .79 & .007 & .02 & ** \\
Q17 & SYS: Client experience integration & 4.17 (0.72) & .89 & .002 & .02 & ** \\
\midrule
Q1  & C2: Clarity & 3.92 (1.24) & .73 & .027 & .05 & * \\
Q2  & C2: Attention distribution & 4.17 (1.03) & .81 & .010 & .03 & ** \\
Q3  & C2: Hotspots guide changes & 3.17 (1.03) & .18 & .781 & .78 & n.s. \\
\midrule
Q4  & C3: Clarity & 3.58 (1.16) & .50 & .131 & .19 & n.s. \\
Q5  & C3: Authenticity & 3.58 (1.31) & .50 & .219 & .25 & n.s. \\
Q6  & C3: Preference understanding & 3.83 (0.94) & .75 & .027 & .05 & * \\
\midrule
Q9  & C4: Identify areas & 3.83 (1.03) & .64 & .036 & .06 & n.s. \\
Q10 & C4: Precision of modifications & 3.50 (1.09) & .45 & .186 & .23 & n.s. \\
Q7  & C4: Trustworthiness & 3.50 (1.09) & .45 & .186 & .23 & n.s. \\
Q8  & C4: Gaze explanation & 3.17 (1.03) & .18 & .781 & .78 & n.s. \\
\midrule
Q11 & LLM: Heatmap explanations & 4.17 (0.94) & .79 & .007 & .02 & ** \\
Q12 & LLM: Design suggestions & 4.17 (0.94) & .79 & .007 & .02 & ** \\
\bottomrule
\end{tabular}

\vspace{2pt}
{\footnotesize
\parbox{0.90\linewidth}{\raggedright
\textit{Note.} SYS = System-level Impressions; C2 = Gaze Heatmap; C3 = Questionnaire Visualizations; C4 = AI-predicted Overlays; LLM = LLM Support. $M$ denotes the mean and $SD$ the standard deviation.
Effect sizes are reported as $r = z/\sqrt{n}$.
Raw $p$ values are from one-sample Wilcoxon signed-rank tests against the neutral midpoint ($\mu = 3$).
Adjusted $p$ values ($p_{\text{adj}}$) were computed using the Benjamini--Hochberg false discovery rate (FDR) correction.
Significance codes: $^{*}p < .05$, $^{**}p < .01$, n.s.\ = not significant.}}

\end{table*}

\subsubsection{The Role of the AIDED System in Real-World Design Phases}

The qualitative interviews provided deeper insights into how designers envisioned the system’s integration into practice. Workflow alignment emerged as a central theme: many participants regarded the system as most valuable in the early to mid stages for clarifying client preferences and prioritizing edits (D007–D012), while overlays (C4) were seen as useful in later stages as a checklist to identify potential oversights (D001). At the same time, persistent skepticism about AI accuracy, rendering quality, and command control (D006–D010) suggested that, although richer information boosted confidence, it did not guarantee satisfaction with the final design outcomes.

Taken together, the findings reveal a clear hierarchy of evaluations. Designers trusted questionnaire data as authentic, relied on LLM mediation to make complex signals interpretable, and regarded the integrated workflow as feasible for real practice. By contrast, unmediated gaze data and opaque AI-predicted overlays held limited standalone value. The overall evaluation thus underscores that effective AI-assisted design depends on balancing authenticity (direct client input) with interpretability (language-based mediation), cultivating both trust and adoption in professional contexts.

\subsection{Designers Are Not Satisfied With Their Edited Design Outcome}
\begin{figure}[t]
    \centering
    \includegraphics[width=0.9\linewidth]{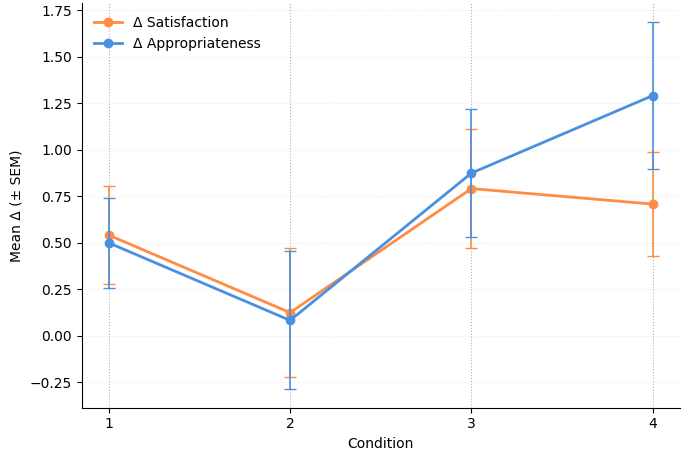}
    \caption{\add{Change in design satisfaction and perceived appropriateness across conditions (designer self-ratings). Line chart showing mean change scores for Satisfaction (orange) and Appropriateness (blue) between original and GAI-edited designs for each condition (C1–C4). Circles mark the condition means, and error bars indicate the standard error of the mean (SEM). Although Conditions 3 and 4 show numerically larger improvements, especially for Appropriateness, Friedman tests revealed no statistically significant differences across modalities.}}
    \Description{This figure is a two-line graph with Condition (1–4) on the x-axis and mean change (delta) on the y-axis. An orange line represents a change in Satisfaction, and a blue line represents a change in Appropriateness. Both lines start around 0.5 at Condition 1, dip near zero at Condition 2, then rise at Conditions 3 and 4. Appropriateness climbs slightly higher than Satisfaction in Conditions 3 and especially 4 (around 1.3). Vertical error bars show variability at each point. Visually, the pattern suggests that richer client data (Conditions 3–4) yielded larger perceived improvements, but the caption notes that these differences were not statistically significant.}
    
    \label{fig:before_after_delta}
\end{figure}

Although Fig.~\ref{fig:before_after_delta} illustrated apparent patterns, such as greater changes in satisfaction in Condition 4 and enhanced changes in appropriateness in Conditions 3 and 4, the results of the Friedman tests did not demonstrate statistically significant differences among the modalities. Specifically, for \textit{Delta\_Satisfaction}, the test yielded $\chi^{2}(3)=0.74$, $p=0.865$, with an effect size of $W=0.02$; for \textit{Delta\_Proper}, the values were $\chi^{2}(3)=4.44$, $p=0.218$, and $W=0.13$. These findings reflect negligible to small effect sizes. 

These findings indicate that designers did not view their edited results as significantly more satisfying or appropriate, regardless of the kind of client information provided. Qualitative feedback illuminates this outcome: participants often noted that although the system provided helpful cues, the final results frequently did not meet professional standards. As one designer stated, \emph{"Even if the AI adjusted the layout according to the client’s needs, I still felt it was just a draft, not something I would show to a client"} (D006). Another pointed out that changes based on overlays or questionnaires sometimes addressed only superficial issues, leaving deeper design problems unresolved and yielding only slight improvements in satisfaction.

This trend highlights a key limitation: having more detailed client data does not necessarily translate into perceived improvements in the quality of the final designs. Instead, designers appeared to rely on their professional judgment to bridge the gap between AI-assisted edits and acceptable deliverables, which may explain the minimal effect sizes observed in the quantitative results.

\subsection{Client Information Availability Shapes Designer-AI Co-Design Outcomes: A Novice Perspective}

\subsubsection{Design Satisfaction Varies with Client Information Availability}

\begin{figure*}[htbp]
\centering
\includegraphics[width=0.70\linewidth]{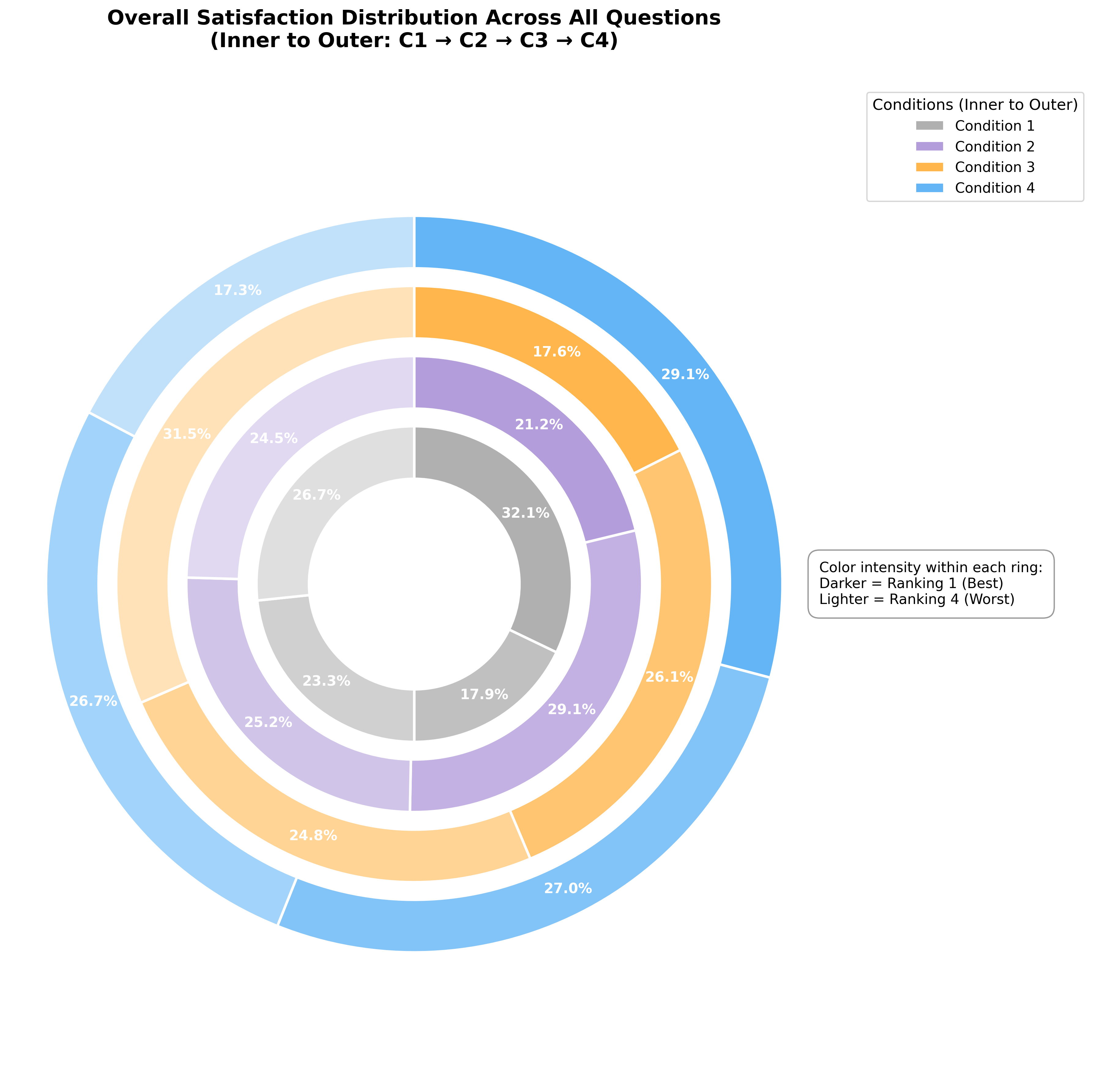}
\caption{\add{Distribution of novice satisfaction rankings across four experimental conditions in the Design Satisfaction Assessment (n = 30). The multi-ring donut plot summarizes rank distributions (1 = most satisfactory, 4 = least satisfactory) for each condition (inner to outer: C1 → C2 → C3 → C4); darker segments indicate higher satisfaction. Overall, Condition 4 (outer blue ring, full client data plus AI-predicted overlays) has more Rank 1 and Rank 2 ratings and fewer Rank 4 ratings than the other conditions, while Conditions 1–3 show a more mixed or lower distribution of rankings. Exact percentages for each rank are labeled on the chart.}}
\Description{The figure is a circular multi-ring donut composed of four concentric rings. From the center outward, the rings represent Conditions 1 (gray), 2 (purple), 3 (orange), and 4 (blue). Each ring is divided into four wedges corresponding to satisfaction ranks 1 through 4, with darker shades indicating higher ranks and lighter shades indicating lower ranks. The outer blue ring for Condition 4 shows larger dark segments for Ranks 1 and 2 and a smaller pale segment for Rank 4, whereas the inner rings for Conditions 1–3 have more evenly split or lighter segments. A side legend maps each color to a condition and notes that color intensity within each ring encodes ranking quality (darker = better).}

\label{fig:to_user_overall_Q1}
\end{figure*}

Analysis of the Design Satisfaction Assessment reveals distinct patterns in how the availability of client information affects perceived design quality among general novices. Fig. \ref{fig:to_user_overall_Q1} presents the satisfaction distribution across all four experimental conditions, based on 11 design sets evaluated by 30 participants. \add{A Friedman test revealed significant differences in ranking distributions across the four conditions ($\chi^2 = 15.38, p = .002$), indicating that the amount of client information systematically shaped how participants evaluated design satisfaction.}

\textbf{Condition 4 achieves the most consistent satisfaction.} When designers had access to complete client profiles (demographics, interviews, questionnaires, and AI-predicted attention maps), the resulting designs showed the most balanced satisfaction profile \add{with the lowest (i.e., most satisfaction) mean rank of 2.32 among all conditions}. The combined proportion of top-tier ratings (Rank 1: 29.1\% + Rank 2: 27.0\% = 56.1\%) indicates that comprehensive information enables designers to create reliably satisfactory outcomes when collaborating with generative AI. \add{Post-hoc pairwise Wilcoxon signed-rank tests with Holm–Bonferroni correction showed that Condition 4 received significantly better satisfaction rankings than Condition 3 $(p = .002, r = .67)$, suggesting that the additional layer of AI‑predicted attention maps improved participants’ perceived overall design quality compared to relying on questionnaire data alone.}
.

\textbf{Limited information produces polarized outcomes.} By contrast, Condition 1, where designers accessed only basic demographics, yielded the highest proportion of Rank 1 ratings at 32.1\% but also recorded 26.7\% Rank 4 ratings, indicating more polarized outcomes. Without detailed information to guide their decisions, designers must rely entirely on professional expertise and creative intuition when crafting AI prompts. While this approach can produce designs that appeal to some users, others may find the same designs unsuitable. The high variance suggests that creative freedom, though capable of producing exceptional designs, comes at the cost of predictable outcomes. \add{Wilcoxon tests indicated that Condition~1 was ranked significantly more favorably than Condition~3 ($p = .049$, $r = .47$), but did not differ significantly from Condition~4 ($p = .51$), implying that designs based on minimal demographics plus professional intuition can sometimes match the satisfaction of fully informed designs, albeit with greater variability.}

\subsubsection{Information Enrichment Progressively Improves Design-Client Alignment} 

\begin{table}[ht]
\centering
\caption{\add{Preference distribution for original versus generative-AI modified designs from the Design Appropriateness Assessment ($n = 26$). Modified designs were created by designers referencing different levels of client information when co-designing with generative AI.}}
\label{tab:q2_appropriateness}

\begin{tabular}{lccc}
\toprule
\add{\textbf{Condition}} & \add{\textbf{Prefer Before}} & \add{\textbf{Prefer After}} & \add{\textbf{\textit{p}-value}} \\
\midrule
\add{C1} & \add{63.2\%} & \add{36.8\%} & \add{$<$.001*} \\
\add{C2} & \add{52.2\%} & \add{47.8\%} & \add{.092} \\
\add{C3} & \add{40.7\%} & \add{59.3\%} & \add{.115} \\
\add{C4} & \add{45.1\%} & \add{54.9\%} & \add{.681} \\
\bottomrule
\end{tabular}
\vspace{2pt}
{\footnotesize
\parbox{0.85\linewidth}{\raggedright
Percentages represent the mean proportion of scenarios where participants preferred the original (Before)
versus the designer–GAI co-designed (After) versions.
Participants evaluated which design better suited the described client profiles across seven scenarios per condition.
\textit{p}-values are from one-sample Wilcoxon signed-rank tests against the neutral baseline of 50\%. $^{*}$ indicates \textit{p} $< .001$.
}}
\end{table}

The Design Appropriateness Assessment demonstrates that comprehensive client information enables designers to create modifications better suited to client profiles. Table \ref{tab:q2_appropriateness} illustrates participants' judgments regarding which designs were more appropriate for the described client scenarios: the original designs or those modified through designer-GAI co-design. This evaluation was based on 7 scenario-based comparisons assessed by 26 participants. \add{A Friedman test on participant‑level preference proportions showed significant differences across conditions $(\chi^2 = 23.12, p < .001)$, indicating that the degree of client information systematically influenced judgments of design appropriateness.}

\textbf{Preference for modifications increases with information availability.} As shown in table \ref{tab:q2_appropriateness}, participants increasingly judged co-designed modifications as more appropriate for clients when designers had access to richer information. The proportion of participants selecting co-designed versions over originals progressed from Condition 1 (36.8\%) to Condition 2 (47.8\%), Condition 3 (59.3\%), and Condition 4 (54.9\%). This shift from minority to majority preference demonstrates that additional layers of client information progressively enhance designers' ability to guide generative AI toward client-appropriate modifications. \add{One‑sample Wilcoxon signed‑rank tests indicated that only Condition 1 significantly deviated from the 50\% neutrality point $(p < .001)$, with participants strongly favoring the original designs when designers had access to only minimal client information.}

\textbf{Qualitative data provides the greatest improvement.} The most substantial gain occurred between Conditions 1 and 2, where adding interview transcripts and eye-tracking visualizations improved preference rates by 11 percentage points. This improvement suggests that understanding clients' verbal reasoning and visual attention patterns fundamentally changes how designers approach prompt crafting. The results indicate that co-designed modifications in Condition 2 more closely reflected the described clients' lifestyles and preferences than those in Condition 1, which provided only basic demographic information. \add{However, pairwise Wilcoxon tests with Holm–Bonferroni correction showed that this improvement did not reach statistical significance $(p = .072, r = .43)$, although the moderate effect size suggests that qualitative insights may offer practically meaningful benefits for aligning design modifications with client characteristics.}

\textbf{Structured preferences enable targeted modifications.} Condition 3 achieved the highest preference for designer-GAI co-designed modifications at 59.3\%, representing an 11.5 percentage point improvement over Condition 2. This result indicates that questionnaire data provide particularly actionable guidance for designer-GAI co-design. Quantified aesthetic preferences across multiple dimensions (comfort, naturalness, complexity, among others) allow designers to formulate more precise prompts that directly address specific client needs and preferences. \add{Pairwise Wilcoxon tests confirmed significant differences between Condition 3 and both Condition 1 $(p = .003, r = .67)$ and Condition 2 $(p = .015, r = .56)$, providing statistical evidence that structured preference data yields substantially better client‑aligned modifications than either minimal demographics or qualitative information alone. These findings imply that structured and quantified preference information provides clearer guidance for prompt formulation than qualitative insights alone.}

\textbf{AI-predicted attention maps show mixed results.} While Condition 4 maintained a majority preference for designer-GAI co-designed modifications at 54.9\%, this represents a 4.4 percentage point decrease from Condition 3. This reduction suggests that AI-predicted attention maps may introduce interpretive complexity not present with direct questionnaire responses. Nevertheless, the sustained majority preference for co-designed modifications over originals indicates that comprehensive information, including AI-predicted visual attention patterns, still enables effective design modifications through designer-GAI co-design. \add{Pairwise comparisons further showed that Condition 4 significantly outperformed Condition 1 $(p = .003, r = .68)$, but did not differ significantly from Condition 3 $(p = .42)$, suggesting that AI‑predicted attention maps neither substantially enhanced nor hindered client‑appropriateness beyond what structured questionnaire data already achieved. This suggests that attention maps may function best as supplementary cues rather than primary drivers of design decisions.}

\subsubsection{Trade-offs Between Creative Freedom and Design Reliability}

Comparing results from Design Satisfaction Assessment and Design Appropriateness Assessment reveals a fundamental trade-off in designer-GAI co-design: creative freedom versus client-specific reliability.

\textbf{Limited client information results in variable design outcomes.} The polarized satisfaction distribution (32.1\% Rank 1, 26.7\% Rank 4) combined with low appropriateness scores (36.8\% preference) indicates that designing without client information yields inconsistent results. While professional expertise and creative intuition can produce designs that some users find highly satisfactory, others deem them unsatisfactory. This variability extends to client appropriateness, where modifications lack the personalization needed to address specific client scenarios.

\textbf{Comprehensive information enables strategic design decisions.} Conditions 3 and 4 demonstrate improved design reliability, with both achieving majority preference for co-designed modifications in the appropriateness assessment (59.3\% and 54.9\% respectively). These modifications consistently suit the described clients but sacrifice the potential for exceptional satisfaction, as evidenced by their lower Rank 1 percentages. The shift from variable creativity to reliable personalization represents the core trade-off when introducing client information. At this phase, designers ceased to rely exclusively on their personal aesthetic preferences; despite the original design's visual appeal, they modified it to better align with client requirements (e.g., budgetary constraints) or contextual factors (e.g., alleviating spatial constraints). 

\textbf{Condition 4 balances satisfaction with appropriateness.} In particular, Condition 4 achieves both strong overall satisfaction (56.1\% combined Rank 1 and 2) and majority preference for co-designed modifications (54.9\%). This suggests that comprehensive information, including AI-predicted attention maps, enables designers to create modifications that satisfy general quality standards while remaining suitable for specific clients. The balanced performance indicates that full information availability may offer an optimal compromise between creative expression and client-centered reliability in designer-GAI co-design. As one designer reflected, “When there are too many choices, I’m unsure which path to take” (D008). Taken together, the results suggest that although all modalities are functionally equivalent with respect to outcome metrics, they entail different trade-offs between creative freedom and reliable personalization.

\section{Discussion}

Our study investigated how multimodal client data influences interior designers’ collaboration with GAI. First, we identify the authenticity–interpretability tradeoff as a central challenge in client-informed generative design: raw signals may be authentic but difficult to interpret, while structured representations increase clarity but risk oversimplification. Second, we find that client data operates as supportive rather than prescriptive input, providing scaffolding for better decisions while preserving designers’ professional control. Third, our results reveal a tension between creative freedom and design reliability: minimal data enables open-ended exploration but yields inconsistent outcomes. In contrast, richer data promotes appropriateness and stability at the cost of some variability. Together, these insights allow us to revisit our research questions and hypotheses, and to situate our contributions within the broader HCI discourse on human–data interaction, co-design, and GAI \cite{CollabDiffusion2023,RoomDreaming2024, UXPerceptions2024}.

\subsection{The Authenticity–Interpretability Tradeoff}

A central insight from this study is that greater availability of comprehensive client data does not, in itself, lead to improved design outcomes. Instead, designers consistently navigated a tradeoff between the authenticity of the data signals and their interpretability. For instance, raw gaze heatmaps (Condition 2) captured highly authentic behavioral information; however, designers found these visualizations ambiguous, as exemplified by one participant’s remark: “I can see where they looked, but I don’t know if it’s good or bad” (D005). In contrast, structured questionnaire visualizations (Condition 3) offered more interpretable and actionable insights, which designers regarded as the most reliable representation of client perspectives. AI-predicted attention maps (Condition 4) showed promise for scalability but elicited skepticism when presented without accompanying explanations, underscoring that interpretability is as critical as authenticity.

These findings directly address RQ1 and provide empirical support for H1 and H2: designers considered structured or AI-augmented data modalities more useful for decision-making, contingent on the presence of interpretability aids, such as explanations generated by LLMs. This trade-off builds on prior research in eye-tracking visualization and personal informatics, which has shown that raw behavioral data are often underutilized without appropriate mediation \cite{VideoBehaviorVis2011, CognitivePI2023, AIDataImages2024}. Our contribution lies in articulating this tradeoff as a foundational principle for client-informed co-design, reframing the challenge from merely acquiring more data to achieving a balance between authenticity and interpretability that effectively facilitates design action \cite{rekimoto2025gazellm, yan2024voila, GazeDashboard2025}.

\subsection{Client Data as Facilitative Rather Than Determinative}

Across experimental conditions, designers characterized client data as supportive frameworks that aided, but did not prescribe, creative decisions. Quantitative analysis revealed that only one task evaluation metric—TE6 (“Information helped me make better design decisions”)—exhibited statistically significant variation across modalities, indicating that client data primarily influenced decision confidence without fundamentally altering the overall design process. Qualitative feedback further highlighted that the data “helped me judge the direction more quickly, but it did not tell me exactly how to change” (D002).

This finding addresses RQ2 and partially corroborates H3. Specifically, client data improved decision-making efficacy while preserving professional autonomy and creative control, consistent with the effect sizes observed in TE13 and TE14. These results contribute to the HCI literature on human–AI co-creation, which has frequently expressed concerns that AI systems may supplant human expertise \cite{UXPerceptions2024, TeamsPrompting2024, MetacognitiveAI2024}. Our study reveals an alternative dynamic: when client inputs are integrated thoughtfully, AI can enhance professional judgment without imposing prescriptive solutions. This perspective aligns with prior research emphasizing explainability and transparency \cite{Explainability2022, DesignWeaver2025}, demonstrating that such mediation promotes user trust without compromising agency.

\subsection{Creative Autonomy versus Design Consistency}

The third key finding concerns the balance between creative autonomy and the reliability of design outcomes. Designers’ self-assessments revealed no statistically significant differences across experimental conditions, suggesting that their satisfaction and perceived alignment with client needs remained relatively consistent irrespective of the availability of client data. This observed stability may indicate underlying variability in the degree of trust or confidence placed in the quality of outputs produced by GAI.
However, novice evaluators' assessments showed a contrasting pattern: designs created under Condition 1 (demographic data only) elicited polarized responses, receiving both the highest and lowest satisfaction rankings. Conversely, conditions involving more comprehensive data modalities (Conditions 3 and 4) yielded designs that novices consistently deemed more appropriate to client profiles, albeit with fewer instances of exceptional ratings. This discrepancy directly addresses \add{R3} and \add{H4}, underscoring that while professional heuristics enable designers to mitigate the limitations of sparse data, external evaluators, novices in this study, place greater value on the consistency and relevance afforded by richer client information. This observation aligns with broader human-computer interaction literature concerning the trade-offs between variability and predictability in generative systems \cite{suh2025storyensemble, FashioningExpertise2024, SketchVsPrompt2024}. Yet it extends these discussions to a professional design context, where the consequences of misalignment are significant. Rather than advocating exclusively for either creative freedom or reliability, our findings suggest that the utility of client data resides in facilitating an effective equilibrium between these two dimensions.

\subsection{Implications for AI-Mediated Spatial Design Tools}

The present study reveals a notable deficiency in contemporary AI-assisted design methodologies: although computational tools effectively simulate physical performance metrics, such as lighting, ventilation, and circulation, they seldom incorporate end-user experiential data \add{\cite{salingaros2025living, gregorians2025integrating, Sjovall2025NDIX}}, including attention, preferences, or affective responses, in a manner that is both genuine and interpretable \cite{AttentionSurvey2024, EyeTrackedColor2022}. The AIDED framework exemplifies how such experiential signals can be systematically converted into actionable design support for professionals. Building upon this foundation, we identify three key implications for the development of future AI systems within creative disciplines:

\begin{enumerate}
    \item \textbf{Achieve a balance between authenticity and interpretability.} While designers benefit from access to authentic experiential signals, such as gaze data, these inputs must be complemented by interpretive frameworks, such as structured questionnaires, visualizations, or natural language explanations, to render them actionable.
    \item \textbf{Incorporate natural language mediation.} A pivotal feature identified in our study was the use of LLM-based interpretation. Designers consistently reported that linguistic explanations of gaze heatmaps and system-generated recommendations transformed abstract client data into concrete insights, thereby directly influencing their intentions to adopt the technology.
    \item \textbf{Maintain professional autonomy.} Participants valued client data as a supportive resource rather than a prescriptive mandate. Consequently, AI systems should present information as scaffolding that facilitates professional judgment rather than as directives that supplant expertise. This approach aligns with established HCI concerns regarding agency and autonomy in human–AI collaboration \cite{MetacogAgents2025, Explainability2022}.
\end{enumerate}

Collectively, these implications reconceptualize the role of GAI in design practice. Rather than serving solely as a “style engine” that produces superficial variations, AIDED demonstrates how end-user experiential data can be systematically integrated into design workflows, enabling AI to function as a mediator of human experience. This paradigm shift suggests the potential for future systems to embed client perspectives more profoundly within generative processes while preserving designer creativity and professional judgment \cite{AnalogyAI2024, DesignWeaver2025}.

\subsection{Limitations and Prospects for Future Research}

\subsubsection{Sample Size and Task Context.}  
The present investigation involved 12 design professionals and examined 4 room categories constrained by predefined stylistic parameters, thereby limiting the generalizability of the findings. Subsequent research should broaden both participant diversity and task variety by recruiting individuals across a spectrum of firm sizes, levels of seniority, cultural backgrounds, and project budgets, and by incorporating a wider array of spatial typologies beyond the four residential scenarios studied here. It would be advantageous to employ experimental paradigms from a “white shell” scenario that simulates early-stage conceptual design from a blank slate.

\subsubsection{Accuracy and Interpretability of AI-Predicted Overlays.}  
The utility of AI-generated attention maps is inherently constrained by the validity of the underlying models and by designers' ability to interpret these visualizations effectively. Future iterations should incorporate mechanisms to communicate model confidence and potential failure modes, such as uncertainty intervals or counterfactual heatmaps, alongside concise “model cards” integrated within the user interface. Additionally, embedding explanatory layers that associate specific regions with concrete aesthetic attributes and training data evidence would enhance interpretability. Incorporating human-in-the-loop calibration, allowing designers to relabel or mask misleading regions, could iteratively refine overlay accuracy. 

\add{Our observations about the LLM-based interpretation feature are based on brief post-study exposure rather than a dedicated condition or controlled comparison, and should therefore be interpreted as exploratory. A more systematic evaluation of how such explanations affect design quality and workflow efficiency is an important direction for future work.}

\subsubsection{Client Sample Bias Toward Low-Budget, Student Demographics.}  
The client cohort predominantly comprised individuals with limited renovation budgets and student status, factors that likely influenced both designers’ strategic approaches and novices’ assessments of design appropriateness. To enhance external validity, future datasets should stratify participants by budget categories, life stages, and household compositions. They should include professional clients subject to more stringent constraints and multi-stakeholder approval processes. Analytically, examining effects within and across these strata (e.g., contrasting low- versus high-budget groups) can elucidate whether specific modalities, such as questionnaires, exhibit universal utility or context-dependent efficacy. 

\subsubsection{Extending Beyond Laboratory Settings.}  
Designers engage with client information differently across various phases of the design process, including briefing, conceptualization, schematic design, and design development. Future investigations should condition workflows on these stages and deploy AI-mediated design tools in authentic projects to observe negotiation dynamics, iteration timing, and barriers to adoption in situ. To facilitate sustainable experiential data collection in practice, it is imperative to develop ethics-oriented toolkits that enable designers to gather perceptual data autonomously. Such toolkits might include consent templates, concise questionnaires aligned with design-relevant dimensions, optional gaze or attention proxies, and ecological momentary assessment (EMA) or diary prompts—all implemented with privacy safeguards and data storage guidelines. Integrating these data-capture routines into familiar platforms (e.g., BIM or scene viewers) can reduce user burden while maintaining a focus on the end-user experience.

\add{Building on these directions, our current deployment primarily supports early-stage concept drafts, but future versions of AIDED could extend toward later design phases and more direct client participation. Rather than capturing client data only at intake, subsequent iterations could support longitudinal data collection (e.g., follow-up walkthroughs, in situ comfort check-ins, or post-occupancy feedback) that progressively refine the experiential profile. Selected visualizations and explanations could be made client-facing, enabling clients to annotate regions, flag mismatches, and negotiate trade-offs with designers. These richer, temporally extended traces could also serve as inputs to GAI models that directly propose design modifications based on evolving client profiles rather than solely on designer-authored prompts, raising new questions about how control and responsibility should be distributed between designers and AI.}

\subsubsection{Generative AI Instability and Control over Edits.}  
The inherent stochasticity of text-to-image editing complicates efforts to conduct fair comparisons and to ensure reproducibility in professional contexts (e.g., fine-tuned with interior design terms). Additionally, interaction techniques must extend beyond prompt-based inputs to include region-of-interest and multi-mask editing, geometry- or material-aware constraints, and before-and-after difference visualizations. These features empower designers to localize modifications and minimize unintended collateral changes, thereby improving efficiency.

\subsubsection{\add{Prompt Interpretation.}}
\add{Although our study captured how designers prompted the GAI system during real-time editing, we did not carry out full behavioral coding or a detailed linguistic analysis of their prompt strategies, as this was beyond the scope of the present work. To provide some transparency, we include several prompt logs in \add{the supplementary material}. A more systematic examination of how designers formulate, adapt, and negotiate prompts across different information modalities remains a topic for future research. Follow-up work could examine prompting patterns, how particular modalities influence how designers express or refine their requests, and how alternative input methods, such as sketching, selecting regions, or mixed-initiative corrections, might support or even replace text-based prompting in professional practice.}

In summary, expanding participant diversity, enhancing the validity and interpretability of AI-predicted heatmaps, diversifying client profiles, stabilizing GAI outputs, and situating research within authentic project stages will collectively strengthen both the scientific rigor and practical applicability of AI-augmented spatial design methodologies.

%GAI 生成的設計 以設計師主導的設計為學習 

\section{Conclusions}

This paper introduces AIDED, a workflow that incorporates multi-modal client data into GAI-assisted interior design. Through an empirical study involving 12 design professionals and four distinct data modalities, we derived three principal insights: the trade-off between authenticity and interpretability, the role of client data as a supportive rather than prescriptive element, and the necessity of balancing creative autonomy with design reliability. These findings contribute to the field of human-computer interaction by advancing understanding of human-data interaction and AI-augmented creativity, demonstrating how experiential signals can be systematically integrated into spatial design tools.
Our work re-conceptualizes GAI not merely as a stylistic generator but as a mediator of human experience. More broadly, we underscore the significance of embedding clients’ physiological and psychological data within architectural design practices. Future research in HCI should explore methodologies to guide designers in collecting, interpreting, and applying such data throughout the design process, fostering environments that more closely align with lived human experience.

\begin{acks}
This research was supported by the National Science and Technology Council (NSTC), Taiwan (NSTC112-2628-E-007-013-MY3). We sincerely thank the designers for their contributions to the architectural and interior design work. We also thank the participants for their time and engagement, and the anonymous reviewers for their valuable comments and constructive suggestions.
\end{acks}

\bibliographystyle{ACM-Reference-Format}
\bibliography{sample-base}

@inproceedings{selvaraju2017grad,
  title={Grad-cam: Visual explanations from deep networks via gradient-based localization},
  author={Selvaraju, Ramprasaath R and Cogswell, Michael and Das, Abhishek and Vedantam, Ramakrishna and Parikh, Devi and Batra, Dhruv},
  booktitle={Proceedings of the IEEE international conference on computer vision},
  pages={618--626},
  year={2017}
}

@inproceedings{lin2025shaping,
  title={Shaping the Future of Architectural Design Tools Through the HCI Paradigm and Collective Human-Machine Intelligence},
  author={Lin, Yang Chen and Chung, Wen-Yen and Chien, Chen-Ying and Su, Chien-Hui and Kuo, Po-Chih},
  booktitle={Proceedings of the Extended Abstracts of the CHI Conference on Human Factors in Computing Systems},
  pages={1--8},
  year={2025}
}

@misc{chien2026incorporatingeyetrackingsignalsmultimodal,
      title={Incorporating Eye-Tracking Signals Into Multimodal Deep Visual Models For Predicting User Aesthetic Experience In Residential Interiors}, 
      author={Chen-Ying Chien and Po-Chih Kuo},
      year={2026},
      eprint={2601.16811},
      archivePrefix={arXiv},
      primaryClass={cs.CV},
      url={https://arxiv.org/abs/2601.16811}, 
}

@inproceedings{DesignMindToolkit, series={eCAADe 2023}, title={The DesignMind toolkit}, volume={1}, ISSN={2684-1843}, url={http://dx.doi.org/10.52842/conf.ecaade.2023.1.051}, DOI={10.52842/conf.ecaade.2023.1.051}, booktitle={Proceedings of the 41st International Conference on Education and Research in Computer Aided Architectural Design in Europe (eCAADe)  [Volume 1]}, publisher={eCAADe}, author={Gath-Morad, Michal and Baur, Raphaël and Hölscher, Christoph}, year={2023}, pages={51–60}, collection={eCAADe 2023} }

@article{CognitiveBIM2022, title={Beyond the shortest-path: Towards cognitive occupancy modeling in BIM}, volume={135}, ISSN={0926-5805}, url={http://dx.doi.org/10.1016/j.autcon.2022.104131}, DOI={10.1016/j.autcon.2022.104131}, journal={Automation in Construction}, publisher={Elsevier BV}, author={Gath-Morad, Michal and Aguilar Melgar, Leonel Enrique and Conroy-Dalton, Ruth and Hölscher, Christoph}, year={2022}, month={mar}, pages={104131} }

@misc{MentalGen2024,
  doi = {10.48550/arXiv.2409.00962},
  url = {https://arxiv.org/abs/2409.00962},
  author = {Liu, Yijiang and Wang, Hui},
  title = {Mental-Gen: A Brain-Computer Interface-Based Interactive Method for Interior Space Generative Design},
  publisher = {arXiv},
  year = {2024},
  copyright = {arXiv.org perpetual, non-exclusive license}
}

@article{Coburn2020, title={Psychological and neural responses to architectural interiors}, volume={126}, ISSN={0010-9452}, url={http://dx.doi.org/10.1016/j.cortex.2020.01.009}, DOI={10.1016/j.cortex.2020.01.009}, journal={Cortex}, publisher={Elsevier BV}, author={Coburn, Alexander and Vartanian, Oshin and Kenett, Yoed N. and Nadal, Marcos and Hartung, Franziska and Hayn-Leichsenring, Gregor and Navarrete, Gorka and González-Mora, José L. and Chatterjee, Anjan}, year={2020}, month={may}, pages={217–241} }

@article{MBTI_AI_2024, title={Adaptive interior design method for different MBTI personality types based on generative artificial intelligence}, volume={3}, ISSN={2731-6726}, url={http://dx.doi.org/10.1007/s44223-024-00066-z}, DOI={10.1007/s44223-024-00066-z}, number={1}, journal={Architectural Intelligence}, publisher={Springer Science and Business Media LLC}, author={Huang, Zhaoxu}, year={2024}, month=jul }

@inproceedings{RoomDreaming2024, series={CHI ’24}, title={RoomDreaming: Generative-AI Approach to Facilitating Iterative, Preliminary Interior Design Exploration}, url={http://dx.doi.org/10.1145/3613904.3642901}, DOI={10.1145/3613904.3642901}, booktitle={Proceedings of the CHI Conference on Human Factors in Computing Systems}, publisher={ACM}, author={Wang, Shun-Yu and Su, Wei-Chung and Chen, Serena and Tsai, Ching-Yi and Misztal, Marta and Cheng, Katherine M. and Lin, Alwena and Chen, Yu and Chen, Mike Y.}, year={2024}, month={may}, pages={1–20}, collection={CHI ’24} }

@inproceedings{AIArchSurvey2023, series={CHI ’22}, title={AI beyond Deus ex Machina – Reimagining Intelligence in Future Cities with Urban Experts}, url={http://dx.doi.org/10.1145/3491102.3517502}, DOI={10.1145/3491102.3517502}, booktitle={CHI Conference on Human Factors in Computing Systems}, publisher={ACM}, author={Mlynar, Jakub and Bahrami, Farzaneh and Ourednik, André and Mutzner, Nico and Verma, Himanshu and Alavi, Hamed}, year={2022}, month={apr}, pages={1–13}, collection={CHI ’22} }

@article{VisibilityWayfinding2024, title={The role of strategic visibility in shaping wayfinding behavior in multilevel buildings}, volume={14}, ISSN={2045-2322}, url={http://dx.doi.org/10.1038/s41598-024-53420-6}, DOI={10.1038/s41598-024-53420-6}, number={1}, journal={Scientific Reports}, publisher={Springer Science and Business Media LLC}, author={Gath-Morad, Michal and Grübel, Jascha and Steemers, Koen and Sailer, Kerstin and Ben-Alon, Lola and Hölscher, Christoph and Aguilar, Leonel}, year={2024}, month=feb }

@inproceedings{ArchDecisionViz2010, series={ICSE ’10}, title={Improving understandability of architecture design through visualization of architectural design decision}, url={http://dx.doi.org/10.1145/1833335.1833348}, DOI={10.1145/1833335.1833348}, booktitle={Proceedings of the 2010 ICSE Workshop on Sharing and Reusing Architectural Knowledge}, publisher={ACM}, author={Shahin, Mojtaba and Liang, Peng and Khayyambashi, Mohammad Reza}, year={2010}, month={may}, pages={88–95}, collection={ICSE ’10} }

@inproceedings{VideoBehaviorVis2011, series={CHI ’11}, title={Evaluating video visualizations of human behavior}, url={http://dx.doi.org/10.1145/1978942.1979155}, DOI={10.1145/1978942.1979155}, booktitle={Proceedings of the SIGCHI Conference on Human Factors in Computing Systems}, publisher={ACM}, author={Romero, Mario and Vialard, Alice and Peponis, John and Stasko, John and Abowd, Gregory}, year={2011}, month={may}, pages={1441–1450}, collection={CHI ’11} }

@article{EyeTrackedColor2022, title={Creative and Progressive Interior Color Design with Eye-tracked User Preference}, volume={30}, ISSN={1557-7325}, url={http://dx.doi.org/10.1145/3542922}, DOI={10.1145/3542922}, number={1}, journal={ACM Transactions on Computer-Human Interaction}, publisher={Association for Computing Machinery (ACM)}, author={Guo, Shihui and Shi, Yubin and Xiao, Pintong and Fu, Yinan and Lin, Juncong and Zeng, Wei and Lee, Tong-Yee}, year={2023}, month={feb}, pages={1–31} }

@article{PrefTunedInterior2024, title={Creating spatial visualizations using fine-tuned interior design style models informed by user preferences}, volume={62}, ISSN={1474-0346}, url={http://dx.doi.org/10.1016/j.aei.2024.102686}, DOI={10.1016/j.aei.2024.102686}, journal={Advanced Engineering Informatics}, publisher={Elsevier BV}, author={Lee, Jin-Kook and Jeong, Hyun and Kim, Youngchae and Cha, Seung Hyun}, year={2024}, month={oct}, pages={102686} }

@inproceedings{CognitivePI2023, series={MobileHCI ’23}, title={The Future of Cognitive Personal Informatics}, url={http://dx.doi.org/10.1145/3565066.3609790}, DOI={10.1145/3565066.3609790}, booktitle={Proceedings of the 25th International Conference on Mobile Human-Computer Interaction}, publisher={ACM}, author={Schneegass, Christina and Wilson, Max L and Maior, Horia A. and Chiossi, Francesco and Cox, Anna L and Wiese, Jason}, year={2023}, month={sep}, pages={1–5}, collection={MobileHCI ’23} }

@inproceedings{AIDataImages2024, series={CHI ’25}, title={Reimagining Personal Data: Unlocking the Potential of AI-Generated Images in Personal Data Meaning-Making}, url={http://dx.doi.org/10.1145/3706598.3713722}, DOI={10.1145/3706598.3713722}, booktitle={Proceedings of the 2025 CHI Conference on Human Factors in Computing Systems}, publisher={ACM}, author={Park, Soobin and Kim, Hankyung and Lim, Youn-kyung}, year={2025}, month={apr}, pages={1–25}, collection={CHI ’25} }

@misc{AttentionSurvey2024,
  doi = {10.48550/arXiv.2402.18673},
  url = {https://arxiv.org/abs/2402.18673},
  author = {Cartella, Giuseppe and Cornia, Marcella and Cuculo, Vittorio and D'Amelio, Alessandro and Zanca, Dario and Boccignone, Giuseppe and Cucchiara, Rita},
  title = {Trends, Applications, and Challenges in Human Attention Modelling},
  publisher = {arXiv},
  year = {2024},
  copyright = {Creative Commons Attribution Non Commercial Share Alike 4.0 International}
}

@inproceedings{GazeNoter2024, series={CHI ’25}, title={GazeNoter: Co-Piloted AR Note-Taking via Gaze Selection of LLM Suggestions to Match Users’ Intentions}, url={http://dx.doi.org/10.1145/3706598.3714294}, DOI={10.1145/3706598.3714294}, booktitle={Proceedings of the 2025 CHI Conference on Human Factors in Computing Systems}, publisher={ACM}, author={Tsai, Hsin-Ruey and Chiu, Shih-Kang and Wang, Bryan}, year={2025}, month={apr}, pages={1–22}, collection={CHI ’25} }

@inproceedings{GazeLog2024, series={ETRA ’25}, title={GazeLog: Optimizing Eye-Tracking with Fixation Keyframes \& LLM Insights}, url={http://dx.doi.org/10.1145/3715669.3726786}, DOI={10.1145/3715669.3726786}, booktitle={Proceedings of the 2025 Symposium on Eye Tracking Research and Applications}, publisher={ACM}, author={Mardanbegi, Diako and Hylands, Nicholas and Sarkar, Neil and Zahirovic, Nino}, year={2025}, month={may}, pages={1–8}, collection={ETRA ’25} }

@article{yan2024voila,
  title={Voila-a: Aligning vision-language models with user's gaze attention},
  author={Yan, Kun and Wang, Zeyu and Ji, Lei and Wang, Yuntao and Duan, Nan and Ma, Shuai},
  journal={Advances in Neural Information Processing Systems},
  volume={37},
  pages={1890--1918},
  year={2024}
}

@misc{GazeDashboard2025,
  doi = {10.48550/arXiv.2509.03741},
  url = {https://arxiv.org/abs/2509.03741},
  author = {Davalos, Eduardo and Zhang, Yike and Jain, Shruti and Srivastava, Namrata and Truong, Trieu and Haque, Nafees-ul and Van, Tristan and Salas, Jorge and McFadden, Sara and Cho, Sun-Joo and Biswas, Gautam and Goodwin, Amanda},
  title = {Designing Gaze Analytics for ELA Instruction: A User-Centered Dashboard with Conversational AI Support},
  publisher = {arXiv},
  year = {2025},
  copyright = {arXiv.org perpetual, non-exclusive license}
}

@inproceedings{DesignWeaver2025, series={CHI ’25}, title={DesignWeaver: Dimensional Scaffolding for Text-to-Image Product Design}, url={http://dx.doi.org/10.1145/3706598.3714211}, DOI={10.1145/3706598.3714211}, booktitle={Proceedings of the 2025 CHI Conference on Human Factors in Computing Systems}, publisher={ACM}, author={Tao, Sirui and Liang, Ivan and Peng, Cindy and Wang, Zhiqing and Palani, Srishti and Dow, Steven P.}, year={2025}, month={apr}, pages={1–26}, collection={CHI ’25} }

@inproceedings{SketchVsPrompt2024, series={CHI ’24}, title={The Impact of Sketch-guided vs. Prompt-guided 3D Generative AIs on the Design Exploration Process}, url={http://dx.doi.org/10.1145/3613904.3642218}, DOI={10.1145/3613904.3642218}, booktitle={Proceedings of the CHI Conference on Human Factors in Computing Systems}, publisher={ACM}, author={Lee, Seung Won and Jo, Tae Hee and Jin, Semin and Choi, Jiin and Yun, Kyungwon and Bromberg, Sergio and Ban, Seonghoon and Hyun, Kyung Hoon}, year={2024}, month={may}, pages={1–18}, collection={CHI ’24} }

@inproceedings{CollabDiffusion2023,
  author    = {Verheijden, Mathias Peter and Funk, Mathias},
  title     = {Collaborative Diffusion: Boosting Designerly Co-Creation with Generative AI},
  booktitle = {CHI EA '23: Extended Abstracts of the 2023 CHI Conference on Human Factors in Computing Systems},
  publisher = {Association for Computing Machinery},
  address   = {New York, NY, USA},
  year      = {2023},
  month     = {apr},
  pages     = {1--8},
  doi       = {10.1145/3544549.3585680},
  url       = {https://dl.acm.org/doi/10.1145/3544549.3585680}
}

@inproceedings{FashioningExpertise2024, series={CHI ’24}, title={Fashioning Creative Expertise with Generative AI: Graphical Interfaces for Design Space Exploration Better Support Ideation Than Text Prompts}, url={http://dx.doi.org/10.1145/3613904.3642908}, DOI={10.1145/3613904.3642908}, booktitle={Proceedings of the CHI Conference on Human Factors in Computing Systems}, publisher={ACM}, author={Davis, Richard Lee and Wambsganss, Thiemo and Jiang, Wei and Kim, Kevin Gonyop and Käser, Tanja and Dillenbourg, Pierre}, year={2024}, month={may}, pages={1–26}, collection={CHI ’24} }

@inproceedings{MVPrompt2024, series={CHI ’25}, title={MVPrompt: Building Music-Visual Prompts for AI Artists to Craft Music Video Mise-en-scène}, url={http://dx.doi.org/10.1145/3706598.3713876}, DOI={10.1145/3706598.3713876}, booktitle={Proceedings of the 2025 CHI Conference on Human Factors in Computing Systems}, publisher={ACM}, author={Lee, ChungHa and Lee, DaeHo and Hong, Jin-Hyuk}, year={2025}, month={apr}, pages={1–21}, collection={CHI ’25} }

@inproceedings{MetacognitiveAI2024, series={CHI ’24}, title={The Metacognitive Demands and Opportunities of Generative AI}, url={http://dx.doi.org/10.1145/3613904.3642902}, DOI={10.1145/3613904.3642902}, booktitle={Proceedings of the CHI Conference on Human Factors in Computing Systems}, publisher={ACM}, author={Tankelevitch, Lev and Kewenig, Viktor and Simkute, Auste and Scott, Ava Elizabeth and Sarkar, Advait and Sellen, Abigail and Rintel, Sean}, year={2024}, month={may}, pages={1–24}, collection={CHI ’24} }

@inproceedings{MetacogAgents2025, series={DIS ’25}, title={Exploring the Potential of Metacognitive Support Agents for Human-AI Co-Creation}, url={http://dx.doi.org/10.1145/3715336.3735785}, DOI={10.1145/3715336.3735785}, booktitle={Proceedings of the 2025 ACM Designing Interactive Systems Conference}, publisher={ACM}, author={Gmeiner, Frederic and Luo, Kaitao and Wang, Ye and Holstein, Kenneth and Martelaro, Nikolas}, year={2025}, month={jul}, pages={1244–1269}, collection={DIS ’25} }

@inproceedings{Explainability2022, series={FAccT ’22}, title={How Explainability Contributes to Trust in AI}, url={http://dx.doi.org/10.1145/3531146.3533202}, DOI={10.1145/3531146.3533202}, booktitle={2022 ACM Conference on Fairness Accountability and Transparency}, publisher={ACM}, author={Ferrario, Andrea and Loi, Michele}, year={2022}, month={jun}, pages={1457–1466}, collection={FAccT ’22} }

@inproceedings{AnalogyAI2024, series={CHI ’24}, title={Beyond Numbers: Creating Analogies to Enhance Data Comprehension and Communication with Generative AI}, url={http://dx.doi.org/10.1145/3613904.3642480}, DOI={10.1145/3613904.3642480}, booktitle={Proceedings of the CHI Conference on Human Factors in Computing Systems}, publisher={ACM}, author={Chen, Qing and Shuai, Wei and Zhang, Jiyao and Sun, Zhida and Cao, Nan}, year={2024}, month={may}, pages={1–14}, collection={CHI ’24} }

@inproceedings{TeamsPrompting2024, series={CHI ’24}, title={When Teams Embrace AI: Human Collaboration Strategies in Generative Prompting in a Creative Design Task}, url={http://dx.doi.org/10.1145/3613904.3642133}, DOI={10.1145/3613904.3642133}, booktitle={Proceedings of the CHI Conference on Human Factors in Computing Systems}, publisher={ACM}, author={Han, Yuanning and Qiu, Ziyi and Cheng, Jiale and LC, RAY}, year={2024}, month={may}, pages={1–14}, collection={CHI ’24} }

@inproceedings{UXPerceptions2024, series={CHI ’24}, title={User Experience Design Professionals’ Perceptions of Generative Artificial Intelligence}, url={http://dx.doi.org/10.1145/3613904.3642114}, DOI={10.1145/3613904.3642114}, booktitle={Proceedings of the CHI Conference on Human Factors in Computing Systems}, publisher={ACM}, author={Li, Jie and Cao, Hancheng and Lin, Laura and Hou, Youyang and Zhu, Ruihao and El Ali, Abdallah}, year={2024}, month={may}, pages={1–18}, collection={CHI ’24} }

@inproceedings{rekimoto2025gazellm,
  title={GazeLLM: Multimodal LLMs incorporating human visual attention},
  author={Rekimoto, Jun},
  booktitle={Proceedings of the Augmented Humans International Conference 2025},
  pages={302--311},
  year={2025}
}

@article{rezwana_designing_2023,
  author  = {Jeba Rezwana and Mary Lou Maher},
  title   = {Designing Creative {AI} Partners with {COFI}: A Framework for Modeling Interaction in Human--{AI} Co-Creative Systems},
  journal = {ACM Transactions on Computer-Human Interaction},
  year    = {2023},
  month   = oct,
  volume  = {30},
  number  = {5},
  pages   = {1--28},
  issn    = {1073-0516, 1557-7325},
  doi     = {10.1145/3519026},
  url     = {https://dl.acm.org/doi/10.1145/3519026}
}

@misc{lin_beyond_2023,
  author        = {Zhiyu Lin and Upol Ehsan and Rohan Agarwal and Samihan Dani and Vidushi Vashishth and Mark Riedl},
  title         = {Beyond Prompts: Exploring the Design Space of Mixed-Initiative Co-Creativity Systems},
  year          = {2023},
  howpublished  = {arXiv:2305.07465},
  archivePrefix = {arXiv},
  eprint        = {2305.07465},
  primaryClass  = {cs.AI},
  doi           = {10.48550/arXiv.2305.07465},
  url           = {https://arxiv.org/abs/2305.07465},
  note          = {Accepted at ICCC 2023}
}

@inproceedings{li_stage-based_2010,
  author    = {Ian Li and Anind Dey and Jodi Forlizzi},
  title     = {A Stage-Based Model of Personal Informatics Systems},
  booktitle = {Proceedings of the SIGCHI Conference on Human Factors in Computing Systems (CHI '10)},
  address   = {Atlanta, GA, USA},
  publisher = {ACM},
  year      = {2010},
  month     = apr,
  pages     = {557--566},
  isbn      = {978-1-60558-929-9},
  doi       = {10.1145/1753326.1753409},
  url       = {https://dl.acm.org/doi/10.1145/1753326.1753409}
}

@article{lavdas_eye-tracking_2024,
  author  = {Alexandros A. Lavdas},
  title   = {Eye-Tracking Applications in Architecture and Design},
  journal = {Encyclopedia},
  year    = {2024},
  month   = sep,
  volume  = {4},
  number  = {3},
  pages   = {1312--1323},
  issn    = {2673-8392},
  doi     = {10.3390/encyclopedia4030086},
  url     = {https://www.mdpi.com/2673-8392/4/3/86}
}

@article{karmann_saliency_2023,
  author  = {Caroline Karmann and Bahar Aydemir and Kynthia Chamilothori and Seungryong Kim and Sabine S{\"u}sstrunk and Marilyne Andersen},
  title   = {Saliency Prediction in 360{\textdegree} Architectural Scenes: Performance and Impact of Daylight Variations},
  journal = {Journal of Environmental Psychology},
  year    = {2023},
  month   = dec,
  volume  = {92},
  pages   = {102110},
  issn    = {0272-4944},
  doi     = {10.1016/j.jenvp.2023.102110},
  url     = {https://linkinghub.elsevier.com/retrieve/pii/S0272494423001585}
}

@inproceedings{fan_emotional_2018,
  author    = {Shaojing Fan and Zhiqi Shen and Ming Jiang and Bryan L. Koenig and Juan Xu and Mohan S. Kankanhalli and Qi Zhao},
  title     = {Emotional Attention: A Study of Image Sentiment and Visual Attention},
  booktitle = {2018 IEEE/CVF Conference on Computer Vision and Pattern Recognition (CVPR)},
  address   = {Salt Lake City, UT, USA},
  publisher = {IEEE},
  year      = {2018},
  month     = jun,
  pages     = {7521--7531},
  isbn      = {978-1-5386-6420-9},
  doi       = {10.1109/CVPR.2018.00785},
  url       = {https://ieeexplore.ieee.org/document/8578883/}
}

@article{yang2022continuous,
  title={Continuous gaze tracking with implicit saliency-aware calibration on mobile devices},
  author={Yang, Songzhou and Jin, Meng and He, Yuan},
  journal={IEEE Transactions on Mobile Computing},
  volume={22},
  number={10},
  pages={5816--5828},
  year={2022},
  publisher={IEEE}
}

@inproceedings{epstein_lived_2015,
  author    = {Daniel A. Epstein and An Ping and James Fogarty and Sean A. Munson},
  title     = {A Lived Informatics Model of Personal Informatics},
  booktitle = {Proceedings of the 2015 ACM International Joint Conference on Pervasive and Ubiquitous Computing (UbiComp '15)},
  address   = {Osaka, Japan},
  publisher = {ACM},
  year      = {2015},
  month     = sep,
  pages     = {731--742},
  isbn      = {978-1-4503-3574-4},
  doi       = {10.1145/2750858.2804250},
  url       = {https://dl.acm.org/doi/10.1145/2750858.2804250}
}

@article{victorelli_understanding_2020,
  author  = {Eliane Zambon Victorelli and Julio Cesar Dos Reis and Heiko Hornung and Alysson Bolognesi Prado},
  title   = {Understanding Human-Data Interaction: Literature Review and Recommendations for Design},
  journal = {International Journal of Human-Computer Studies},
  year    = {2020},
  month   = feb,
  volume  = {134},
  pages   = {13--32},
  issn    = {1071-5819},
  doi     = {10.1016/j.ijhcs.2019.09.004},
  url     = {https://linkinghub.elsevier.com/retrieve/pii/S1071581919301193}
}

@inproceedings{victorelli_human-data_2020,
  author    = {Eliane Zambon Victorelli and Julio Cesar Dos Reis},
  title     = {Human-Data Interaction Design Guidelines for Visualization Systems},
  booktitle = {Proceedings of the 19th Brazilian Symposium on Human Factors in Computing Systems (IHC '20)},
  address   = {Diamantina, Brazil},
  publisher = {ACM},
  year      = {2020},
  month     = oct,
  pages     = {1--10},
  isbn      = {978-1-4503-8172-7},
  doi       = {10.1145/3424953.3426511},
  url       = {https://dl.acm.org/doi/10.1145/3424953.3426511}
}

@inproceedings{sailaja_human-data_2021,
  author    = {Neelima Sailaja and Joseph Lindley and Lachlan Urquhart and Derek McAuley and Ian Forrester},
  title     = {Human-Data Interaction Through Design: An Explorative Step from Theory to Practice Using Design as a Vehicle},
  booktitle = {Extended Abstracts of the 2021 CHI Conference on Human Factors in Computing Systems (CHI EA '21)},
  address   = {Yokohama, Japan},
  publisher = {ACM},
  year      = {2021},
  month     = may,
  pages     = {1--5},
  isbn      = {978-1-4503-8095-9},
  doi       = {10.1145/3411763.3441344},
  url       = {https://dl.acm.org/doi/10.1145/3411763.3441344}
}

@article{higuera-trujillo_cognitive-emotional_2021,
  author  = {Juan Luis Higuera-Trujillo and Carmen Llinares and Eduardo Macagno},
  title   = {The Cognitive-Emotional Design and Study of Architectural Space: A Scoping Review of Neuroarchitecture and Its Precursor Approaches},
  journal = {Sensors},
  year    = {2021},
  month   = mar,
  volume  = {21},
  number  = {6},
  pages   = {2193},
  issn    = {1424-8220},
  doi     = {10.3390/s21062193},
  url     = {https://www.mdpi.com/1424-8220/21/6/2193}
}

@inproceedings{schneegass2023future,
  title={The Future of Cognitive Personal Informatics},
  author={Schneegass, Christina and Wilson, Max L and Maior, Horia A and Chiossi, Francesco and Cox, Anna L and Wiese, Jason},
  booktitle={Proceedings of the 25th International Conference on Mobile Human-Computer Interaction},
  pages={1--5},
  year={2023}
}

@inproceedings{aseniero2024experiential,
  title={Experiential views: Towards human experience evaluation of designed spaces using vision-language models},
  author={Aseniero, Bon Adriel and Lee, Michael and Wang, Yi and Zhou, Qian and Shahmansouri, Nastaran and Goldstein, Rhys},
  booktitle={Extended Abstracts of the CHI Conference on Human Factors in Computing Systems},
  pages={1--7},
  year={2024}
}

@article{duran2025review,
  title={A review on artificial intelligence applications for facades},
  author={Duran, Ayca and Waibel, Christoph and Piccioni, Valeria and Bickel, Bernd and Schlueter, Arno},
  journal={Building and Environment},
  volume={269},
  pages={112310},
  year={2025},
  publisher={Elsevier}
}

@inproceedings{gunduz2024proposal,
  title={A proposal of AI-powered HCI system to enhance spatial design creativity: InSpace},
  author={G{\"u}nd{\"u}z, Esra Nur and Karatoyun, Alper and Uyan, Bet{\"u}l and Erhan, Halil},
  booktitle={International Conference on Human-Computer Interaction},
  pages={307--323},
  year={2024},
  organization={Springer}
}

@article{CetinEr2024PlayPotentials,
  author  = {Çetin Er, Cansu and Özcan, Oğuzhan},
  title   = {Learning From Users: Everyday Playful Interactions to Support Architectural Spatial Changes},
  journal = {Proceedings of the ACM on Human-Computer Interaction},
  volume  = {8},
  number  = {CHI PLAY},
  year    = {2024},
  articleno = {320},
  doi     = {10.1145/3677085}
}

@article{lee2024creating,
  title={Creating spatial visualizations using fine-tuned interior design style models informed by user preferences},
  author={Lee, Jin-Kook and Jeong, Hyun and Kim, Youngchae and Cha, Seung Hyun},
  journal={Advanced Engineering Informatics},
  volume={62},
  pages={102686},
  year={2024},
  publisher={Elsevier}
}

@article{lee2021socio,
  title={Socio-spatial comfort: using vision-based analysis to inform user-centred human-building interactions},
  author={Lee, Bokyung and Lee, Michael and Zhang, Pan and Tessier, Alexander and Saakes, Daniel and Khan, Azam},
  journal={Proceedings of the ACM on Human-Computer Interaction},
  volume={4},
  number={CSCW3},
  pages={1--33},
  year={2021},
  publisher={ACM New York, NY, USA}
}

@article{paananen2021investigating,
  title={Investigating human scale spatial experience},
  author={Paananen, Ville and Oppenlaender, Jonas and Goncalves, Jorge and Hettiachchi, Danula and Hosio, Simo},
  journal={Proceedings of the ACM on Human-Computer Interaction},
  volume={5},
  number={ISS},
  pages={1--18},
  year={2021},
  publisher={ACM New York, NY, USA}
}

@article{jung2022domain,
  title={How domain experts work with data: Situating data science in the practices and settings of craftwork},
  author={Jung, Ju Yeon and Steinberger, Tom and King, John L and Ackerman, Mark S},
  journal={Proceedings of the ACM on human-computer interaction},
  volume={6},
  number={CSCW1},
  pages={1--29},
  year={2022},
  publisher={ACM New York, NY, USA}
}

@inproceedings{kim2023help,
  title={" help me help the ai": Understanding how explainability can support human-ai interaction},
  author={Kim, Sunnie SY and Watkins, Elizabeth Anne and Russakovsky, Olga and Fong, Ruth and Monroy-Hern{\'a}ndez, Andr{\'e}s},
  booktitle={proceedings of the 2023 CHI conference on human factors in computing systems},
  pages={1--17},
  year={2023}
}

@article{schelble2022see,
  title={I see you: Examining the role of spatial information in human-agent teams},
  author={Schelble, Beau G and Flathmann, Christopher and Musick, Geoff and McNeese, Nathan J and Freeman, Guo},
  journal={Proceedings of the ACM on Human-Computer Interaction},
  volume={6},
  number={CSCW2},
  pages={1--27},
  year={2022},
  publisher={ACM New York, NY, USA}
}

@inproceedings{nouraei2024thinking,
  title={Thinking Outside the Box: Non-Designer Perspectives and Recommendations for Template-Based Graphic Design Tools},
  author={Nouraei, Farnaz and Siu, Alexa and Rossi, Ryan and Lipka, Nedim},
  booktitle={Extended Abstracts of the CHI Conference on Human Factors in Computing Systems},
  pages={1--9},
  year={2024}
}

@inproceedings{earle2025dreamgarden,
  title={DreamGarden: A Designer Assistant for Growing Games from a Single Prompt},
  author={Earle, Sam and Parajuli, Samyak and Banburski-Fahey, Andrzej},
  booktitle={Proceedings of the 2025 CHI Conference on Human Factors in Computing Systems},
  pages={1--19},
  year={2025}
}

@inproceedings{hilgard2021learning,
  title={Learning representations by humans, for humans},
  author={Hilgard, Sophie and Rosenfeld, Nir and Banaji, Mahzarin R and Cao, Jack and Parkes, David},
  booktitle={International conference on machine learning},
  pages={4227--4238},
  year={2021},
  organization={PMLR}
}

@article{chang5357388inside,
  title={Inside Out: Genai and Human Perspectives on the Quality of Indoor and Outdoor Urban Environments},
  author={Chang, Ahyoung and Choi, Seung Jun and Jiao, Junfeng},
  journal={Available at SSRN 5357388}
}

@inproceedings{lee2019actoviz,
  title={ActoViz: a human behavior simulator for the evaluation of the dwelling performance of an atypical architectural space},
  author={Lee, Yun Gil},
  booktitle={International Conference on Human-Computer Interaction},
  pages={361--365},
  year={2019},
  organization={Springer}
}

@article{Sjovall2025NDIX,
  title   = {The Neurodesign/Neuroarchitecture Index (NDIX): Development of a Method to Evaluate the Impact of the Built Environment on Health, Cognitive Performance, and Wellbeing},
  author  = {Sj{\"o}vall, Isabelle and Chatterjee, Anjan and K{\"u}hn, Simone and Meyer-Lindenberg, Andreas and Eyre, Harris and Carlsson, Raul and Kounios, John and others},
  journal = {PsyArXiv},
  year    = {2025},
  month   = oct,
  doi     = {10.31234/osf.io/8369k_v2},
  url     = {https://doi.org/10.31234/osf.io/8369k_v2},
  note    = {Preprint}
}

@article{salingaros2025living,
  title={Living geometry, AI tools, and Alexander's 15 fundamental properties. Remodel the architecture studios!},
  author={Salingaros, Nikos A},
  journal={Frontiers of Architectural Research},
  year={2025},
  publisher={Elsevier}
}

@inproceedings{gregorians2025integrating,
  title={Integrating Human-Centric Knowledge into Architectural Design Practice: Insights from Industry},
  author={Gregorians, Lara and Aguilar Melgar, Leonel and Anklesaria, Freyaan and Jamaluddin, Azrin and Mavros, Panagiotis and Trivic, Zdravko and Zhong, Yuqin and H{\"o}lscher, Christoph},
  booktitle={Environmental Design Research Association (EDRA) 56th Conference},
  year={2025}
}

@article{postle2025llm,
  title={LLM and Pattern Language Synthesis: A Hybrid Tool for Human-Centered Architectural Design},
  author={Postle, Bruno and Salingaros, Nikos A},
  journal={Buildings},
  volume={15},
  number={14},
  pages={2400},
  year={2025},
  publisher={MDPI}
}

@inproceedings{suh2025storyensemble,
  title={StoryEnsemble: Enabling dynamic exploration \& iteration in the design process with AI and forward-backward propagation},
  author={Suh, Sangho and Lai, Michael and Pu, Kevin and Dow, Steven P and Grossman, Tovi},
  booktitle={Proceedings of the 38th Annual ACM Symposium on User Interface Software and Technology},
  pages={1--36},
  year={2025}
}

@inproceedings{naqvi2025catalyst,
  title={Catalyst for Creativity or a Hollow Trend?: A Cross-Level Perspective on The Role of Generative AI in Design},
  author={Naqvi, Syeda Masooma and He, Ruichen and Kaur, Harmanpreet},
  booktitle={Proceedings of the 2025 CHI Conference on Human Factors in Computing Systems},
  pages={1--16},
  year={2025}
}

@inproceedings{weisz2024design,
  title={Design principles for generative AI applications},
  author={Weisz, Justin D and He, Jessica and Muller, Michael and Hoefer, Gabriela and Miles, Rachel and Geyer, Werner},
  booktitle={Proceedings of the 2024 CHI Conference on Human Factors in Computing Systems},
  pages={1--22},
  year={2024}
}

\appendix
\section*{A. Formative Study Interview Protocol}

{Overview.} This protocol outlines the questions used to conduct expert interviews in our formative study. 
\begin{itemize}
\item \textbf{Welcome Script.} Thank the participant for joining the study, confirm consent and recording, and remind them: “There are no right or wrong answers; you may skip any question at any time. Please feel free to share your thoughts, experiences, and comments openly.” Clarify that this interview aims to gain deeper insights into four areas, including authentic design workflows, client communication practices, and professional perspectives on data- and AI-supported tools.

\end{itemize}

\subsection*{a. Background \& Workflow}
\begin{enumerate}
  \item \textbf{Role \& scope.} Please describe your role and years of experience. What types of projects do you take on (residential/commercial/public)? 
  \item \textbf{Typical workflow.} Walk me through a recent project from concept to client sign-off. \textit{Probe: deliverables at each stage; tools (2D/3D/CAD/T2I/LLMs).}
  \item \textbf{Bottlenecks.} At which stage do you most often encounter friction? \textit{Probe: why; examples; time lost.}
\end{enumerate}

\subsection*{b. Client Communication}
\begin{enumerate}
  \item \textbf{Eliciting needs.} Early in a project, how do you gather and interpret client needs/feelings? \textit{Probe: mood boards, reference albums, interviews, site visits.}
  \item \textbf{Clarity \& misalignment.} How clear are clients’ preferences in practice? \textit{Probe: common misunderstandings; examples of rework.}
  \item \textbf{Useful inputs.} Which client materials are most helpful? Any you once thought trivial but later found useful?
  \item \textbf{Residential priorities.} In homes, what do clients care about most (style, budget, materials, functionality)? \textit{Probe: how you surface trade-offs.}
\end{enumerate}

\subsection*{c. Iteration \& Feedback}
\begin{enumerate}
  \item \textbf{When clients can’t articulate.} What do you do if a client struggles to describe preferences or feelings? \textit{Probe: options testing; A/B renderings; living-with mockups.}
  \item \textbf{Reading signals.} What cues do you rely on (verbal, facial, body language, reactions to images/models)?
  \item \textbf{Trust calibration.} How do you balance stated preferences vs.\ professional judgment? \textit{Probe: when you override; how you justify.}
  \item \textbf{Feedback collection.} How do you collect, structure, and store feedback? \textit{Probe: forms, questionnaires, revision logs.}
\end{enumerate}

\subsection*{d. Attitudes of New Data Modalities \& AI}
\begin{enumerate}
  \item \textbf{Hypothetical data.} Imagine pairing 3D renders with (i) eye-tracking heatmaps and (ii) a brief questionnaire (comfort/brightness/preference). How would this change your revisions? \textit{Probe: prioritizing vs.\ validating changes.}
  \item \textbf{Use cases.} Where would such data help most: identifying issues, explaining choices to clients, or prompting GAI?
  \item \textbf{AI in workflow.} Which stage would you want AI to assist (ideation, variation, detailing, documentation)? Biggest benefits and current limitations?
  \item \textbf{Beyond CAD.} In what ways does AI-assisted design differ from traditional CAD for you? What feels genuinely new?
  \item \textbf{Ideal feature.} If you could design one AI feature, what would it do? \textit{Probe: inputs, outputs, constraints, control.}
\end{enumerate}

\subsection*{Wrap-Up}
\begin{itemize}
  \item Anything we didn’t ask that matters for your practice? Thanks and next steps.
\end{itemize}

\section*{B. Post-Study Interview Protocol}
\textbf{Overview.} Evaluate system experience, decision basis across conditions, and thoughts on using AIDED for real design workflows.
\begin{itemize}
  \item Script: 'We will review your experience using AIDED. Please think aloud and reference concrete moments.'
\end{itemize}

\subsection*{a. Overall Experience}
\begin{enumerate}
  \item \textbf{Impressions.} How would you summarize your experience using the system?
  \item \textbf{Workflow fit.} Which steps felt smooth vs.\ cumbersome or unintuitive? \textit{Probe: where you slowed down and why.}
  \item \textbf{Information load.} Was the amount of information too much, too little, or about right? \textit{Probe: specific panels or charts.}
  \item \textbf{Media quality.} Any issues with base images or video renders that affected decisions?
\end{enumerate}

\subsection*{b. Comparing Conditions}
(Conditions: C1 none, C2 eye gaze, C3 questionnaire charts, C4 AI-predicted overlays \& LLM features)
\begin{enumerate}
  \item \textbf{Impact on design.} Which condition most influenced your edits? Why?
  \item \textbf{Understanding needs.} Which condition best helped you infer client needs/feelings?
  \item \textbf{Decision efficiency.} Under which condition did you decide the fastest? \textit{Probe: concrete example of a change you made.}
  \item \textbf{Heatmaps.} Did eye gaze help you locate modification areas faster? In what way? When was it ambiguous? Why?
\end{enumerate}

\subsection*{c. Real-World Placement \& Value}
\begin{enumerate}
  \item \textbf{Best stage fit.} At which real project stage would you use this (concept, schematic, pre-final)? Why?
  \item \textbf{Precision \& rework.} Did AIDED help you identify change areas faster and make more precise revisions? Did it reduce unnecessary changes?
  \item \textbf{Design communication.} Did system information help you articulate client needs/feelings to GAI (prompting) and to clients?
  \item \textbf{Decision integration.} How easy was it to integrate the system’s information into your final choices?
  \item \textbf{Feature requests.} What would you add or change to better fit professional work?
  \item \textbf{Recommendation.} If recommending AIDED to peers, how would you describe its value?
\end{enumerate}

\subsection*{d. Designer Experience Dimensions}
\begin{enumerate}
  \item \textbf{Cognitive load.} Which parts required the most mental effort? \textit{Probe: specific panels/interactions.}
  \item \textbf{Creativity.} Did the system spark new ideas or variations? Example?
  \item \textbf{Trust.} Which information formats increased your willingness to rely on them? Which reduced trust?
  \item \textbf{Understanding needs.} Which modality best captured true client needs for you?
  \item \textbf{Outcome confidence.} Which revisions were you most confident in, and why? Which were less so?
  \item \textbf{Sense of control.} When did AI feel like it was “following” your intent vs.\ “drifting”? \textit{Probe: prompt adjustments you made.}
\end{enumerate}

\section*{C. Preliminary Survey}

\subsection*{a. Participant Identification} 

\begin{enumerate}
    \item On average, how many iterations of communication with clients are necessary to complete a single project? 
    \begin{itemize}
        \item 1–3 exchanges
        \item 3–5 exchanges
        \item 5–10 exchanges
        \item More than 10 exchanges
        \item Other
    \end{itemize}
    \item Typically, what is the duration between submitting a design proposal and receiving feedback from the client? 
    \begin{itemize}
        \item 1–2 days
        \item Approximately one week
        \item More than two weeks
        \item Other
    \end{itemize}
    \item Before initiating a design, which methods do you employ to ascertain the client’s requirements? (Select two) 
    \begin{itemize}
        \item Observation
        \item Interview
        \item Questionnaire
        \item Other
    \end{itemize}
    \item During or following the design process, which approaches do you utilize to comprehend client feedback? (Select two) 
    \begin{itemize}
        \item Observation
        \item Interview
        \item Questionnaire
        \item Other
    \end{itemize}
    \item Please indicate your level of agreement with the following statement:\begin{quote} The speed at which clients comprehend the design proposal is a critical factor. 
    \end{quote}

    (1 = Strongly Disagree, 5 = Strongly Agree)
    
    \item Visualization tools can reduce the frequency of client communications. 

    \item Producing multiple design variants within constrained timeframes is essential. 
 
    \item Clients generally require comprehensive visual aids to facilitate decision-making. 
    
    \item I am open to adopting new design tools to enhance workflow efficiency. 
   
    \item The client’s budget significantly influences the quality of the design outcome. 
 
    \item Relying exclusively on my design experience is adequate for making sound design decisions. 

    \item I am accustomed to designing in the absence of client data. 

    \item I can intuitively assess the client’s experience with a design. 

    \item In most cases, the client’s experience aligns with my design expectations. 
    
   \textit{Note. Question (6)-(14) were rated on a 5-point Likert scale.}
\end{enumerate}

\subsection*{b. Experience with Generative AI (GAI)}

\begin{enumerate}
        \item Which of the following Generative AI (GAI) tools have you utilized? (Select all that apply) 
    \begin{itemize}
        \item Text-to-Image (T2I) tools (e.g., Midjourney, Stable Diffusion, DALL·E)
        \item Large Language Models (LLMs) (e.g., ChatGPT, Claude)
        \item Generative AI tools specific to interior design (e.g., ReRoom, RoomDreaming, RoomGPT)
        \item None
    \end{itemize}
    \item In which design-related activities have you incorporated GAI within your projects? (Select all that apply) 
    \begin{itemize}
        \item Concept ideation
        \item Rapid rendering
        \item Client communication
        \item Narrative generation
        \item Other
    \end{itemize}
    \item Please indicate your level of agreement with the following statements regarding GAI:  
    \begin{enumerate}
        \item I consider GAI to be beneficial for design tasks. 
        \item I am willing to continue employing GAI in future projects. 
        \item GAI enhances my work efficiency. 
        \item GAI poses a threat to the professional value of designers. 
        \item GAI facilitates communication with clients. 
    \end{enumerate}
    (1 = Strongly Disagree, 5 = Strongly Agree)
\end{enumerate}

\section*{D. Add-on LLM-generated interpretations feature}

For the LLM-generated interpretations, we prompted the model to act as an interior design assistant that helps designers interpret client psychological representation data (e.g., gaze heatmaps, questionnaire scores). The input consists of an image of the interior scene and a corresponding heatmap visualizing the AI-predicted gaze under a specific subjective dimension (e.g., “boring/interesting”). The prompt specifies the client’s rating on that dimension (e.g., leaning toward “boring”) and asks the model to produce a 150–200-word design report in Mandarin. The report is structured into three sections: Scene overview (summarizing atmosphere and functionality), AI-predicted Overlay interpretation (explaining 3–4 main heatmap regions and how they relate to the target dimension) and Design suggestions (three concrete action-oriented suggestions grounded in Pattern Language / Living Geometry principles, each with a brief rationale). The output is returned in a JSON format (see in supplementary material) that includes metadata fields (scene, style, task, client preference) and a nested object for the three report sections.

\section*{E. Sample Questions for Evaluating Design Outputs (Novice)}

\subsection*{1. Design Satisfaction Assessment} 
\begin{figure}[ht]
  \centering
  \includegraphics[width=0.65\linewidth]{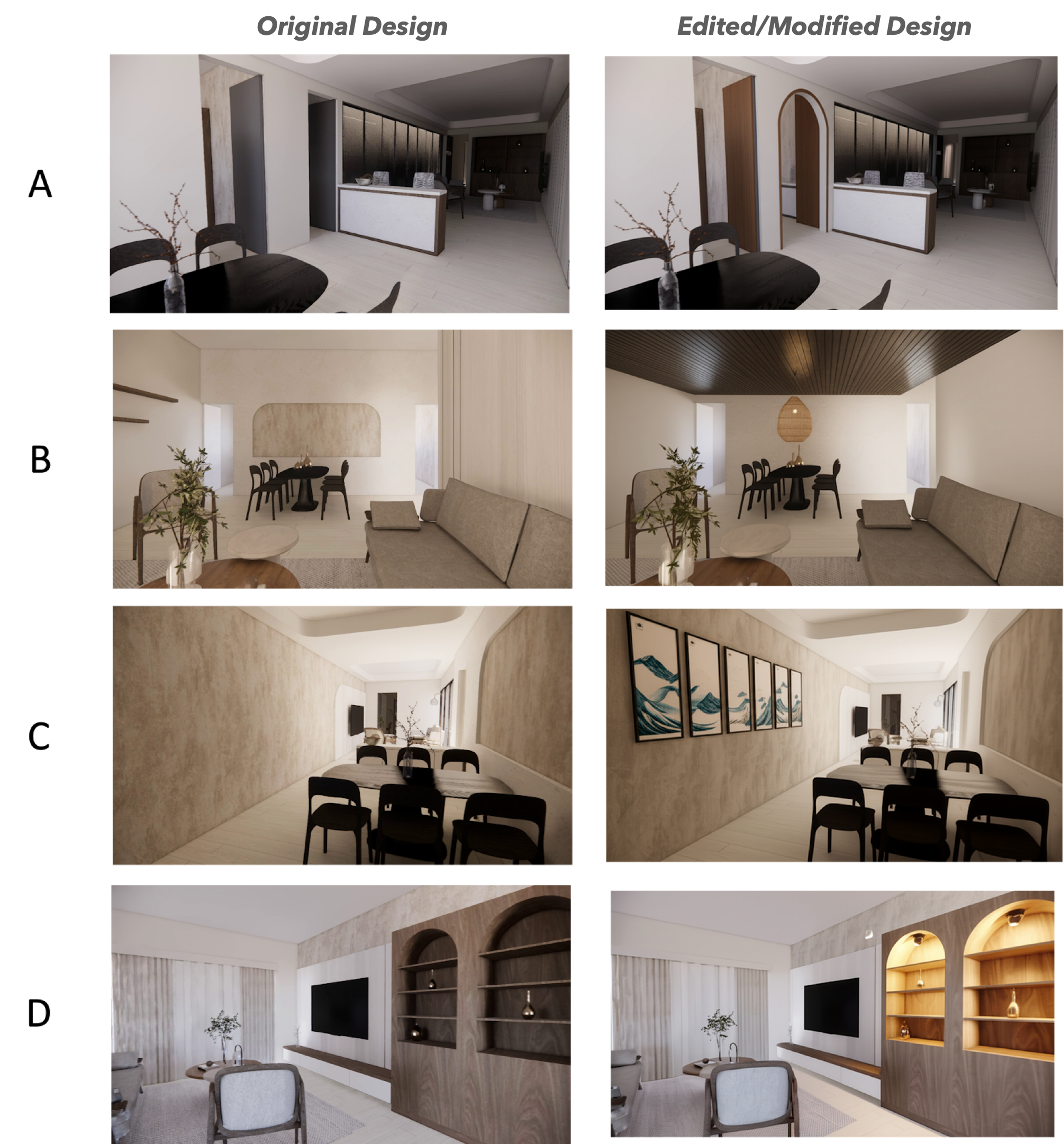}
  \caption{Sample questions used in Design 
  Satisfaction Assessment.}
  \label{fig:novice_DO1_example}

  \Description{The figure shows representative interior design images used in the study, illustrating how participants were asked to assess edited designs with respect to satisfaction.}
\end{figure}

See Fig.~\ref{fig:novice_DO1_example} as an example. Please rank your satisfaction with the edited designs for the following scene only. Use ranks A, B, C, and D, assigning 1 to the most satisfying and 4 to the least enjoyable.
\textit{Note. Participants do not know the condition.}

\newpage
\subsection*{2. Design Appropriateness Assessment}
\begin{figure}[ht]
  \centering
  \includegraphics[width=0.65\linewidth]{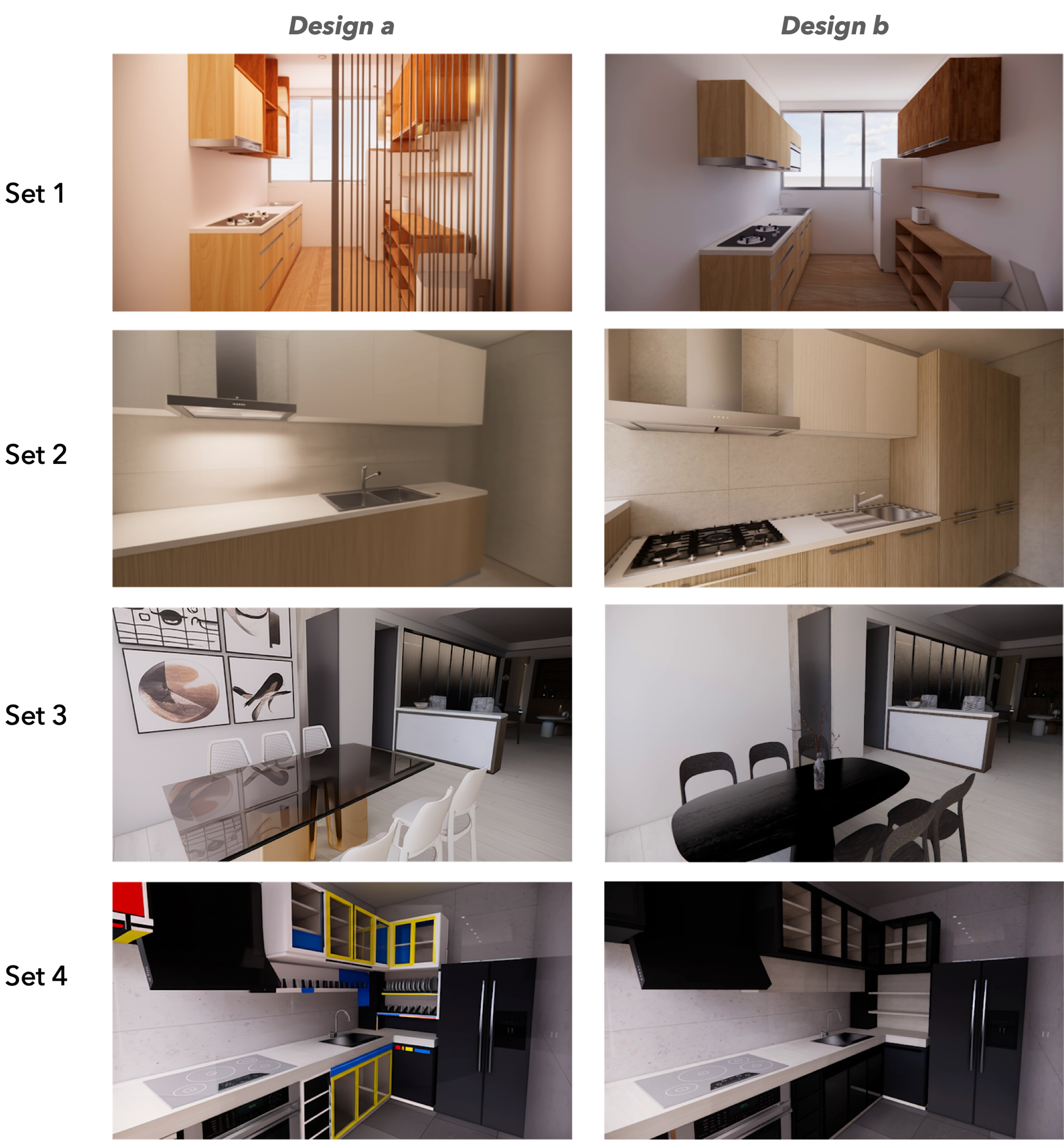}
  \caption{Sample questions used in Design Appropriateness Assessment.}
  \label{fig:novice_D02_example}
  \Description{The figure shows representative interior design images used in the study, illustrating how participants were asked to assess edited designs for suitability with a given client profile.}
  
\end{figure}

See Fig.~\ref{fig:novice_D02_example} as an example. Please read the following client descriptions and, for each case, select which interior design is more suitable. 

\textit{\textbf{Description.}}
\textit{Client 1 (Anonymous)} A 27-year-old male student with a moderate budget. He prefers wooden interiors, natural light, a spacious study area, and built-in appliances.  

\begin{itemize}
  \item Design a is more suitable for the client
  \item Design b is more suitable for the client
\end{itemize}

\textit{Note. Participants are provided with client profiles and asked to select the design (a or b) that best suits the client's needs. Each question contains four sets of identical scene categories. Participants are unaware of the condition or which image (a or b) has been edited.}

\end{document}